%% file: article_smg_overdensity.tex
\documentclass[twocolumn]{aastex701}
\usepackage{amsmath}
\usepackage{graphicx}
\usepackage{booktabs} 
\usepackage{threeparttable}
\usepackage{tabularx} 
\usepackage{pifont}
\usepackage{rotating}
\usepackage{float}
\usepackage{placeins}
\usepackage{caption}
\usepackage{ragged2e}
\usepackage{verbatim}
\usepackage{booktabs}

\usepackage{newtxtext,newtxmath}
\usepackage{bm}
\newcommand{\CIIline}{[C\,\textmd{\textsc{ii}}]}

\begin{document}

\title{Do Submillimeter Galaxies Trace Megaparsec Large-scale Structures? -- An Overdensity Analysis of 449 Submillimeter Galaxies in COSMOS}

\author[orcid=0009-0005-2363-8626]{Canze Deng}
\affiliation{Purple Mountain Observatory, Chinese Academy of Science, 10 Yuanhua Road, Nanjing 210023, China}
\affiliation{School of Astronomy and Space Sciences, University of Science and Technology of China, Hefei 230026, China}
\email{czdeng@pmo.ac.cn}

\correspondingauthor{czdeng@pmo.ac.cn, dzliu@pmo.ac.cn}
\author[orcid=0000-0001-9773-7479]{Daizhong Liu}
\affiliation{State Key Laboratory of Radio Astronomy and Technology, Purple Mountain Observatory, Chinese Academy of Sciences, 10 Yuanhua Road, Nanjing 210023, China}
\affiliation{Purple Mountain Observatory, Chinese Academy of Science, 10 Yuanhua Road, Nanjing 210023, China}
\email{dzliu@pmo.ac.cn}

\author[orcid=0000-0001-7147-3575]{Helmut Dannerbauer}
\affiliation{Instituto de Astrof\'{i}sica de Canarias (IAC), E-38205, La Laguna, Tenerife, Spain}
\affiliation{Universidad de La Laguna, Dpto. Astrof\'{i}sica, E-38206, La Laguna, Tenerife, Spain}
\email{helmut@iac.es}

\author[orcid=0000-0001-6831-0687]{Gianluca Castignani}
\affiliation{INAF-Osservatorio di Astrofisica e Scienza dello Spazio di Bologna, via Gobetti 93/3, I-40129, Bologna, Italy}
\email{gianluca.castignani@unibo.it}

\author[orcid=0000-0002-1462-9462]{Rosa Calvi}
\affiliation{INAF-Osservatorio Astronomico di Capodimonte, Salita Moiariello 16, 80131, Napoli, Italy}
\email{rosa.calvi@unife.it}

\author[orcid=0000-0002-3473-6716]{Lauro Moscardini}
\affiliation{Dipartimento di Fisica e Astronomia ``A. Righi'', Alma Mater Studiorum Universit\`{a} di Bologna, Via Gobetti 93/2, 40129 Bologna, Italy}
\affiliation{INAF-Osservatorio di Astrofisica e Scienza dello Spazio di Bologna, via Gobetti 93/3, I-40129, Bologna, Italy}
\affiliation{INFN-Sezione di Bologna, Viale Berti Pichat 6/2, 40127, Bologna, Italy}
\email{}

\author[orcid=0000-0002-6777-6490]{Benjamin Magnelli}
\affiliation{Universit\'{e} Paris-Saclay, Universit\'{e} Paris Cit\'{e}, CEA, CNRS, AIM, 91191, Gif-sur-Yvette, France}
\email{}

\author[orcid=0009-0008-5723-6969]{Sylvia Adscheid}
\affiliation{Argelander-Institut f\"{u}r Astronomie, Universit\"{a}t Bonn, Auf dem H\"{u}gel 71, 53121, Bonn, Germany}
\email{}

\author[orcid=0000-0002-3933-7677]{Eva Schinnerer}
\affiliation{Max-Planck-Institut f\"{u}r Astronomie, K\"{o}nigstuhl 17, D-69117, Heidelberg, Germany}
\email{}

\author[orcid=0009-0005-8842-424X]{Hui Li}
\affiliation{Purple Mountain Observatory, Chinese Academy of Science, 10 Yuanhua Road, Nanjing 210023, China}
\affiliation{School of Astronomy and Space Sciences, University of Science and Technology of China, Hefei 230026, China}
\email{}


\author[orcid=0000-0002-0569-5222]{Rafael Arango-Toro}
\affiliation{Aix Marseille Univ, CNRS, CNES, LAM, Marseille, France}
\email{}

\author[orcid=0000-0002-0930-6466]{Caitlin Casey}
\affiliation{Department of Physics, University of California Santa Barbara, Santa Barbara, CA, USA}
\email{}

\author[orcid=0009-0009-5330-3287]{Jacky S.W. Chan}
\affiliation{Purple Mountain Observatory, Chinese Academy of Science, 10 Yuanhua Road, Nanjing 210023, China}
\affiliation{School of Astronomy and Space Sciences, University of Science and Technology of China, Hefei 230026, China}
\email{}

\author[orcid=0000-0002-9382-9832]{Andreas Faisst}
\affiliation{IPAC, California Institute of Technology, MC 100-22, 1200 E. California Blvd Pasadena, CA 91125, USA}
\email{}

\author[orcid=0000-0002-3560-8599]{Maximilien Franco}
\affiliation{Universit\'{e} Paris-Saclay, Universit\'{e} Paris Cit\'{e}, CEA, CNRS, AIM, 91191, Gif-sur-Yvette, France}
\email{}

\author[orcid=0000-0002-6467-8078]{Zohreh Ghaffari}
\affiliation{INAF Osservatorio Astronomico di Trieste,
Via G.B. Tiepolo 11 e via Bazzoni 2. 34100, Trieste, Italy \label{inaft}}
\affiliation{IFPU: Institute for Fundamental Physics of the Universe, Via Beirut, 2, 34151, Trieste, Italy \label{ifpu}}
\email{zohreh.ghaffari@inaf.it}

\author[orcid=0000-0003-0129-2079]{Santosh Harish}
\affiliation{Laboratory for Multiwavelength Astrophysics, School of Physics and Astronomy, Rochester Institute of Technology, 84 Lomb Memorial Drive, Rochester, NY 14623, USA}
\email{}

\author[orcid=0000-0001-9187-3605]{Jeyhan Kartaltepe}
\affiliation{Laboratory for Multiwavelength Astrophysics, School of Physics and Astronomy, Rochester Institute of Technology, 84 Lomb Memorial Drive, Rochester, NY 14623, USA}
\email{}

\author[orcid=0009-0005-3133-1157]{Greta Toni}
\affiliation{Dipartimento di Fisica e Astronomia ``A. Righi'', Alma Mater Studiorum Universit\`{a} di Bologna, Via Gobetti 93/2, 40129 Bologna, Italy}
\affiliation{INAF-Osservatorio di Astrofisica e Scienza dello Spazio di Bologna, via Gobetti 93/3, I-40129, Bologna, Italy}
\affiliation{Zentrum f\"{u}r Astronomie, Universit\"{a}t Heidelberg, Philosophenweg 12, 69120, Heidelberg, Germany}
\email{}

\author[orcid=0000-0002-7051-1100]{Jorge Zavala}
\affiliation{National Astronomical Observatory of Japan, 2-21-1 Osawa, Mitaka, Tokyo 181-8588, Japan}
\affiliation{University of Massachusetts Amherst, 710 North Pleasant Street, Amherst, MA 01003-9305, USA}
\email{}

\author[orcid=0009-0000-6036-7320]{Xin Zeng}
\affiliation{Purple Mountain Observatory, Chinese Academy of Science, 10 Yuanhua Road, Nanjing 210023, China}
\affiliation{School of Astronomy and Space Sciences, University of Science and Technology of China, Hefei 230026, China}
\email{}

\author[orcid=0000-0003-3728-9912]{Xianzhong Zheng}
\affiliation{Tsung-Dao Lee Institute and State Key Laboratory of Dark Matter Physics, Shanghai Jiao Tong University, Shanghai, 201210, China}
\email{xzzheng@sjtu.edu.cn}


\begin{abstract}

    We present an environmental analysis and overdensity catalog for a large sample of submillimeter galaxies (SMGs) in the COSMOS field in order to understand whether SMGs are signposts of megaparsec-scale overdensity structures. We apply the Poisson Probability Method (PPM) protocluster finder to characterize the significance ($\sigma$) and overdensity value ($\delta$) of any overdense structures surrounding specific `beacon' galaxies. These beacons include 449 SMGs from A$^{3}$COSMOS (including 168 with spectroscopic redshifts, 52 of which have CO or \CIIline{} 158$\mu$m line detections), as well as a control sample of over 5000 main-sequence star-forming galaxies (SFGs) selected from the 40-band photometric-redshift catalog in COSMOS2025.
    We successfully reproduce 15 known protoclusters and clusters in COSMOS (including \textit{Hyperion} large-scale structure, CL~J1001+0220 at $z \sim 2.5$, and AzTEC-3 protocluster at $z \sim 5.3$), 
    and report 26 new PPM-based high-overdensity ($\delta \geq 5$), high-significance ($\sigma \geq 5$), robust candidates.  
    We measure an overdensity fraction of 11.5--71.2\% (1.3--61.0\%) in the CO/\CIIline{} 52 SMG (full 449 SMG) sample with different criterion categories, confirming that SMGs are signposts of overdensities compared to the control SFG sample. We also notice that the fraction is lower than 100\%, indicating that not all SMGs are associated with overdensities. We thus inspect the JWST morphology and argue that the triggering of starburst in SMGs must have non-environmental mechanisms, for instance, merger and violent disk instability.
    Finally, we further conduct a spatial distribution analysis, finding that the location of SMGs in overdensities has a redshift evolution, with SMGs being more concentrated in the inner regions of overdensities at higher redshift.

\end{abstract}

\keywords{\uat{High-redshift galaxy clusters}{2007}, \uat{Galaxy clusters}{584}, \uat{Starburst galaxies}{1570}, \uat{Galaxy evolution}{594}, \uat{Large-scale structure of the universe}{902}}

\section{Introduction}\label{Introduction}

\setcounter{footnote}{0}

Cosmological simulations predict that large scale structures (LSS) evolve substantially with cosmic time \citep[e.g.,][]{Springel2006,Overzier2016}. Within the dense regions of the cosmic web, early galaxies form non-virialized protoclusters at redshift $z>2$, and these protoclusters evolve into virialized, massive clusters at $z<1$, with some of them being the progenitor of the most massive clusters we see today, like the Coma Cluster. 
How to observationally map the overdensity of galaxies and constrain the large scale structures at high redshift is a key question for understanding our cosmology and galaxy evolution.

Searching for high-redshift protoclusters is very challenging due to the sparse distribution of a limited number of member galaxies over several megaparsecs~\citep[e.g.,][]{2013ApJ...779..127C,Muldrew2015,2025A&A...698A.280A}. Currently, several observational methods are developed, including (a) using narrow band or spectroscopic survey to search for Lyman-alpha or other emission or absorption lines~\citep[e.g.,][]{2003ApJ...586L.111S,2012ApJ...750..137T}, (b) using overdensities of color-selected galaxies or galaxy samples with accurate photometric redshifts~\citep[e.g.,][]{2009MNRAS.395.1845V,Diener2013,2016ApJ...828...56W,Toni2025}, and (c) using different types of galaxies, e.g., radio galaxies~\citep[e.g.,][]{2014MNRAS.445..280H,2017ApJ...846L..31D}, bright SMGs~\citep[e.g.,][]{2009ApJ...694.1517D,2011Natur.470..233C,2012Natur.486..233W,2018Natur.556..469M}, and QSOs~(Quasi-Stellar Objects)~\citep[e.g.,][]{2014ApJ...788..125S,2015Sci...348..779H}, as beacons to trace potential protocluster candidates.

Among the previously identified protoclusters, many have found associated SMGs (typically defined as observed-frame $850\mathrm{\mu m}$ flux $> 1$ mJy, e.g., \citealt{2002PhR...369..111B}) or dusty star-forming galaxies (DSFGs, broadly defined as high-$z$ galaxies having far-IR or (sub-)mm detections; \citealt{2014ApJ...796...95C}). 

However, to understand whether SMGs are reliable signposts of protoclusters, we also need to know whether all SMGs are associated with overdensity structures. Numerous studies have shown a physical positional correlation between high-density galaxy regions and SMGs \citep[e.g.,][]{2014MNRAS.439.1193C,2016MNRAS.461.1719C,2014A&A...570A..55D,Casey2015,2016A&A...585A..54F,2016ApJ...826..130H,2016ApJ...828...56W,2017A&A...597A...4S,2018ApJ...856...72O,2018Natur.556..469M,2022A&A...660A.137G}. For instance, \cite{2015MNRAS.452..878M} claim that most of the strongest overdensities do not host SMGs. In their review of infrared observations of protoclusters, \cite{2022Univ....8..554A} systematically explored this issue and ultimately concluded that massive galaxies may serve as signposts for identifying overdensity regions. 

For statistically understanding the link between the SMG population and the overdensity of cosmic web, an efficient method to compute the overdensity of a large sample of SMGs is needed. A recent study by \cite{Calvi2023} was the first to use the Poisson Probability Method \cite[PPM, focusing to search for galaxy overdensities at megaparsec scales~($\sim$ 1 pMpc), which is the typical size of protocluster cores;][]{2014ApJ...792..113C,2019A&A...623A..48C} to search for and analyze the characteristics of a sample of 12 spectroscopically confirmed SMGs with redshifts of $z \sim 2-5.3$ selected from the GOODS-N field. The results showed that 11 out of the 12 SMGs (i.e., $92\% \pm 8\%$) are associated with galaxy overdensities; this not only confirmed three previously discovered protoclusters but also led to the discovery of five new ones around SMG J123634~\cite[$z = 1.225$,][]{Bothwell2013}, ID.19~\cite[$z = 2.047$,][]{Decarli2014}, SMM J123607~\cite[$z = 2.487$,][]{Bothwell2013}, SMM J123606~\cite[$z = 2.505$,][]{Bothwell2013}, and GN10~\cite[$z = 5.303$,][]{Riechers2020,Sun2024}. Nevertheless, to draw a definitive conclusion on the hypothesis that SMGs are excellent tracers of distant protoclusters, further in-depth studies with larger sample sizes are still required. To address this issue, the goal of this paper is to conduct such an in-depth study using a larger sample size. To this end, we selected samples of SMGs and galaxies in the COSMOS-Web field from the A$^3$COSMOS Data Version 20220606\footnote{\url{https://sites.google.com/view/a3cosmos}; \url{http://vizier.cds.unistra.fr/viz-bin/VizieR?-source=J/A+A/685/A1}}and the COSMOS-Web~\citep{Casey2023} DR1 COSMOS2025 photo-$z$ Catalog~\citep{Shuntov2026}\footnote{\url{https://cosmos2025.iap.fr/catalog.html}}(hereafter A$^3$COSMOS and COSMOS2025, respectively). We adopted a time-dependent star-forming main sequence criterion~\citep{2014ApJS..214...15S} to select a set of control samples of star-forming main sequence galaxies.

The structure of this paper is organized as follows: Section~\ref{sec:data} introduces the data used in this study, including the selection of the SMG sample and SFG samples; Section~\ref{sec:method} elaborates on the main steps and parameter configurations of the PPM, and explains the definition of various traced overdensity environments based on the parameters output by the PPM; Section~\ref{sec:Results} and Section~\ref{sec:discussion} present and discuss the results; Section~\ref{sec:summary} presents the conclusions. This paper adopts the flat $\Lambda$ Cold Dark Matter ($\Lambda$CDM) cosmological model with $\Omega_m$ = 0.30, $\Omega_\Lambda$ = 0.70, and $H_0$ = 70 km~s$^{-1}$ $\mathrm{Mpc^{-1}}$~(\citealt{Spergel2003,Spergel2007}).

\begin{figure}[ht]
\includegraphics[width=0.45\textwidth]{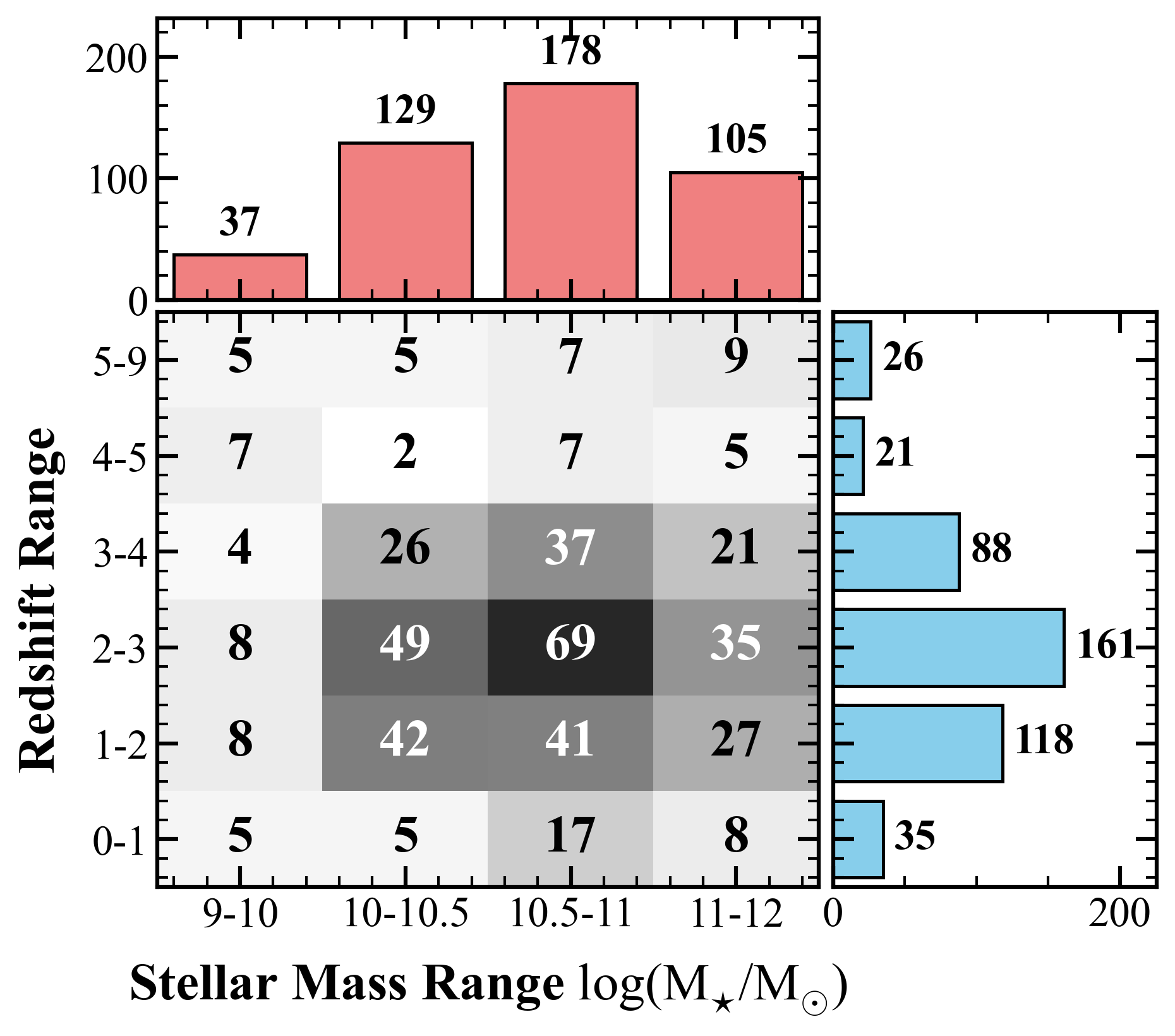}
\caption{The results of binning 449 SMG samples by mass and redshift show that the peak of the SMG number distribution falls within the ranges of redshift 1--3 and $\log(\rm M_{\star}/\rm M_{\odot})$ 10.5--11.5. For the SFG samples, random sampling and matching with a comparable number of SMG samples are conducted based on the redshift and stellar mass of each bin, respectively.
\label{fig:image1}}

\end{figure}

\section{DATA AND SAMPLES} \label{sec:data}
\subsection{A$^3$COSMOS SMG Samples}

We select our parent SMG sample from the latest A$^3$COSMOS public data release \citep{2019ApJS..244...40L,2024A&A...685A...1A}. This database provides approximately 1,400 (sub-)mm sources covering a broad redshift range, along with physical parameters such as star formation rates, gas masses, and dust masses derived from SED fitting. It also includes continuum maps, blind-detection catalogs, prior-based photometric catalogs, and value-added galaxy catalogs.

In total, the A$^3$COSMOS v20220606 catalog has 449 unique galaxies in COSMOS-Web (hereafter considering them as SMGs). We then use the latest JWST-based photometric redshifts~(hereafter photo-$z$) from COSMOS2025~\citep{Shuntov2026} and the largest compilation of spectroscopic redshift (hereafter spec-$z$) in COSMOS~\citep{Khostovan2026} to update the prior redshifts for A$^3$COSMOS SMGs. 
Furthermore, we also cross-match with the latest compilation of high-$z$ CO/\CIIline{} line-emitting galaxies~(including \citealt{2013ARAA..51..105C, 2018ApJ...867...92S, 2019ApJ...880...15K, 2020AA...641A.155V, 2025AA...699A..80H}; a full reference list is provided in D. Liu et al., in prep.), for secure redshifts and cold gas properties.

After cross-matching with these catalogs, we find that 52 SMGs have literature CO/\CIIline{} line detections, 116 have reliable spec-$z$'s (with a quality flag~$\geq$~2~, i.e., Confidence Level \(\geq 80\%\)), and 281 only have photo-$z$'s~(Summary of Redshift Sources for the All SMGs Sample in Appendix~\ref{sec:Redshift Sources for the All SMGs Sample} Table~\ref{tab:ztype}). Therefore, we construct three SMG samples for the further studies:
\begin{enumerate}
    \item ``CO/\CIIline{} spec-$z$ SMG'' sample (with 52 galaxies)
    \item ``All spec-$z$ SMG'' sample (with 168 galaxies, including the CO/\CIIline{} spec-$z$ sample)
    \item ``All SMGs'' sample (with 449 galaxies, including the above two spec-$z$ samples and all photo-$z$ SMGs)
\end{enumerate}

We derive stellar masses for all sources from the COSMOS2025 catalog, all 449 SMGs were selected as our final sample. As shown in Figure \ref{fig:image1}, we divided the SMG samples into bins based on redshift and stellar mass, the peak of the SMG number distribution falls within the ranges of redshift 1--3 and $\log(\rm M_{\star}/\rm M_{\odot})$ 10.5--11.5. facilitating subsequent matching with SFG samples.

\subsection{COSMOS2025 Galaxy Catalog}

COSMOS2025 contains photometric data, morphological information, photometric redshifts, and physical parameters for over 780,000 galaxies in the COSMOS-Web field. This catalog is based on the The COSMOS-Web Survey \citep[PIs: J. Kartaltepe and C. M. Casey; ][]{Casey2023}, is a Cycle 1 Treasury Program conducted using JWST. Within a region of ~0.54 deg$^2$ in the central COSMOS field, this program provides deep near-infrared imaging via four filters (NIRCam F115W, F150W, F277W, F444W; \citealt{Franco2025}) and one mid-infrared camera \cite[MIRI F770W;][]{Harish2025}. These data are combined with ground-based and space-based observational data. For sources detected by the NIRCam, photometric measurements are conducted using two methods: fixed-aperture photometry (on the space-based bands) and profile fitting techniques across all 37 bands (covering the wavelength range of 0.3$\mu \rm{m}$ to 8$\mu \rm{m}$). Morphological characteristics of all sources are obtained through complementary techniques such as profile fitting and machine learning classification. Meanwhile, photometric redshifts, physical parameters, and non-parametric star formation histories are derived via SED fitting. This catalog has been extensively validated against previous COSMOS releases and external survey data. As the latest iteration of the COSMOS-Web catalog \citep{Shuntov2026}, COSMOS2025 features comprehensive documentation and full redshift probability distributions, providing a robust foundation for scientific analysis.\

For the photometric redshifts, we use the LEPHARE~(\citealt{Arnouts2002,Ilbert2006}) $z_{phot}$ results from COSMOS2025. According to its documentation, we set \verb|warn_flag == 0| to select the most reliable sources—this directly excludes sources with poor quality caused by hot pixels, inconsistencies between ground-based and space-based data, excessively small radii, single NIRCam detections, and small flux ratios. We then set \verb|type == 0| to filter out stars and QSOs which may out-shine faint surrounding galaxies and prohibit accurate environmental analysis. Finally, we inspect all key physical parameters to remove sources with missing data (e.g., those with \verb|stellar mass == -99|). Through these steps, we obtain a complete and reliable sample of 672,655 galaxies.\

\subsection{Selecting Main Sequence Control Sample}
\label{sec:Selecting Main Sequence Control Sample}

\begin{figure}[ht]
\includegraphics[width=0.48\textwidth]{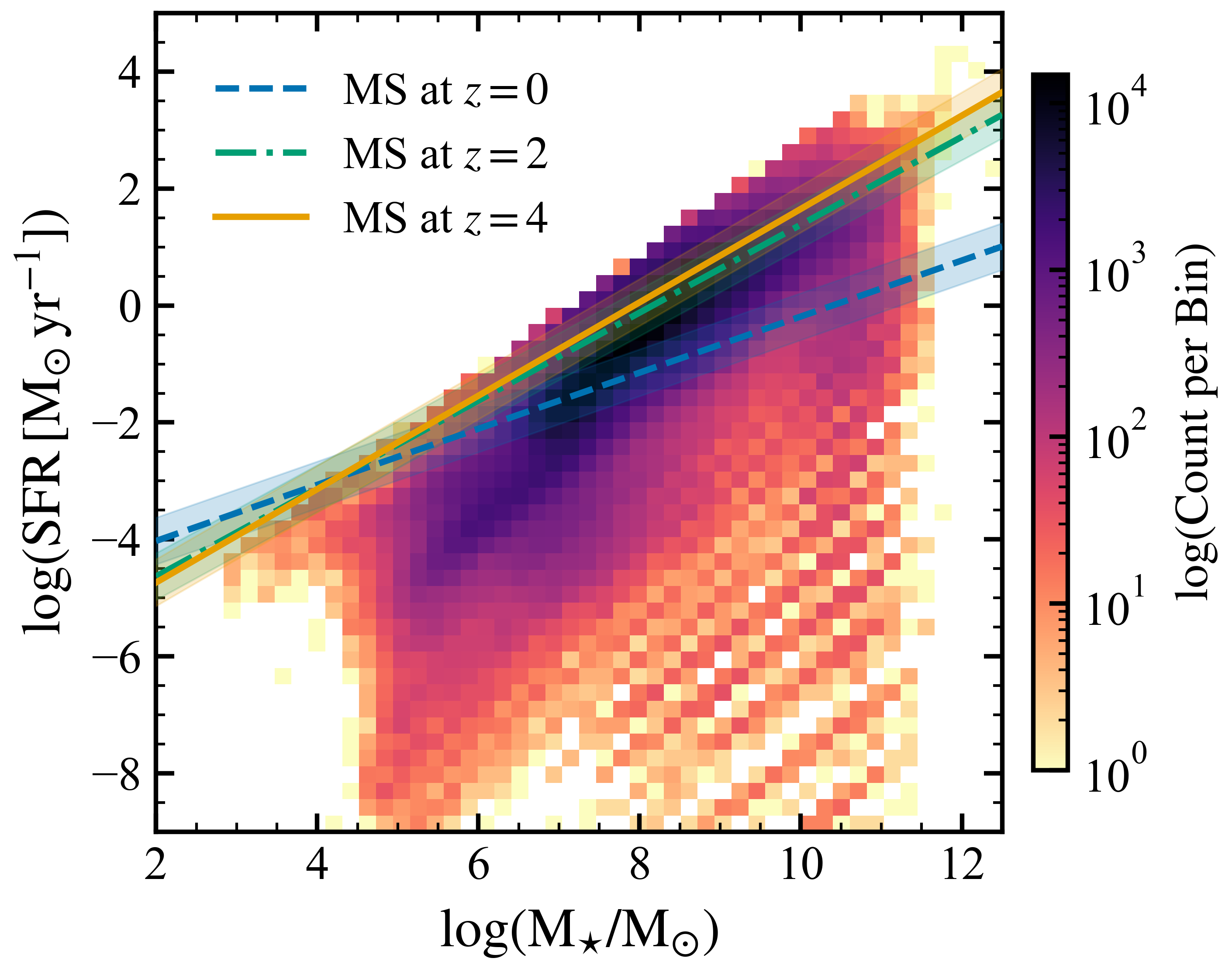}
\caption{Results of star-formation main sequence sample selection. Dashed lines of different colors represent the main sequence relationships at different redshifts~\citep{2014ApJS..214...15S}. The filled regions in the corresponding colors indicate the $\Delta \rm{MS} \sim \pm 0.4$ dex range at the respective redshifts, which is used as the criterion for selecting star-forming galaxies.
\label{fig:image2}}

\end{figure}

\begin{figure}[ht]
\includegraphics[width=0.45\textwidth]{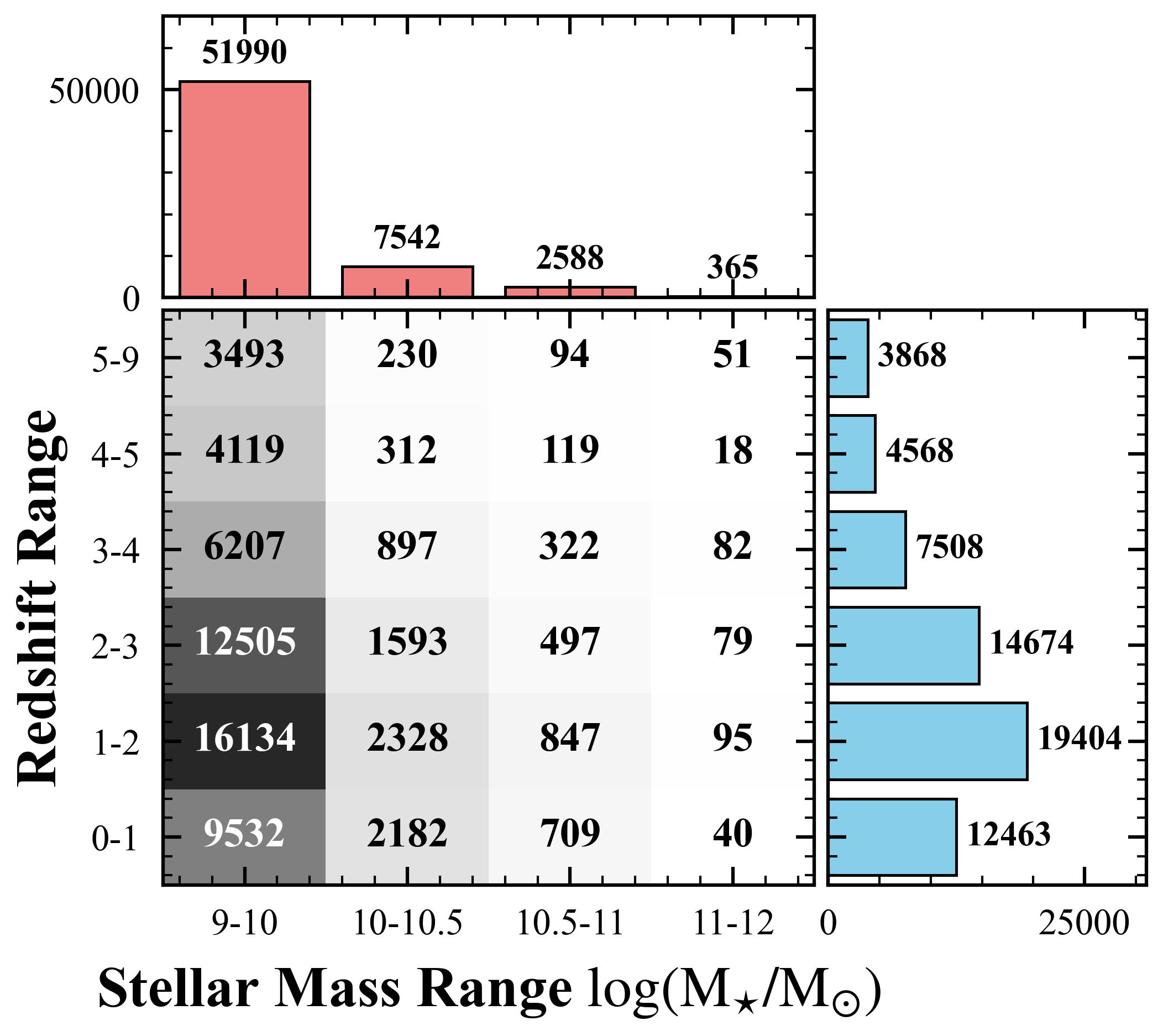}
\caption{The stellar mass-redshift distribution of star-forming main-sequence galaxies with $9 \leq \log(\rm M_{\star}/\rm M_{\odot}) \leq 12$ and $0 \leq z \leq 9$. showing that most sources are distributed in the stellar mass-redshift range of $9 \leq \log(\rm M_{\star}/\rm M_{\odot}) \leq 10$ and $0 \leq z \leq 3$. For example, to construct the redshift-mass matched SFG sample, we randomly draw main-sequence SFGs with replacement from each stellar mass-redshift ($\rm M_\star-z$) bin. The number of SFGs drawn from each bin is strictly determined to match the corresponding distribution of the primary sources shown in Figure~\ref{fig:image1}.
\label{fig:image2.5}}

\end{figure}

We use the \citet{2014ApJS..214...15S} star-forming main sequence~(MS) to select a set of control samples of star-forming main sequence galaxies~(SFGs). The logarithmic form expressed as :\
\begin{multline}
    \log \mathrm{SFR}(M_{\ast},t) = (0.84 \pm 0.02 - 0.026 \pm 0.003 \times t) \\ \log M_{\ast}
    - (6.51 \pm 0.24 - 0.11 \pm 0.03 \times t),
\end{multline}
where $\textit{t}$ is the age of the universe in Gyr. The observable dispersion $\Delta \rm{MS}$ $\sim$ $\pm 0.4$ dex was used as the criterion for SFGs. The results are presented in Figure~\ref{fig:image2}. The vast majority of the galaxies lies on the star-forming main sequence. Their stellar masses are primarily distributed in the range log($\rm{M}_\star$/$\rm{M}_\odot)=7-10$, and their star formation rates~(SFRs) span log (SFR/[$\rm{M}_\odot~\rm{yr}^{-1}$])$=-2~\rm{to}~1$. In total, we obtained 317,484 SFGs (accounting for 46.9\%). We use SFGs as control samples, and their stellar mass-redshift distribution is shown in Figure \ref{fig:image2.5}, most sources are distributed in the redshift-stellar mass range of redshift 0--9 and log($\rm{M}_{\star}$/$\rm{M}_{\odot}$) 9--12. Next, based on the previously obtained redshift-mass binning results of SMGs (see Figure \ref{fig:image1}), we randomly selected sources from the SFG samples, such that the number of sources in each redshift-mass bin matched that of the SMG sample. 

Taking the selection of SFGs as an example: specifically, we first retained all samples within the redshift range $\sim$[0, 9]. Then, we divided the stellar mass range log($\rm{M}_{\star}$/$\rm{M}_{\odot}$) $\sim$ [9, 12] into 4 bins. To be consistent with the bins division of SMG samples, we set the first bin as [9, 10] and the last bin as [11, 12], while the remaining 2 bins were divided with a step size of 0.5. Through this process, all samples within the entire range were assigned to different bins.

Subsequently, for each bin, we set a random seed to extract samples such that the number of extracted samples matched that of the corresponding SMG bin. Finally, we extracted 5 sets of mass match samples. Similarly, we then retained all samples within the stellar mass range 9 $<$ log($\rm{M}_{\star}$/$\rm{M}_{\odot}$) $<$ 12 and divided the redshift range 0 $<z<$ 9 into 6 bins. The division of each redshift bin and the number of samples extracted were determined based on the corresponding SMG bins. Eventually, 5 sets of redshift-matched samples were obtained. 

We construct the redshift‑mass matched samples as follows. For each bin defined in the SMG sample, we randomly draw a subset of equal size from all available SFG sources within the corresponding redshift‑mass bin. Each complete set of these subsets, which comprises a total of 449 sources to exactly match the total number of SMGs, is retained. We repeat this random sampling process five times to generate five independent sample sets. Furthermore, because the parent SFG catalog is cross-matched beforehand and includes all of our SMGs, the star-forming main-sequence SFG pool derived from random resampling inherently contains a small population of main-sequence SMGs. Since these SMGs constitute only a minor fraction of the total control sample ($\sim$2.2\%; 116/5296), we do not remove them from the SFG pool prior to sampling. This approach simulates a natural random sampling process within the COSMOS field and ensures that our subsequent analysis remains rigorously unbiased.

Together with the mass-match and redshift-match samples, we have a total of 15 sets of control samples. Each of the 15 sets of control samples contained 449 sources~(5,296 unique objects after cross-matching). And we later use the Monte Carlo method to derive the uncertainty of the fraction of SFGs with significant overdensities from the 15 sets of calculations.

\begin{figure*}[ht]
    \centering  
\begin{minipage}{0.98\linewidth}
    \centering
    \hspace{-0cm}
    \includegraphics[width=\textwidth]{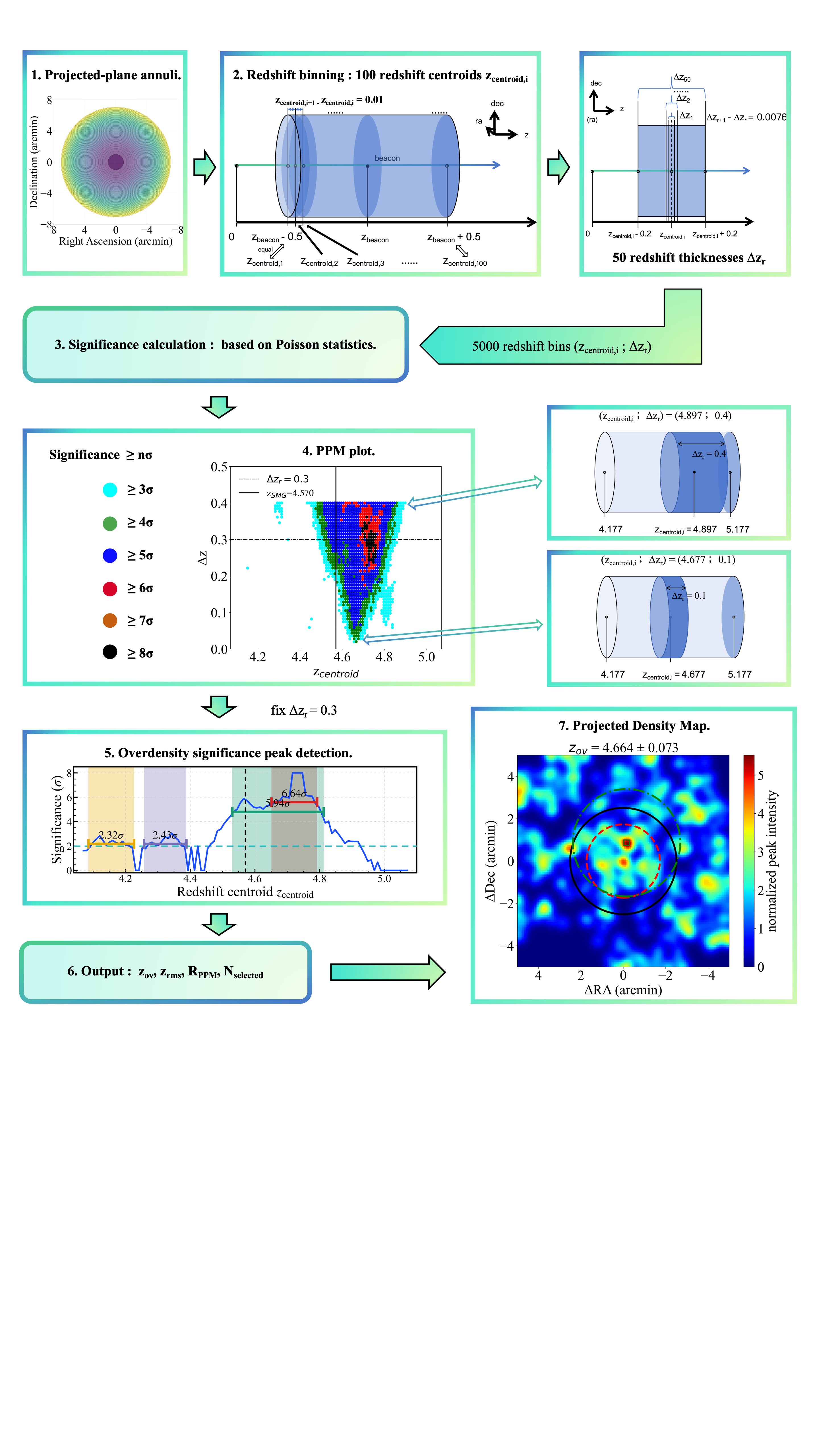}
    \begin{center}
    \end{center}
\end{minipage}
\caption{Workflow of our method (see section~\ref{sec:method}). Showing the general steps of the PPM method. The PPM plot and the projected density plot are calculated by a randomly selected SMG (ID: 867532).}
\label{fig4}
\end{figure*}

\section{Method} \label{sec:method}

\subsection{The Poisson Probability Method For Overdensity Identification}\label{sec:ppm}

The PPM is an effective and robust approach for analyzing the significance of overdensity around target galaxies. It was first introduced by \cite{2014ApJ...792..113C,2014ApJ...792..114C} to search for overdensities of galaxies around radio galaxies. Its purpose is to detect the presence of galaxy clusters within a given observation area and specific projected coordinate range by utilizing photometric redshift information.

\citet{Calvi2023} applied the PPM to a sample of 12 spectroscopically confirmed SMGs with redshifts ranging from $z\sim 1.2$~to 5.3 in the GOODS-N field, aiming to find protoclusters around them. Through this method, they detected overdensities around 11 of these SMGs, verifying the effectiveness of the PPM in detecting environmental overdensities when using SMGs as beacons. We emphasize that with the PPM we will be able to search for galaxy overdensities at megaparsec scales, which is the typical
size of protocluster cores, whereas the overall extent of galaxy protoclusters could be up to 10--20 Mpc (see, e.g., \citealt{Muldrew2015,Casey2015}). We now apply this method to the analysis of a 37$\times$ larger sample of SMGs to obtain results with greater statistical significance. In the following, we elaborate on the general steps of the PPM and its output parameters. We also present the workflow in Figure~\ref{fig4}, which illustrates each step of our work.

\begin{enumerate}
    \item \textbf{Projected-plane annuli.} We use the projected coordinates and redshift of each SMG as input beacons for the PPM. For each beacon target, we draw 50 concentric annuli, making sure each annulus has the same projected area of $3.14~\rm{arcmin^2}$. This is done in order to have the same average field density for each of the regions. The innermost radius is $1.0'$, and the outermost $\sim 7.07'$ (corresponding to $0.47$ and $3.34~\rm{Mpc}$ at $z=3$, respectively).
    \item \textbf{Redshift binning.} The redshift interval of our interest is \([z_{\text{beacon}}-0.5, z_{\text{beacon}}+0.5]\). We uniformly divide this interval with a step size of 0.01, yielding 101 redshift centroids \(z_{\text{centroid},i}\) where \(i = 0, 1, ..., 100\). For each \(z_{\text{centroid},i}\), we extract a slice centered at this redshift with a thickness of \(\Delta z_r\), where \(r = 0, 1, ..., 50\). (The values of \(\Delta z_r\) span the interval \([0.02, 0.4]\), which is uniformly discretized with a step size of 0.0076 to generate 50 distinct thickness values of \(\Delta z_r\).) In this way, we obtain a total of 5000 combinations of \((z_{\text{centroid},i}, \Delta z_r)\), the considered ranges are densely spanned in terms of both the redshift bin and the redshift centroid.

    \item \textbf{Significance calculation.} For each area and given redshift bin, we calculate the null hypothesis (no overdensity) probability of more galaxies than observed using Poisson statistics and the average number density from the background. From the beacon’s coordinates, we select only the first consecutive overdense regions with a null hypothesis probability $\leq 30\%$, merge these regions, and recalculate the probability with the single-region method. We define the detection significance of the number count excess as the complementary probability, and exclude overdensities first detected at $r \geq 132 \arcsec$ ($\sim$1 Mpc at $z = 3$) from the beacon.

    \item \textbf{PPM plot.} 
	After completing the significance calculations for the 5000 pairs of \((z_{\text{centroid},i}, \Delta z_r)\) for every area, we visually present the overdensity distributions across different redshifts and slice thicknesses using a PPM plot. For example, we present the PPM plot in Figure~\ref{fig4}~(step 4), where the color of each pixel represents the overdensity significance corresponding to a specific \((z_{\text{centroid},i}, \Delta z_r)\) pair, the vertical solid line marks the redshift value of the beacon~(SMG, ID: 867532), and the colored dots represent significance levels: $\geq$ 3$\sigma$ (cyan), 4$\sigma$ (green), 5$\sigma$ (blue), 6$\sigma$ (red), 7$\sigma$ (brown), and 8$\sigma$ (black). We further apply a Gaussian filter to eliminate high-frequency noise. A conservative smoothing scale of 0.02 in the redshift direction is adopted, which is approximately equal to or smaller than the typical photometric redshift uncertainty.

    \item \textbf{Overdensity significance peak detection.} Since the high-significance patterns in the PPM plot are stable with respect to the \(\Delta z\) axis, we search for overdensities along the line of sight of the SMG by fixing the redshift thickness at \(\overline{\Delta z} = 0.3\), thus simplifying the problem to a one-dimensional (1D) case. This choice follows the original PPM prescription, in which the fiducial redshift thickness is chosen to be comparable to the typical 2$\sigma$ photometric-redshift uncertainty of the input galaxy catalog \citep{2014ApJ...792..113C}, and is the same fiducial value adopted in the SMG-based PPM analysis of \citet{Calvi2023}. For the COSMOS2025 DR1 catalog used here, the spectroscopically calibrated photometric-redshift accuracy is \(\sigma_{\rm NMAD}=0.011{-}0.030\), depending on \(m_{\rm F444W}\) \citep{Shuntov2026}; at \(z\simeq4.6\), this corresponds to \(\sigma_z\simeq0.06{-}0.17\), making \(\overline{\Delta z}=0.3\) a reasonable fiducial width. We apply the peak finding algorithm to the PPM plot to search for overdensity significance peak intervals around the beacon. An overdensity region is identified when it spans an interval of at least $\delta_{z_{\mathrm{ centroid}}}$ = 0.1 along the $z_{\mathrm{centroid}}$ axis of the redshift and has a significance threshold $\geq$ 2$\sigma$. If the spacing between overdensity regions along the $\delta_{z_{\mathrm{ centroid}}}$ axis is less than 0.01, these regions are merged. By gradually increasing the significance threshold, we obtain the peak intervals of overdensity significance that meet the threshold. In the example shown in Figure~\ref{fig4}~(step 5), we identify four intervals that satisfy the criteria, denoted by distinct colored line segments and filled regions, centered at $z_{\mathrm{centroid}} = 4.156$, $4.323$, $4.671$, and $4.722$, with average significance values of $2.3\sigma$, $2.4\sigma$, $5.9\sigma$, and $6.6\sigma$, respectively. Compared to \citet{2014ApJ...792..113C}, we additionally allow an outer interval that completely contains a candidate interval of higher significance to be retained as an independent candidate, provided that the central redshift difference between the two overdensity peaks exceeds $0.02$; the outer interval is retained because its central redshift can be closer to that of the beacon. As shown in Figure~\ref{fig4}~(step~5), we consider the outer candidate interval with $z_{\mathrm{centroid}} = 4.671$, which is the closest to the beacon redshift $z = 4.57$, as the overdensity environment in which the beacon is embedded.

    \item \textbf{Output parameters.} Among the detected peak intervals, we select the one whose central redshift $z_{\mathrm{centroid}}$ is closest to the beacon~(SMG or SFG) redshift as the overdensity associated with the beacon, since this peak represents the (proto)cluster nearest to the beacon. Centered on this $z_{\mathrm{centroid}}$, we adopt a redshift slice of width $\Delta z = 0.3$ (i.e., $z_{\mathrm{centroid}} \pm 0.15$). Within this slice, we consider all sources that lie in the radial annulus bounded by the median of the minimum distances and the median of the maximum distances from the coordinates of the beacon~(SMG or SFG) to the location of the detected overdensity region in the projected space. The overdensity redshift $z_{\mathrm{ov}}$ is then estimated as the mean redshift of these selected sources. The redshift uncertainty $z_{\mathrm{rms}}$ is estimated at the $1\sigma$ level using the root-mean-square dispersion of the redshifts of the selected sources. Meanwhile, the number of these sources is taken as the estimate of the (proto)cluster richness $N_{\rm{select}}$, and the average of the maximum distances over the peak interval of this (proto)cluster is used as the estimate of its projected radius $R_{\rm{PPM}}$ (see Figure~\ref{fig4}~(step~5): we select the peak at $z_{\mathrm{centroid}} = 4.671$ closest to the beacon, yielding $z_{\rm{ov}} = 4.664$ with a significance of $5.9\sigma$).

    \item \textbf{Projected Density Map.} To robustly determine the spatial association of a beacon (SMG or SFG) and an overdensity found in the PPM, we further examine the 2D projected density maps. For each overdensity region detected at redshift $z_{\rm{ov}}$, we select galaxies with photometric redshift values within the range of $z_{\rm{ov}}$-$z_{\rm{rms}}$ to $z_{\rm{ov}}$+$z_{\rm{rms}}$ for analysis. Using these galaxies, we generate two-dimensional distribution maps with pixel sizes of 1/32 $\rm{Mpc}$ and apply a Gaussian filtering with a kernel of $\sigma$ = 3/32 $\rm{Mpc}$~\footnote{\textcolor{black}{This is slightly simpler than the method adopted by \cite{2019A&A...623A..48C} which applies a comprehensive wavelet analysis with the {\tt{MR/1}} software package to filter out high-spatial-frequency components and reveal smoother density distribution. Despite that our Gaussian smoothing does a similar work, we might be affected by strong high-spatial-frequency features and thus would need future improvement (e.g., Python implementation of the wavelet spatial filtering and efficient processing of a large number of targets). However, we conducted the analysis with both methods for the 52 CO/\CIIline{} spec-$z$ SMGs, thanks to the original authors, and found only small differences that would not alter the main results of this study. Therefore, for the larger spec-$z$ and full photo-$z$ SMG samples, we keep using our simpler Python-based Gaussian smoothing results (e.g.,} \textcolor{black}{in Figures~\ref{fig:4.0} and \ref{fig:4.00}), but whenever the {\tt{MR/1}} wavelet analysis is done, we use them as the better results (e.g., in Figures~\ref{fig:zov_zsmg} and \ref{fig:5.0}).}}. From each map, we identify the highest peak point within a circular area of radius $R_{\rm{PPM}}$ centered on the beacon's projection coordinate. The projected distance (i.e., offset) between the beacon coordinates and the overdensity peak point is calculated and defined as $\theta_{ov}$. This method aims to locate the projection spatial position of the beacon's corresponding overdensity peak. To achieve this, the search area is confined to a circular region of radius $R_{\rm{PPM}}$. We then refine the definition of overdensity size by estimating the minimum projected distance of peaks, which corresponds to the position where the overdensity value $\delta$ at the peak drops to one percent of the peak value. This size is defined as $R_{w}$. For the projected density map displayed in Figure~\ref{fig4}~(step~7), the black solid circle and red dashed circle are centered at the projected spatial coordinates of the beacon. The former has a physical radius of $1~\rm{Mpc}$ (estimated based on $z_{\rm{ov}}$), and the latter uses $R_{\rm{PPM}}$ as the radius to represent the overdense region detected by PPM. The green rectangular box indicates the peak point with the highest $\delta$. The center of the green dashed circle is at the detection peak of the density plot, and its radius $R_{w}$ is the position where the overdensity value drops to one percent of the peak value. Here we define the overdensity value~($\delta$) for each pixels in the 2D density map as $\rho_{pixel}=\frac{N_{pixel}}{S_{pixel}},{~} \rho_{bg}=\frac{N_{tot}}{S_{tot}},{~}\delta=\frac{\rho_{pixel}-\rho_{bg}}{\rho_{bg}}$, where the $\rho_{pixel}$ and $\rho_{bg}$ correspond to the galaxy number density of every pixels and the background, respectively.

\end{enumerate}

\input{table1}

\subsection{Overdensity Criteria Categories} \label{Overdensity Criteria Categories}

To identify reliable overdensity candidates, we classify the samples into five categories (A--E) based on the robustness of the spatial association, overdensity and significance parameters from the PPM. 
For the criteria of Category E, we follow~\cite{Calvi2023}; specifically, the difference between the detected overdensity redshift and the input beacon redshift is within 0.3, and the signal-to-noise ratio (S/N) of the detection $\geq 2\sigma$. From Category E to Category A, we progressively tighten the association between the PPM-detected overdensities and their beacon SMGs/SFGs. Specifically, the offset in redshift space is gradually restricted to within the PPM-estimated redshift error of the overdensity ($\le z_{\rm rms}$), while the projected spatial separation is limited to within the estimated radius of the overdensity peak ($\le R_w$). Their exact criteria are listed in Table~\ref{tab:table1}.
 
Concurrently, we systematically raise the thresholds for the PPM detection S/N and the projected galaxy overdensity up to a significance of $\ge 5\sigma$ and $\delta \ge 7$, respectively, in order to obtain the most robust overdensities that are most closely associated with the beacons. By establishing these multiple categories, we can systematically evaluate how varying selection thresholds impact our overall results. Furthermore, this tiered approach allows us to verify the sensitivity and reliability of our findings across different levels of stringency. Our PPM analysis result is comprised of 4 main files, including a maximum redshift of $z\sim8$, with a total of more than 3000 overdensity candidates that meet at least the Category E criteria. Interested readers can access the results of our analysis for all samples online\footnote{\url{https://github.com/chance-deng/PPM-analysis-for-COSMOS-Web-field.git}}.

\section{Results} \label{sec:Results}

\begin{figure}[ht]
\includegraphics[width=0.465\textwidth]{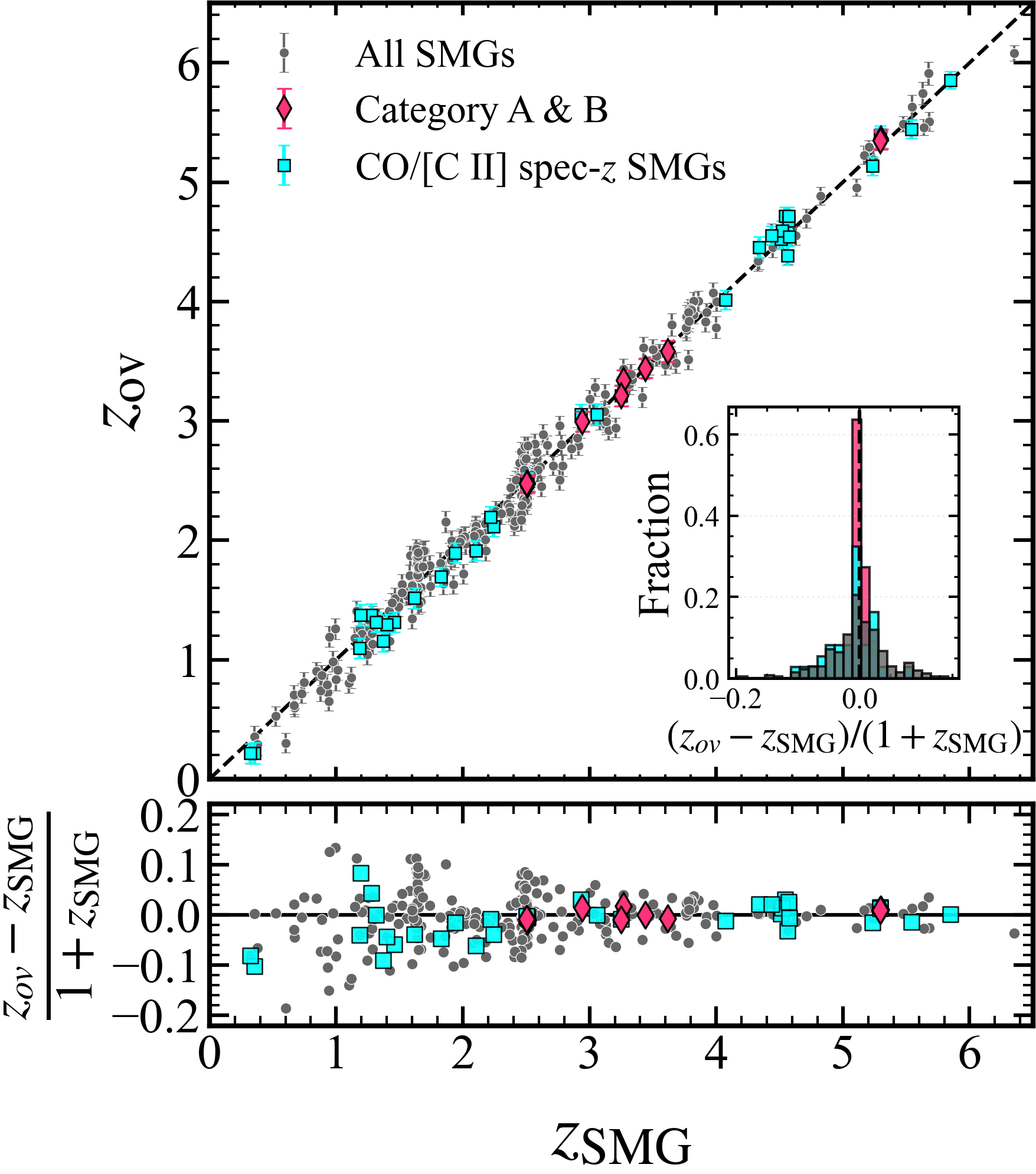}
\caption{The upper panel illustrates the redshifts of the overdensities, estimated with the PPM photo-$z$ catalogs against the redshifts of the All SMGs sample~(gray squares), the CO/\CIIline{} spec-$z$ SMGs samples~(cyan circles) and the subset satisfying the criteria for Category A and B (magenta diamonds). The normalized residuals histograms in the inset show the normalized number distributions of the three samples. In the bottom panel we show the residuals.
\label{fig:zov_zsmg}}
\end{figure}

\subsection{Overdensity redshift accuracy}

Figure~\ref{fig:zov_zsmg} presents a comparison between the overdensity redshift~($z_{\rm ov}$) and the SMG redshift~($z_{\rm SMG}$) for all structures satisfying the Category E criteria. We compare the redshifts of the overdensities, estimated with the PPM photo-$z$ catalogs against the redshifts of the All SMGs sample~(gray squares), the CO/\CIIline{} spec-$z$ SMGs samples~(cyan circles) and the subset satisfying the criteria for Category A and B (magenta diamonds). All $z_{\rm ov}$ detected by the PPM match the input SMG redshifts well. For the entire SMG sample, the accuracy is $\sigma(\frac{z_{\rm ov} - z_{\rm SMG}}{1 + z_{\rm SMG}}) = 0.047$, which is highly consistent with the findings of \cite{Calvi2023}, alongside a negligible bias of $\langle z_{\rm ov} - z_{\rm spec} \rangle = -0.002$. The CO/\CIIline{} spec-$z$ SMG sample exhibits a higher accuracy of $\sigma(\frac{z_{\rm ov} - z_{\rm SMG}}{1 + z_{\rm SMG}}) = 0.038$, with a negligible bias of $\langle z_{\rm ov} - z_{\rm spec} \rangle = -0.006$. Consistent with \cite{Calvi2023}, we also find the highest accuracy ($\sigma(\frac{z_{\rm ov} - z_{\rm SMG}}{1 + z_{\rm SMG}}) = 0.017$) and a negligible bias ($\langle z_{\rm ov} - z_{\rm spec} \rangle = 0.025$) for SMGs at $z = 4$--$5$ that have overdensities detected in their vicinity. As outlined by the criteria for our five defined categories (see Table~\ref{tab:table1}), the Category A--D overdensities possess even higher redshift matching accuracy. As an example, the $z_{\rm ov}$ versus $z_{\rm SMG}$ results for Category A and B overdensities are presented in Figure~\ref{fig:zov_zsmg}. A total of 11 sources satisfy the criteria for Category A and B; these achieve an accuracy of $\sigma(\frac{z_{\rm ov} - z_{\rm SMG}}{1 + z_{\rm SMG}}) = 0.011$ and a negligible bias of $\langle z_{\rm ov} - z_{\rm spec} \rangle = 0.004$, demonstrating that these sources are tightly embedded within the detected high-S/N overdensity structures in redshift space. In subsequent sections, we focus on the most reliable Category A and B overdensities.

\subsection{Validation with the COSMOS-Web galaxy group catalog}\label{Compare to the COSMOS-Web deep galaxy group catalog}

Recently, \cite{Toni2025}\footnote{\url{https://cdsarc.cds.unistra.fr/viz-bin/cat/J/A+A/697/A197}} applied an Adaptive Matched Identifier of Clustered Objects (AMICO; \citealt{Bellagamba2018,Maturi2019}) algorithm to the COSMOS-Web photometric galaxy catalog and obtained the largest-to-date COSMOS-Web deep galaxy group catalog. The catalog contains 1678 candidate groups and protocluster cores up to $z=3.7$. Of our total 54 Category A \& B overdensity candidates, three lie beyond this redshift limit ($z_{\rm ov} > 3.7$). Consequently, we cross-match only the remaining 51 candidates that fall within the catalog's redshift coverage ($z_{\rm ov} \leq 3.7$) to validate our PPM method. Adopting the matching criteria detailed in Section~\ref{sec:(Proto)cluster candidates}, we find that all 51 PPM-identified overdensity candidates successfully match with the COSMOS-Web deep galaxy group catalog, yielding a matching rate of 100\% (51/51).

\begin{figure}[!htbp]
\includegraphics[width=0.485\textwidth]{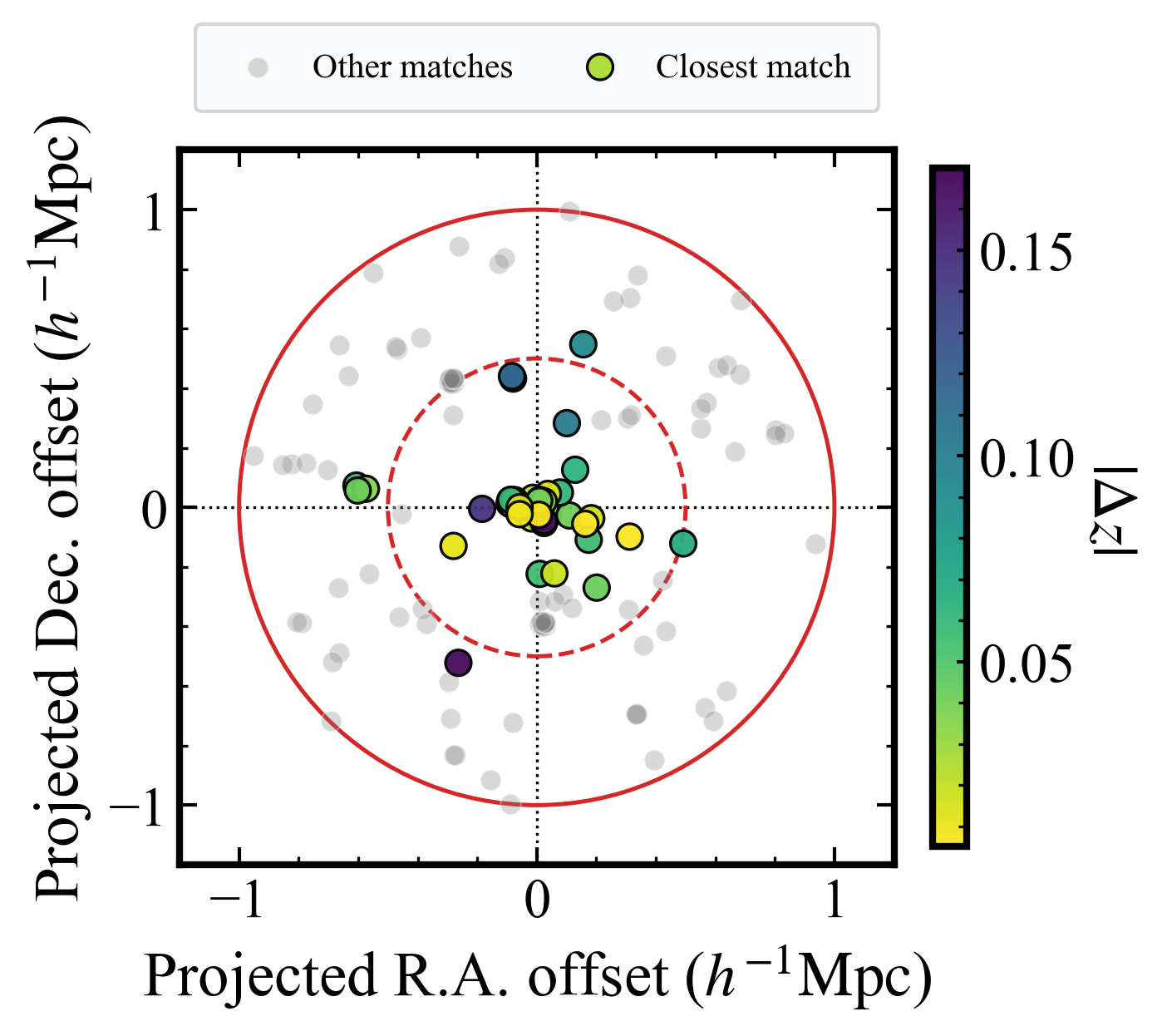}
\caption{Projected spatial offsets (in units of $h^{-1}\text{Mpc}$) of the matched galaxy groups~\citep{Toni2025} relative to the centers of their corresponding Category A \& B overdensity candidates identified by the PPM. The closest matched group for each candidate is indicated by a large circle, color-coded according to the absolute redshift difference, $|\Delta z|$, between the group and the overdensity candidate. Gray circles represent additional, secondary matches for candidates with multiple group associations. The red solid and dashed circles denote projected radii of 1.0 $h^{-1}$Mpc and 0.5 $h^{-1}$Mpc, respectively.
\label{fig:Toni2025match}}
\end{figure}

We do not exclude the possibility of multiple matches; therefore, several overdensity candidates are matched to multiple groups. The maximum number of matches occurs for PC 33 (SFG:560232), which corresponds to six sources (Group IDs: 300, 110, 1152, 1813, 1068, and 723). Figure~\ref{fig:Toni2025match} shows the projected offsets between the COSMOS-Web group positions and the PPM positions from this work.

Colored data points are primary matches, with color-coding by the redshift differences between the COSMOS-Web group catalog and this work, and gray points are secondary matches. 
For the 51 closest matches, the average projected separation is $\overline{\rm sep} = 0.175~{\rm pMpc}$, and the average absolute redshift difference is $\overline{|\Delta z|} = 0.058$. 
We present all matching results in Table~\ref{table4} of Appendix~\ref{sec:Spec-z member of overdensity candidates}.

Since our PPM method needs a prior position to search for overdensities, we do not fully cover the COSMOS-Web area for a complete study of all SFGs. Nevertheless, we applied PPM method to our control SFG sample, therefore, we can cross-match this control sample with the COSMOS-Web galaxy group catalog to further validate our PPM analysis. For 573 common galaxies in both our control SFG sample and the COSMOS-Web group catalog with \verb|AssocProb > 0.5|~(where an AMICO membership probability $> 50\%$ defines a formal group galaxy, see \citealt{Toni2025} for details), our PPM analysis identifies 75.6\% as Category E overdensities. We visually checked that the missed galaxy groups are either small-number groups or those with large photo-$z$ distributions, which could not be tight enough to be captured by our method. As our PPM overdensity structures probe typically larger scales than galaxy groups, we caution that the comparison here may not be a one-to-one comparison.

\subsection{Category A: Identifying the densest $z\sim 2$ and $z \sim 5$ protoclusters in COSMOS} \label{sec:Type A SMGs}
\begin{figure*}[hp]
    \centering
\begin{minipage}{0.465\linewidth}
    \centering
    \includegraphics[width=\textwidth]{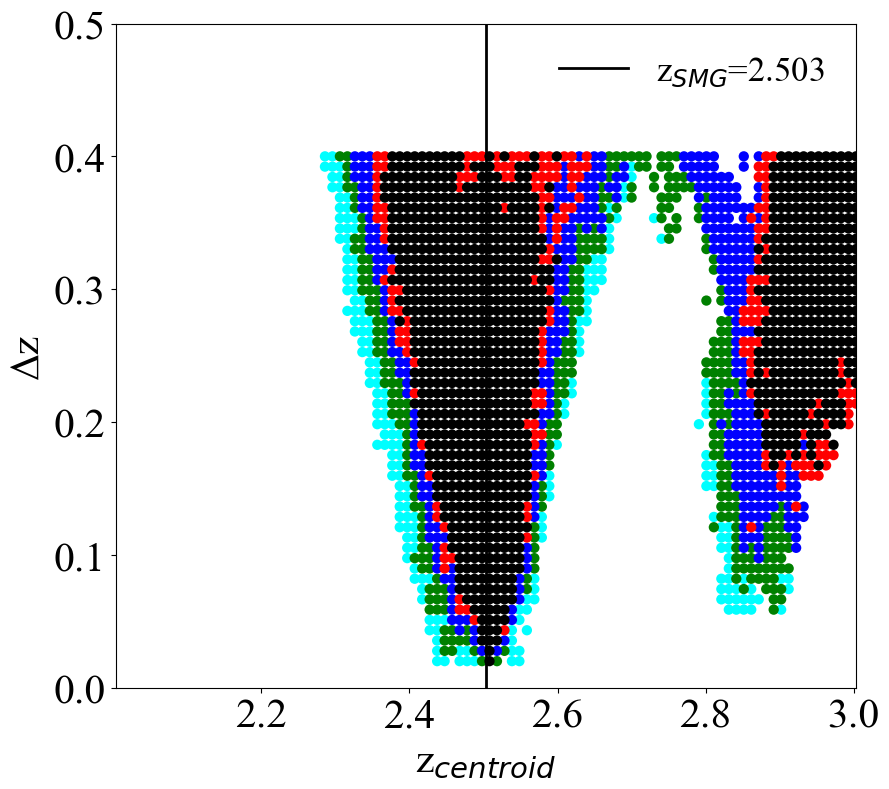}
\end{minipage}
\begin{minipage}{0.525\linewidth}
    \centering
    \includegraphics[width=\textwidth]{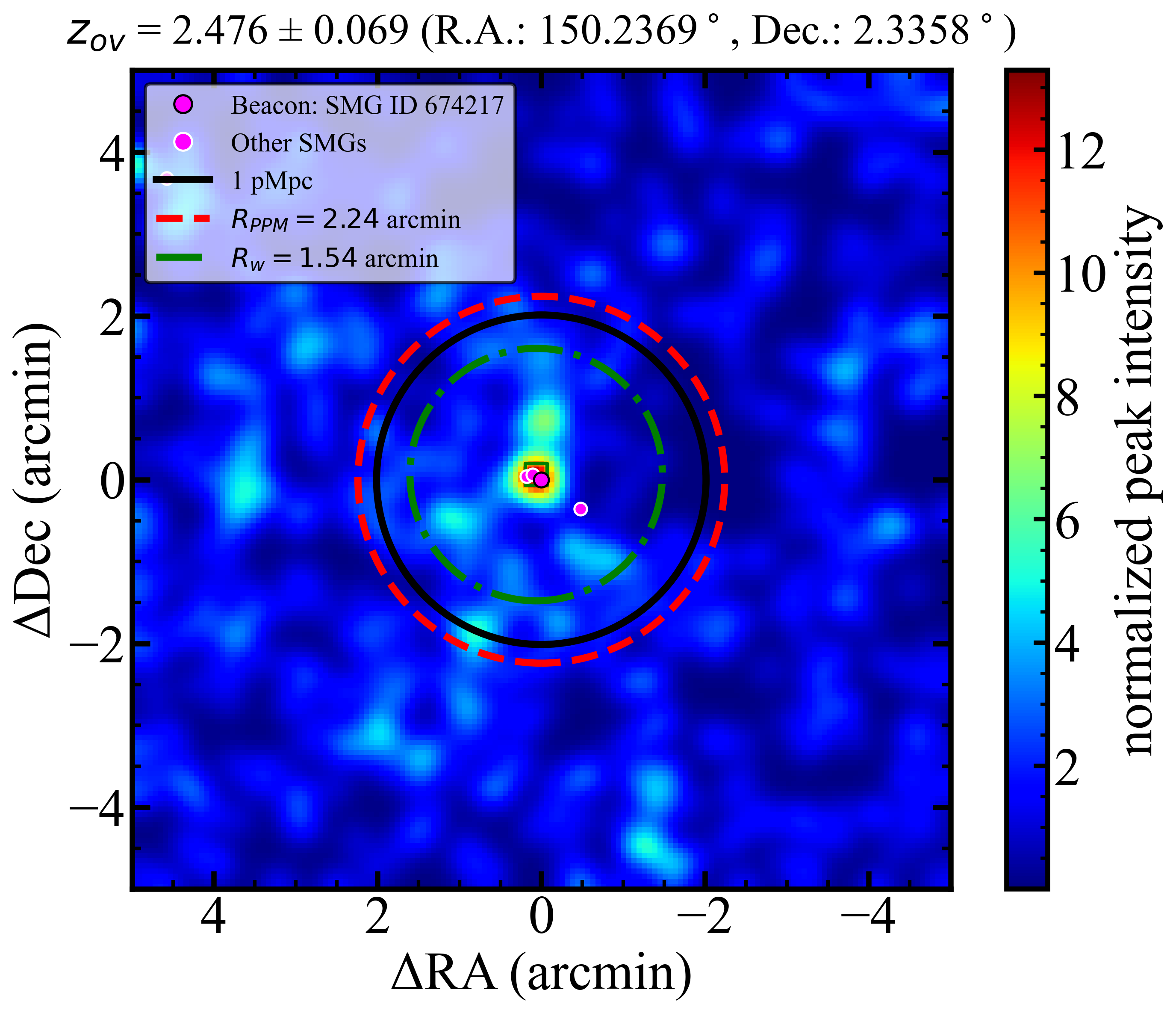}
\end{minipage}
\begin{minipage}{0.90\linewidth}
    \centering
    \includegraphics[width=\textwidth]{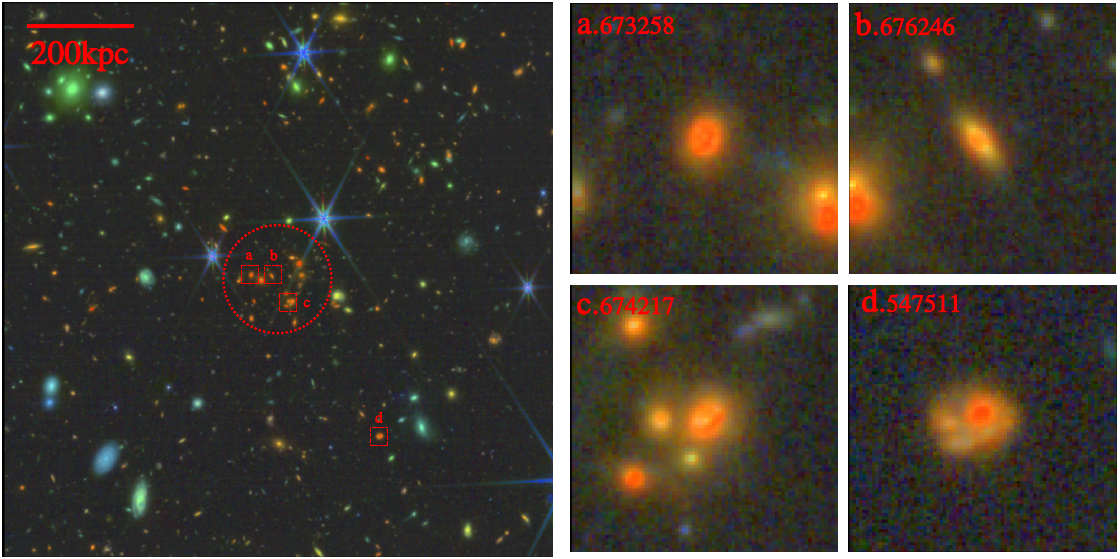}
\end{minipage}
\caption{PPM diagnostic plots of CL~J1001+0220 protocluster. 
\textbf{Top-left}: PPM redshift-overdensity plot (Section~\ref{sec:ppm} Step~4). Colored dots represent different significance levels: $\geq$ 3$\sigma$ (cyan), 4$\sigma$ (green), 5$\sigma$ (blue), 6$\sigma$ (red), 7$\sigma$ (brown), and 8$\sigma$ (black). The `inverted triangle' shape indicates the peak overdensity redshift $z_{\rm{ov}}$ and aligns well with the beacon's redshift indicated by the black vertical line. 
\textbf{Top-right}: 2D density plot (Section~\ref{sec:ppm} Step~7), with color indicating the overdensity value $\delta$. Magenta dots represent the SMGs within the PPM peak's redshift range, and a total of 4 SMGs are found in the core region. The black solid circle and red dashed circle are centered at the projected spatial coordinates of the SMG. The former has a physical radius of $1~\rm{Mpc}$ (estimated based on $z_{\rm{ov}}$), and the latter uses $R_{\rm{PPM}}$ as the radius to represent the overdense region detected by PPM. The green rectangular box indicates the peak point with the highest $\delta$. The center of the green dashed circle is at the detection peak of the density plot, and its radius $R_{w}$ is the position where the overdensity value drops to one percent of the peak value. Our reproduced results are in perfect agreement with the position and structure of the protocluster presented by \cite{2016ApJ...828...56W}. 
\textbf{Bottom}: JWST RGB image of the protocluster with a FoV$\sim$1~$\rm{pMpc^2}$ (\textbf{bottom-left}) and the four protocluster SMG members labelled $a$--$d$ with FoV$\sim 5\arcsec \times 5\arcsec$ (\textbf{bottom-right}). The red dashed circle is centered at the projected center of the overdensity with a radius of $100~\mathrm{kpc}$.
\label{fig:4.0}
}
\end{figure*}

Among the 449 SMG samples, 6 SMGs belong to Category A (see Table~\ref{tab:table1}), that is, the most robust overdensities. Four of them originate from the same overdensity structure --- CL~J1001+0220, a protocluster at $z = 2.506$ discovered by \citet{2016ApJ...828...56W} which was identified by the multi-wavelength spectroscopic observations using IRAM-NOEMA (CO~line), VLT/KMOS ($ H \alpha $ line), and JVLA (CO~(1--0) line) as shown in Figure~\ref{fig:4.0}. 

To illustrate our results, Figure~\ref{fig:4.0} presents the detailed PPM analysis for the field surrounding SMG~674217 (within the CL~J1001+0220 structure; see Table~\ref{table:3} for details). The redshift-overdensity plot in the top-left panel reveals a distinct density peak, where the peak overdensity redshift ($z_{\rm{ov}}$) aligns closely with the systemic redshift of the SMG beacon. The overdensity detection in this region reaches a maximum significance of $8\sigma$. We also observe another prominent overdensity structure at $z \sim 2.95$. As described in Section~\ref{sec:method}, the PPM algorithm searches for overdensity signals centered on a beacon, with the requirement that the overdensity signal must begin to appear within a maximum radius of $2.2\arcmin$. Therefore, this secondary feature represents an independent structure falling within the line of sight that satisfies this detection condition. This peak corresponds precisely to the overdensity detected around PC~5~(SMG:845982) at a projected center of 10:00:57.48, +02:17:02.76, the detailed properties of which are listed in Table~\ref{table:3}. The top-right panel displays the corresponding 2D projected density map, with the color scale tracing the local overdensity value ($\delta$). Within the redshift range of the PPM peak, we identify four SMGs highly clustered in the core region. To delineate the spatial extent of the structure, two boundaries are defined around the central SMG: a physical radius of $1~\rm{Mpc}$ (evaluated at $z_{\rm{ov}}$) and the PPM-detected overdense region with a radius of $R_{\rm{PPM}}$. Furthermore, we characterize the inner core of the structure using $R_{w}$, defined as the radius from the density peak where the overdensity drops to 1\% of its maximum value. Our PPM calculation gives a peak overdensity of $\delta = 12.21$ at the redshift of $z_{\rm{ov}}$ = 2.476 $\pm$ 0.69, with a maximum significance radius of $R_{\rm{ppm}} = 2.24'$, and a maximum significance of $\sigma = 8.0$. All $\Delta z$ values are smaller than $z_{\rm{rms}}$, meaning that the structure is tight in the redshift space. The reproduced results are in perfect agreement with those measured by \citet{2016ApJ...828...56W}. The radius at 1/100-th peak overdensity, $R_w$, is $1.54'$, agreeing with that defined by \citet{2016ApJ...828...56W}. We show the JWST RGB image of the CL~J1001+0220 protocluster with a FoV$\sim$1~$\rm{pMpc^2}$ (bottom-left) which is perfect agreement with our reproduction, and JWST RGB image of the four protocluster SMG members labelled $a$--$d$ with FoV$\sim 5\arcsec \times 5\arcsec$ (bottom-right).

The other two Category A SMGs belong to the same AzTEC-3 protocluster, including the SMG AzTEC-3 \citep{2011Natur.470..233C} and the Lyman-break galaxy (LBG) LBG-1/HZ6 \citep{2014ApJ...796...84R}, at $z \sim 5.3$. This region was the first SMG-associated environment observed at  $z > 5$  and featuring significant overdensity \citep{2011Natur.470..233C}. 

\begin{figure*}[htb]
    \centering
\begin{minipage}{0.465\linewidth}
    \centering
    \includegraphics[width=\textwidth]{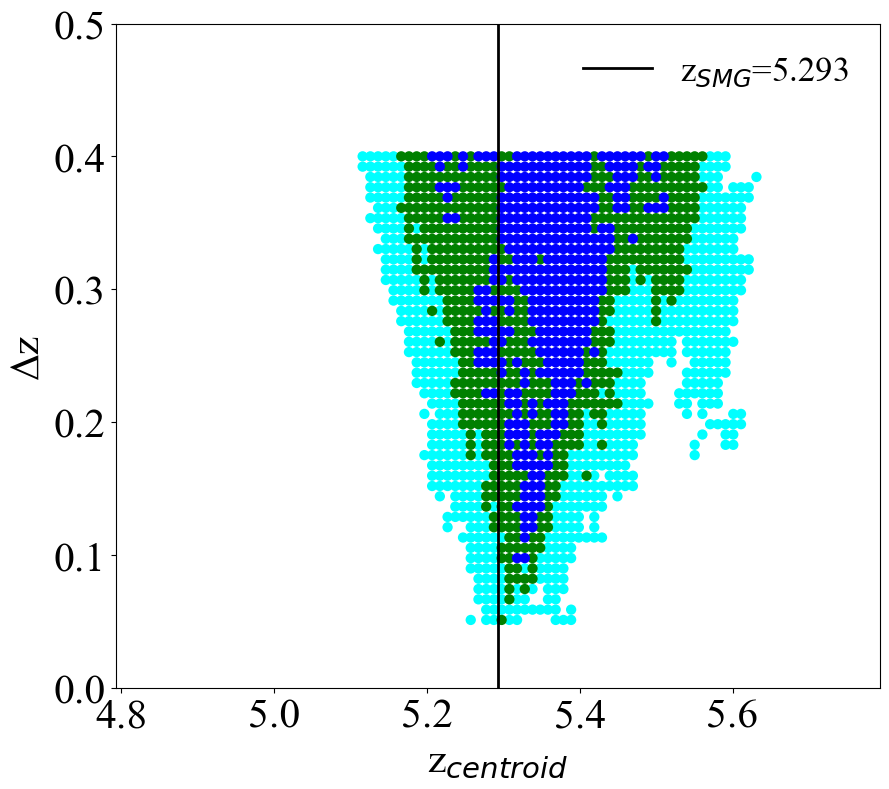}
\end{minipage}
\begin{minipage}{0.525\linewidth}
    \centering
    \includegraphics[width=\textwidth]{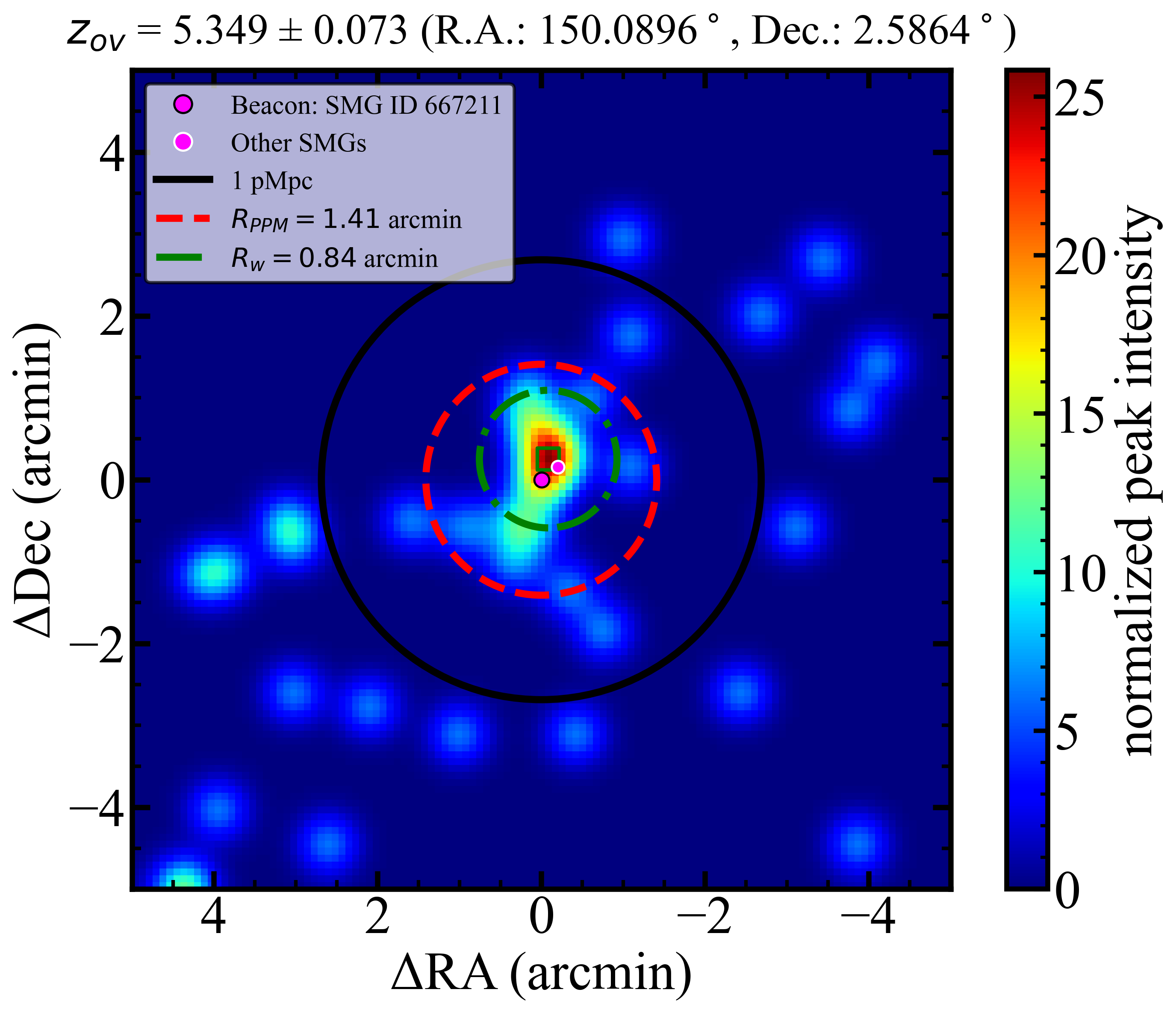}
\end{minipage}
\begin{minipage}{0.90\linewidth}
    \centering
    \includegraphics[width=\textwidth]{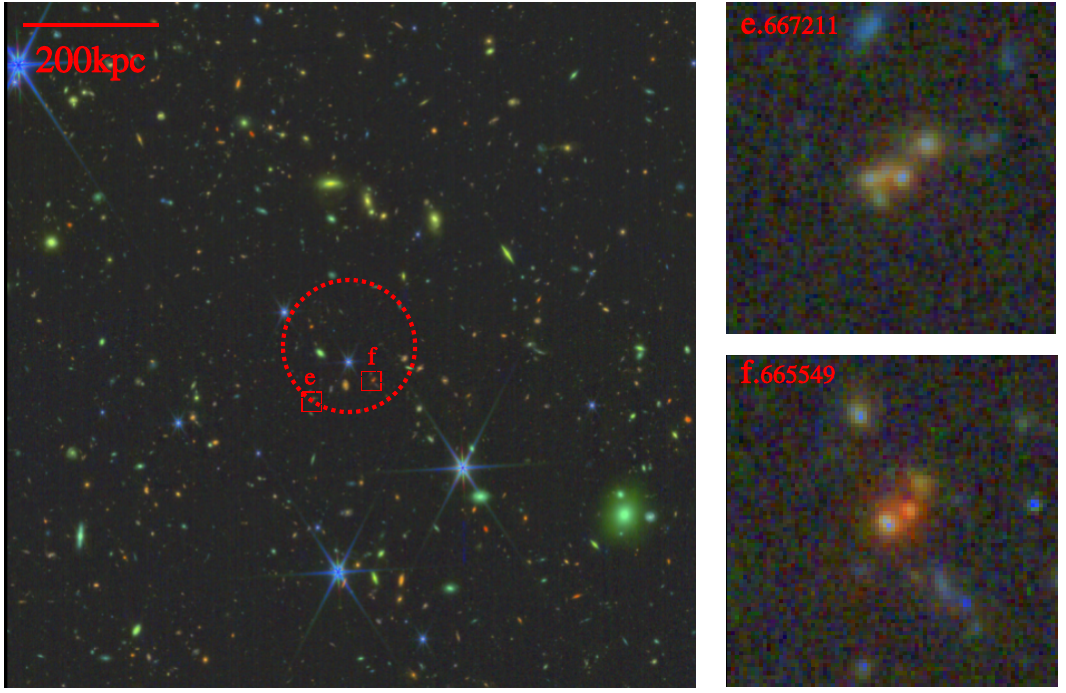}
\end{minipage}
\caption{
Similar to Figure~\ref{fig:4.0}, but for AzTEC-3 protocluster, which is another example of our identification of the previously known densest $z \gtrsim 5$ protoclusters (Section~\textcolor{red}{4.3}). 
\label{fig:4.00}
}
\end{figure*}

Our PPM results of SMG~(ID:667211) are shown in the top-left panel of Figure~\ref{fig:4.00}, same as the result of CL~J1001+0220, the `inverted triangle' shape indicates the peak overdensity redshift $z_{\rm{ov}}$ and aligns well with the beacon's redshift. 2D density map~(top right) shows the high peak overdensity~($\delta = 24.79$) at $z_{\rm{ov}} = 5.349 \pm 0.73$, $R_{\rm{ppm}} = 1.41'$, and a significance of $\sigma = 5.17$, all agreeing well with the previous findings. 

In the following subsection, we further analyze the cross-matches between all Category A and B overdensity candidates and known protoclusters in the COSMOS field to better evaluate the extent to which they successfully recover previously identified overdense regions.

\subsection{Identification of known COSMOS $z \sim 2 \text{--} 5$ protoclusters} \label{sec:(Proto)cluster candidates}

Here we perform a detailed analysis of the cross-matches between the full sample of 54 Category A and B overdensity candidates associated with SMGs/SFGs and known protoclusters. We primarily conduct the matching based on Table~1 of \cite{Toni2025}, which provides a compilation of protoclusters and high-redshift clusters and groups known in the literature for the COSMOS field. This compilation includes known protoclusters from $z \approx 2$ to $z \approx 3.7$ (including \citealt{Diener2013, Forrest2023,2018AA...619A..49C}; see \cite{Toni2025} for a full reference list). This analysis aims to determine the extent to which the PPM parameters successfully recover previously identified overdense regions. Our cross-matching criteria are as follows: following the methodology of \cite{Toni2025}, for each Category A \& B overdensity candidate detected by PPM, we attribute a successful match to any known object that lies within a cylinder of radius $dr = 1.0$ Mpc/$h$ and redshift depth $dz = 0.05(1 + z)$. These values are consistent with the scatter values adopted in the literature for matching clusters and groups at high redshift (e.g., \citealt{Sarron2021}). The matching results for Category A and B overdensity candidates associated with SMGs are summarized in Table~\ref{table:3}. For successfully matched protoclusters, we mark the Name (SMG:ID) with superscripts ($^{a-f}$). Among the 11 overdensity candidates, 7 successfully match with 6 known protoclusters (including the AzTEC-3 protocluster \citep{2011Natur.470..233C}; Elent\'{a}ri-P 8 \citep{Forrest2023}; zCOSMOS-ID 26 \citep{Diener2013}; and Hyperion-P 4 and P 5 \citep{2018AA...619A..49C}, yielding a matching rate of 63.6\%.

We perform the same analysis on the full sample of 43 Category A and B overdensity candidates associated with SFGs, the results of which are presented in Table~\ref{table:sfg} of Appendix~\ref{sec:Category A and B overdensity candidate (Beacon:SFG)}. In this sample, 19 overdensity candidates successfully match with 12 known protoclusters, representing a success rate of 44.2\%. Consequently, more than half of our total sample of 54 Category A and B overdensity candidates associated with SMGs/SFGs match with known protoclusters. Notably, the matching rate for sources in Category A and above reaches 78.9\% (15/19). These results confirm that our PPM parameters effectively recover previously identified overdensities. We therefore designate all overdensity candidates belonging to Category A \& B as reliable overdensity candidates. We also note that approximately half of the Category A \& B candidates do not match any known protocluster; we further analyze the reliability of these new sources in the Section~\ref{New PPM-identified Overdensity Candidates}.

\subsection{New PPM-based Category A\&B Overdensity Candidates}\label{New PPM-identified Overdensity Candidates}

Although our Category A and B candidates exhibit a 100\% matching rate with the AMICO group catalog up to $z = 3.7$ (\citealt{Toni2025}), general galaxy groups and assembling protoclusters typically represent distinct physical scales and evolutionary stages within the cosmic web. To firmly establish the novelty of our findings, we systematically cross-match all 54 candidates against a comprehensive compilation of previously reported protoclusters and large-scale structures in the COSMOS-Web field~(see Section~\ref{sec:(Proto)cluster candidates} for details). Candidates lacking spatial and redshift associations with these historically confirmed structures are formally classified as new PPM-based overdensities. 

After checking duplicated sources in the literature, we find 26 new overdensity candidates at $z \sim 0.8-5.3$ out of 54 Category A\&B sources identified by our PPM analysis. Eleven of 54 overdensity candidates with the SMGs association~(PC~1-11) are listed in Table~\ref{table:3}, including their associated SMG IDs, the cross-matched known protoclusters as table notes, if any, and the beacon and overdensity redshifts ($z$ and $z_{\rm{ov}}$), respectively, overdensity richness $N_{\textrm{select}}$, significance ($\sigma$), maximum radius at 1\%$\delta$ ($R_{w}$), peak overdensity values ($\delta$) and peak positions. The remaining 43 Category A\&B overdensity candidates~(PC~12-54) detected using control sample SFGs as beacons are presented in Table~\ref{table:sfg} of Appendix~\ref{sec:Category A and B overdensity candidate (Beacon:SFG)}. 

We examine their JWST images using the COSMOS-Web survey \citep{Casey2023, Franco2025, Harish2025}, and visually confirm that 13 new overdensity candidates (PC~12--15, 30, 38, 41, 45, 49--52 and 54) have multiple sources of similar color in the RGB image. The JWST RGB cutouts of all overdensity candidates similar to Figures~\ref{fig:4.0}~and~\ref{fig:4.00} are available in our online repository\footnote{The data is available on Zenodo under an open-source 
Creative Commons Attribution license: 
\dataset[doi:10.5281/zenodo.22234688]{https://doi.org/10.5281/zenodo.22234688}.}.

We further check the latest COSMOS Spectroscopic Redshift Compilation Catalog by \cite{Khostovan2026}\footnote{\url{https://github.com/cosmosastro/speczcompilation}}, which contains 482,579 spectroscopic redshift of 266,284 unique objects in the 2-deg$^2$ COSMOS field from 138 individual spectroscopic programs, as well as private communication with many spectroscopic surveys (see references therein). 
We search for spec-$z$ counterparts for each of our new overdensity candidates to see whether we can confirm the existence of more than one spec-$z$ sources within each overdensity candidate.

To search for spec-$z$ members, we first select all spec-$z$ sources with a Quality Flag~$\geq$~2 (i.e., Confidence Levels \(\geq 80\%\)). We then perform a cross-match between these sources and all overdensity candidates based on their projected coordinates, requiring an angular separation \(\leq R_{\rm PPM}\)~(an estimate for the (proto-)cluster core size), to obtain the number of projected associated galaxies (\(N_{\rm projected}\)) for each overdensity candidate. Finally, we impose an additional constraint that the redshift of each source lies within the redshift range of the corresponding overdensity candidate (i.e., \(|z_{\rm spec} - z_{\rm ov}| \leq z_{\rm rms}\)), where \(z_{\rm ov}\) is the redshift of the overdensity candidate and \(z_{\rm rms}\) is its redshift uncertainty. This yields the number of spec-$z$ member galaxies (\(N_{\rm spec-z~member}\)) for each overdensity candidate.

All of our Category A and B overdensity candidates have associated projected high-$z$ sources. The largest number of projected associated galaxies is observed for PC 18 (SFG: 731169) with \(N_{\rm projected} = 904\) and $R_{\rm PPM} = 3.32'$, while the smallest number is for PC 51 (SFG: 250840) with \(N_{\rm projected} = 51\) and $R_{\rm PPM} = 1.21'$. After applying the redshift constraint (\(|z - z_{\rm ov}| \leq z_{\rm rms}\)), we find that \(\sim 58\%\) (15/26) of the new overdensity candidates contain at least one spec-$z$ member galaxy~(\(N_{\rm spec-z~member} \geq 1\)), compared to 100\% (28/28) of the overdensity candidates matched to known protoclusters. In total, about 80\% (43/54) of our Category A and B overdensity candidates are confirmed to have at least one spec-$z$ member galaxy~(\(N_{\rm spec-z~member} \geq 1\)). Among these, PC 13 (SFG: 61114) has the highest number of spec-$z$ member galaxies, with \(N_{\rm spec-z~member} = 85\), and all other overdensity candidates except for PC 52 (SFG: 352238) have more than one spec-$z$ members~(\(N_{\rm spec-z~member} \ge 2\)). ~\(\sim 54\%\) (14/26) new overdensity candidates have \(N_{\rm spec-z~member} \ge 2\).

The spec-$z$ cross-matching results are presented in Table~\ref{table4} of Appendix~\ref{sec:Spec-z member of overdensity candidates}, including the name of Category A \& B overdensity candidates; the estimate of the projected radius of overdensity candidates~($R_{\rm PPM}$); the redshift of the overdensity candidates from PPM analysis~($z_{\rm ov}$); the redshift uncertainty~($z_{\rm rms}$); the number of projected associated galaxies for each overdensity candidate~($N_{\rm projected}$) and the number of spec-$z$ member galaxies for each overdensity candidate~($N_{\rm spec-z member}$).

Above all, the high reliability of our 54 Category A and B overdensity candidates is robustly supported by JWST visual inspections and spectroscopic cross-matching, which confirms that $\sim 78\%$ of the total sample hosts at least two spec-$z$ member galaxy. Furthermore, the effectiveness of our PPM analysis in identifying previously unknown structures is strongly validated, as over half ($\sim 54\%$) of the 26 newly discovered candidates already have multiple confirmed spec-$z$ members.

\input{table2.tex}

\subsection{3D Distribution of Category A\&B overdensity candidates}

We use the RA, Dec and the comoving distance from the redshift ($z_{\rm{ov}}$) to reconstruct the 3D spatial distribution of the overdensity candidates identified in this work, to see whether they can indicate even larger-scale cosmic web structures. The large-scale structures at \(z \sim 2.5\) have a volume of hundreds of megaparsecs on each side—resulting in a total volume of \(\sim 10^5\) $\rm{Mpc}^3-$and extend over 500 Mpc along the line of sight. In contrast, the large-scale structures at \(z \sim 3.3\) are significantly more compact. The 3D spatial distribution analysis maps corresponding to these candidates are all included in Appendix~\ref{sec:Protocluster candidates catalog}. For example, in Figure~\ref{fig:3d_distribution}, each circle is an overdensity candidate, For the two sources connected by the black dashed line: their distance on the projection plane (RA, DEC plane) is $\leq$ 10 pMpc (i.e., the classic size of a protocluster), and their redshift difference is less than twice the maximum $z_{\rm{rms}}$ value between them ($\Delta z$ $\leq$ 2$z_{\rm{rms}}$).
For the two sources connected by the thick cyan solid line: their distance on the projection plane (RA, DEC plane) is $\leq$ 1 pMpc (i.e., the classic core size of a protocluster), and their redshift difference is less than the maximum $z_{\rm{rms}}$ value between them ($\Delta z$ $\leq$ $z_{\rm{rms}}$), and they further show large overdensity within the large projected volume of the COSMOS field. This relates to the hierarchical formation of the Universe and requires furthermore detailed comparison with simulations, which we defer to later works.

\begin{figure*}[hpt]
    \centering
\begin{minipage}{0.75\linewidth}
    \centering
    \includegraphics[width=\textwidth]{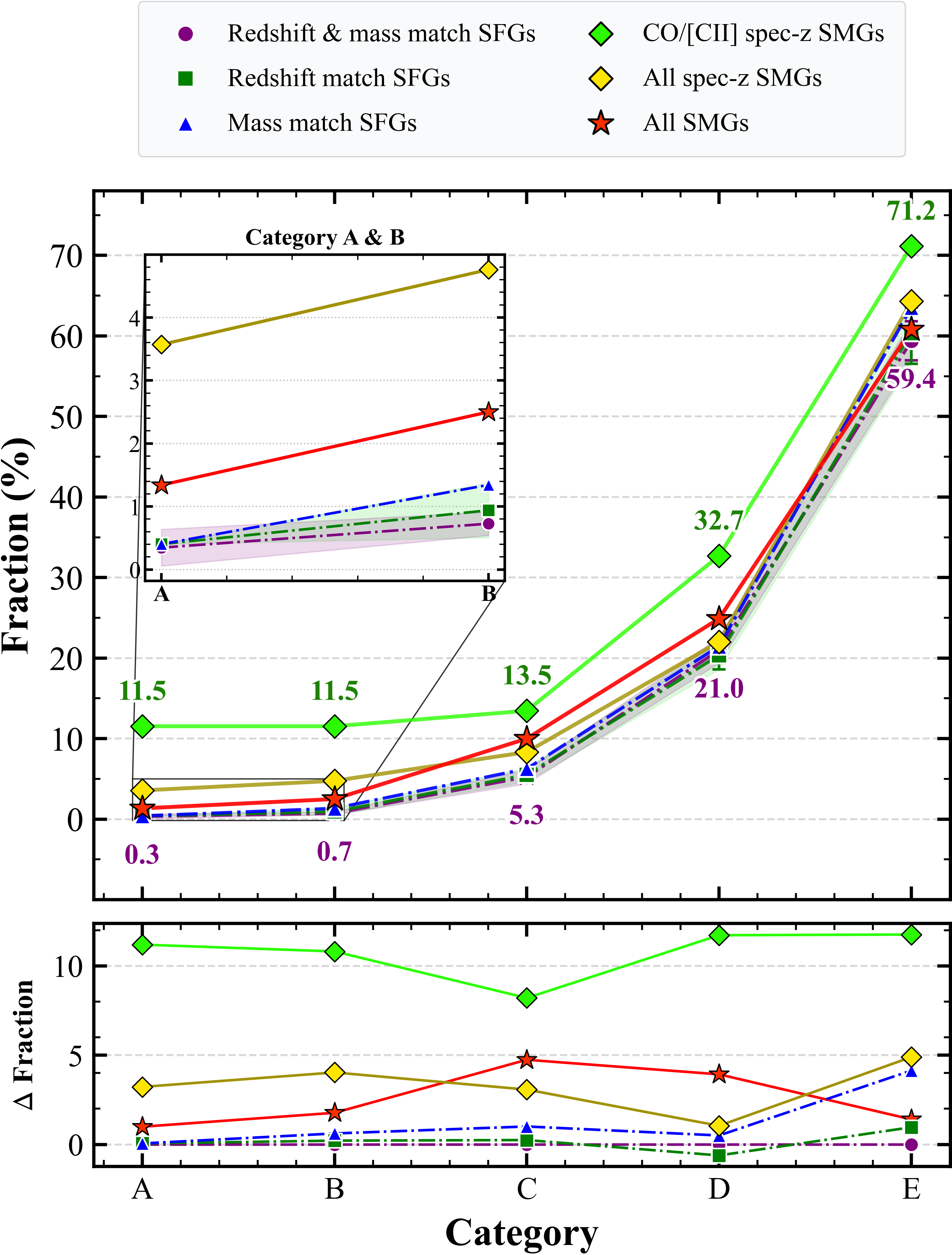}
    \begin{center}
    \end{center}
\end{minipage}
\caption{Statistical results of the sample fractions across the five environment categories (A--E; see Table~\ref{tab:table1}). \textbf{Top panel:} The absolute fractions for each sample type. The purple circles, green squares, and blue triangles represent the average fractions of sources in each category for the control samples: the redshift--mass-matched SFGs, redshift-matched SFGs, and mass-matched SFGs, respectively. Their corresponding dispersions are indicated by the shaded regions in matching colors. These shaded envelopes represent the $1\sigma$ standard deviation across the five independent iterations used to construct the SFG control samples. In contrast, the SMG samples (denoted by the green diamonds, gold diamonds, and red stars for the CO/\CIIline{} spec-$z$ SMG, all spec-$z$ SMG, and all SMG samples, respectively) are fixed, single observational datasets and are therefore plotted as exact fractions without uncertainty envelopes. An inset is provided in the upper left to show a zoomed-in view of the lower-fraction Category A and B. \textbf{Bottom panel:} The relative differences in fraction ($\Delta$ Fraction) for all samples, plotted with respect to the redshift--mass-matched SFG control sample (purple line at $\Delta = 0$). Overall, the fractions for the all SMG and all spec-$z$ SMG samples are generally higher than those of the control samples. Furthermore, the fractions of the CO/\CIIline{} spec-$z$ SMGs are significantly higher than the control samples across all categories. This demonstrates the superiority of SMGs as robust tracer beacons for (proto)clusters.}
\label{fig:5.0}
\end{figure*}

\subsection{Overdensity fractions between the SMG and SFG samples} \label{sec:Comparison with SFGs}

To understand whether SMGs trace overdensities, we quantified the fractions of overdensities across categories A--E for both the SMG and SFG samples. As shown in Figure~\ref{fig:5.0}, the fractions of overdensities found in each sample are indicated by the y--axis, and the five different criterion categories are listed as the x--axis. Under all five categories, the spec-$z$ SMG sample exhibits the highest overdensity fraction, and the control SFG samples shows the lowest overdensity fraction.

We construct the control SFG sample in three ways: redshift--mass matching, only redshift--matching, and only mass--matching to the SMG sample in redshift and stellar mass bins~(see Section~\ref{sec:Selecting Main Sequence Control Sample} for details). We randomly select a matched number of SFGs in each redshift and stellar mass bin, and repeat this for 5 times so as to obtain a scatter for the overdensity fraction (as indicated by the shaded area in Figure~\ref{fig:5.0}).

Figure~\ref{fig:5.0} clearly demonstrates that the fraction of the CO/\CIIline{} spec-$z$ SMGs in every category far exceeds those of the SFG control samples, with significance consistently above $3\sigma$~\footnote{This significance is computed as the difference between the fraction of SMGs and the mean fraction of the SFGs control sample, divided by the standard deviation of the SFGs sample. The standard deviation vanishes only for the redshift-matched SFGs in category~A; for this case we adopt $0.1$ as the error to prevent mathematical divergence. All other control-sample dispersions~($\geq 0.17$) are well above this limit and remain unaffected. For a Gaussian distribution, a 3$\sigma$ deviation corresponds to a 99.73\% containment probability.}. In particular, its fractions in Categories A--D far exceed those of all control samples at a significance level of $>10\sigma$. For instance, the fraction of this sample in Category A is approximately 33 times that of the redshift--mass matched SFGs, corresponding to a significance of nearly $100\sigma$. The fractions of the All spec-$z$ SMG sample in Categories A--C also surpass those of all control samples at a $\ge 3\sigma$ significance level, whereas in Category D and E, they show no significant difference from the SFG control samples. Similarly, the fractions of the All SMG sample in Categories A--D exceed all control samples at $\ge 3\sigma$. However, in Category E, the fraction of the All SMG sample is more than $3\sigma$ lower than those of the mass-matched and redshift-matched SFGs, while exhibiting little difference from the redshift--mass matched SFGs.

As shown in the bottom panel of Figure~\ref{fig:5.0}, we establish the redshift- and mass-matched SFGs (purple dash-dotted line with circles) as the zero-point baseline (i.e., $y=0$). This panel illustrates the absolute percentage excess (or deficit) of the remaining samples within the same overdensity categories, effectively decoupling the overall increasing trend. The $\Delta$ values for the all SMG (red stars) and all spec-$z$ SMG (gold diamonds) samples remain consistently positive across all categories, fluctuating roughly between +1\% and +5\%. Strikingly, the $\Delta$ curve for the CO/\CIIline{} spec-$z$ SMGs (green diamonds) lies significantly above those of all other samples. This provides robust statistical evidence that, compared to typical SFGs, SMGs generally exhibit a stronger propensity to reside in these environments, demonstrating a significant overdense tendency.
 
Our large--sample, statistically controlled analysis thus validates that SMGs are effective tracers of Mpc--scale large--scale structures. The overall overdensity detection capability, ordered from strongest to weakest, is: CO/\CIIline{} spec-$z$ SMGs (strongest) \(>\) All spec-$z$ SMGs \(>\) All SMGs \(>\) SFG control samples. Compared to the SFG control samples, SMGs show a substantial advantage in tracing denser and more significant overdense structures (e.g., Category A --D overdensities).

Interestingly, in Category E—where sources are at least tentatively associated with overdensities—the fraction of All SMGs and All spec-$z$ SMGs is 60.80\% and 64.29\%, respectively. Notably, even among the CO/\CIIline{} spec-$z$ SMGs only 71.15\% are found to reside within reliable overdensity structures, comparable to the fraction for radio galaxies at $z \sim 1-2$~\citep{2014ApJ...792..114C}. This raises the question: What triggers the intense star formation bursts in SMGs? A detailed discussion is provided in Section~\ref{sec:Mergers or secular evolution}.

We also find that differences exist between the overdensity fractions of CO/\CIIline{} spec‑$z$ SMGs and all SMGs. In Section~\ref{sec:Understanding the different overdense fraction in the spec-z SMG and photo-z SMG samples.}, we discuss whether these differences arise because photometric‑redshift errors dilute the overdensity signal, or because the CO/\CIIline{} spec‑$z$ sample contains a larger fraction of faint/bright, lower/higher‑mass SMGs --- given that overdensity may correlate with galaxy luminosity and stellar mass.

\begin{figure*}[ht]
\centering
\begin{minipage}{1\linewidth}
    \centering
    \includegraphics[width=0.9\textwidth]{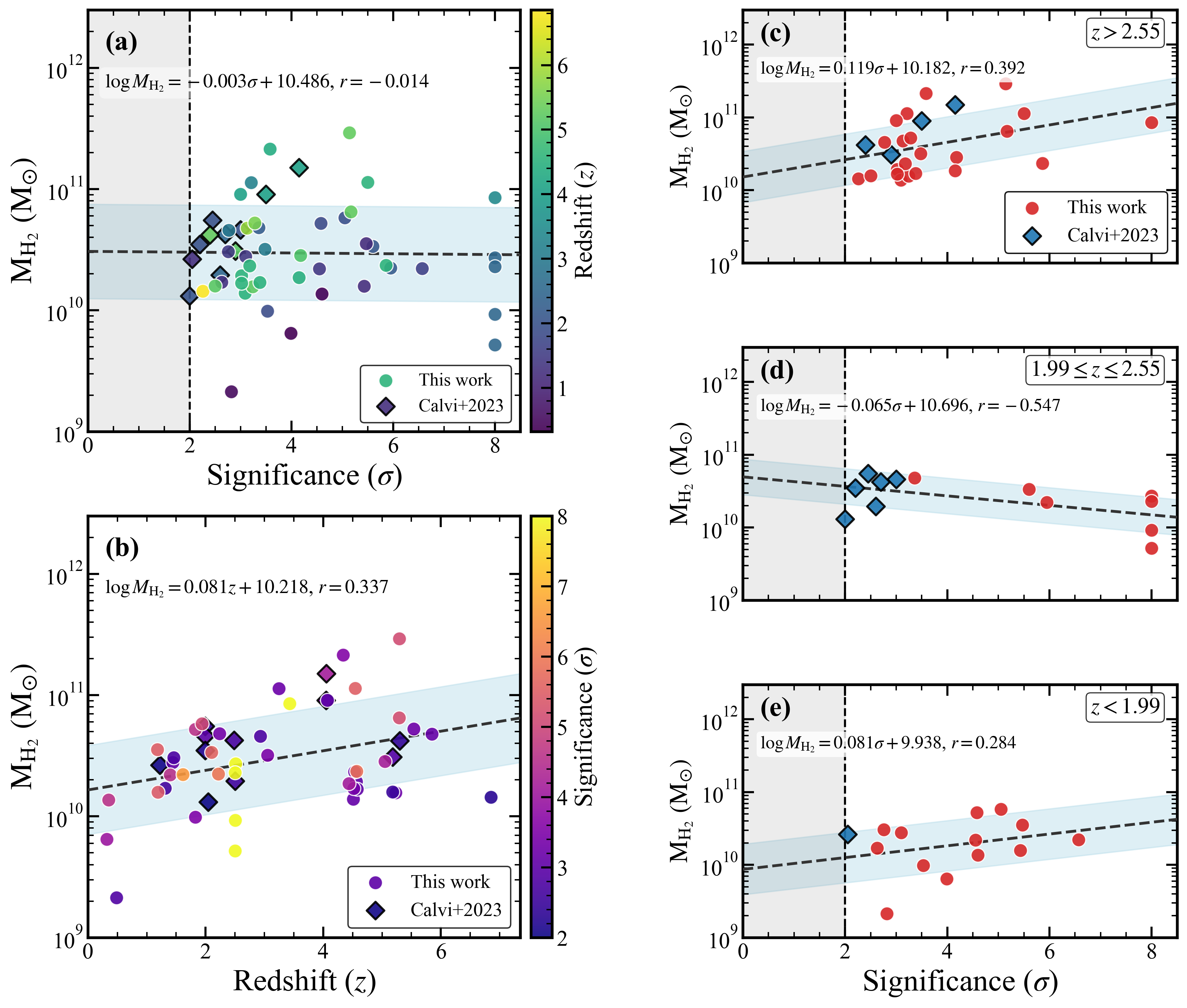}
    \begin{center}
    \end{center}
\end{minipage}
\caption{The relationship between molecular gas mass ($\rm M_{\rm H_2}$), overdensity significance ($\sigma$), and redshift ($z$) for the CO/\CIIline{} spec-$z$ SMGs sample. \textbf{Top-left panel~(a):} $\rm M_{\rm H_2}$ as a function of significance, color-coded by redshift. The global sample shows no significant overall correlation ($r = -0.014$). \textbf{Bottom left panel~(b):} $\rm M_{\rm H_2}$ as a function of redshift, color-coded by significance, revealing a distinct positive evolutionary trend ($r = 0.337$). \textbf{Right panels, from top to bottom~(c) --(e):} $\rm M_{\rm H_2}$ versus significance evaluated across three distinct redshift bins: $z > 2.55$, $1.99 \le z \le 2.55$, and $z < 1.99$, respectively. In all panels, circles represent the SMGs from this work, while diamonds denote the literature sample from~\cite{Calvi2023}. The black dashed lines indicate the best-fit log-linear relations, with the corresponding fitting equations and Pearson correlation coefficients ($r$) provided in each panel. The light blue shaded bands represent the $1\sigma$ scatter of the fits. The vertical dashed line and the gray shaded region mark the $\sigma < 2$ regime.}
\label{fig10}
\end{figure*}

\subsection{Molecular gas mass versus overdensity correlation} \label{sec:Molecular Gas Analysis}

We compile CO and \CIIline{} 158$\mu$m lines fluxes for our SMG sample from the literature (including, e.g., \citealt{2013ARAA..51..105C, 2018ApJ...867...92S, 2019ApJ...880...15K, 2020AA...641A.155V, 2025AA...699A..80H}, etc., see a full reference list in D. Liu et al. in prep.). We find that 47 of the 52 CO/\CIIline{} spec-$z$ SMGs are detected with S/N~$>3$ in at least one CO and \CIIline{} 158$\mu$m lines. We convert the line fluxes to molecular gas masses using empirical conversion factors as follows:
\begin{enumerate}
    \item \textbf{Calculating the CO/\CIIline{} line luminosity.} For sources with measurements in multiple $\mathrm{CO}(J\rightarrow{}J-1)$ transitions, we adopt the lowest-J transition, as lower-J CO lines are excited at lower gas densities and temperatures \citep{2021ApJ...909...56L}. We calculate the CO and \CIIline{} line luminosity using the formalism of \cite{2005ARA&A..43..677S}:
    \begin{multline}
    L^{\prime}_{\mathrm{CO}(J\rightarrow{}J-1)} = 3.25 \times 10^{7}S_{\mathrm{CO}(J\rightarrow{}J-1)}\Delta{v}~\nu_{obs}^{-2}\\(1+z)^{-3}D_L^2,
    \label{equation4}
    \end{multline}
    \begin{multline}
    L_{\mathrm{\CIIline}} = 1.04 \times 10^{-3} S_{\mathrm{\CIIline}} \Delta{v}~\nu_{rest}\\(1+z)^{-1}D_L^2
    \label{equation5}
    \end{multline}
    where the $L^{\prime}_{\mathrm{CO}(J\rightarrow{}J-1)}$ is measured in $\rm{K~km~s}^{-1}~\rm{pc}^2$ and the $L_{\mathrm{\CIIline}}$ in $L_{\odot}$; the velocity-integrated flux, $S_{\rm CO(J\rightarrow{}J-1)} \Delta v$ and $S_{\mathrm{\CIIline}} \Delta v$, in $\rm Jy\ km\ s^{-1}$; the rest frequency, $\nu_{\rm rest} = \nu_{\rm obs}(1 + z)$, in GHz; and the luminosity distance, $D_{\rm L}$, in Mpc.
    For SMGs with the $\mathrm{CO}(J\rightarrow{}J-1)$ emission line, we calculate their $L^{\prime}_{\mathrm{CO}(1-0)}$ based on the excitation ratio $r_{J1}$ = $L^{\prime}_{\mathrm{CO}(J\rightarrow{}J-1)}/L^{\prime}_{\mathrm{CO}(1\text{--}0)}$ = 1.0, 0.9, 0.63, 0.34, 0.36 and 0.27 for $J = $~1, 2, 3, 4, 5 and 6 from \cite{2021MNRAS.501.3926B}.
    
    \item \textbf{Deriving the mass of molecular gas.} For SMGs with the CO$(J\rightarrow J-1)$ emission line, the molecular gas mass estimated assuming a starburst CO-to-H$_{2}$ conversion factor $\alpha_{\mathrm{CO}} = 0.8 \, \mathrm{M_{\odot} (K~km~s^{-1}~pc^2)^{-1}}$ \citep[e.g.,][]{1998ApJ...507..615D, 2021MNRAS.501.3926B}. The \CIIline-to-H$_{2}$ conversion factor $\alpha_{\mathrm{\CIIline}} = 30 \; \mathrm{M_{\odot}/L_{\odot}}$ is taken from \cite{2018MNRAS.481.1976Z}.
\end{enumerate}
Finally, we obtain the CO(1--0) and \CIIline{} line luminosity-based molecular gas mass (M$_{\mathrm{H_2}}$) for all 47 SMGs. These results, along with their spectroscopic redshifts, emission line information, IDs from the literature, referenced studies, and the overdensity significance ($\sigma$) calculated by us based on PPM analysis, are presented in Tables \ref{tab:CO} and \ref{tab:cII} of Appendix~\ref{sec:spectroscopy match SMG}.

In Figure~\ref{fig10}, we present the $\rm M_{\rm H_2}$ as a function of significance, color-coded by redshift. Diamonds outlined in black correspond to the 11 SMGs located in overdensity environments from \cite{Calvi2023}. The light-gray shaded region marks the minimum overdensity significance threshold adopted in this work (2$\sigma$). The black dashed lines indicate the best-fit log-linear relations, with the corresponding fitting equations and Pearson correlation coefficients ($r$) provided in each panel. The light blue shaded bands represent the $1\sigma$ scatter of the fits. Shown as panel (a), the $\rm M_{\rm H_2}$ exhibits no overall correlation with the environmental density ($r = -0.014$); instead, $\rm M_{\rm H_2}$ correlates significantly with redshift (panel b; $r = 0.337$), this result is consistent with the findings of cold molecular gas evolution (e.g., \citealt{2018ApJ...853..179T, 2019ApJ...887..235L, 2020ARA&A..58..157T}). This suggests that SMGs at earlier cosmic epochs inherently possess more massive molecular gas reservoirs.

We evaluate the $\rm M_{\rm H_2}$ --$\sigma$ relation across three distinct redshift bins, the selection of these redshift intervals is matched to the distribution of the~\cite{Calvi2023} sample. At high redshifts ($z > 2.55$, panel c), a positive correlation emerges ($r = 0.392$), suggesting that dense protocluster environments facilitate efficient cold gas accretion and allow galaxies to maintain larger gas reservoirs. Strikingly, this relationship reverses into a strong negative correlation ($r = -0.547$) during the cosmic noon ($1.99 \le z \le 2.55$, panel d) and a weak positive trend reappears ($r = 0.284$) at lower redshifts ($z < 1.99$, panel e).

\section{discussion} \label{sec:discussion}

\subsection{Do SMGs trace Mpc Large-Scale Structures?} \label{sec:Do SMGs trace}

In individual (proto)cluster cases, SMGs are often found in high-$z$ (e.g., $z>2$) overdensity structures. For instance, we detect the overdensity at the peaks of two typical protoclusters in COSMOS field---CL J1001+0220 and the AzTEC-3 regions---both of which have extremely high overdensity values. Combined with the seven overdensity peaks and two massive galaxy groups of Hyperion successfully reproduced in this work, SMGs are found within the $R_{w}$ (the radius where the peak density drops to $1\%$) of $5/10$ overdensities. As shown in Figure~\ref{fig8} in Appendix~\ref{sec:PPM Reproduction Results}, in contrast, SMGs are distributed across the entire $5' \times 5'$ ($\sim 6 \times 6~h^{-2}\,\mathrm{Mpc^2}$ at $z \sim 2.5$) field of view in all 10 overdensities. We note that the scale of an entire protocluster structure often exceeds $15~h^{-1}\,\mathrm{Mpc}$ before it reaches virialization \citep{Muldrew2015}.

\cite{Casey2016} studied several high-$z$ protoclusters in detail~(including the Hyperion and AzTEC-3 protoclusters) and found that these high-redshift environments are rich in rare (most-luminous) galaxies—including DSFGs/SMGs and ultraluminous AGNs. These are in line with our findings that protoclusters at high-$z$ ($z>2$) usually habors SMGs. 

Based on a large statistical sample, our study investigates the overdensity fraction and reveals that overdensities are significantly stronger around SMG beacons compared to the control SFG beacons. It is worth noting that when selecting our control main-sequence SFGs, we did not exclude the small fraction of SMGs that are on the main sequence ($\sim 2.2\%$; 116/5296, see Section~\ref{sec:Selecting Main Sequence Control Sample}). This helps to compare the overdensity fraction of the whole main-sequence galaxy population. However, if all SMGs (e.g., by SFR or 850$\mu$m flux selection) were excluded from the control sample, the contrast in the overdensity fraction between SMGs and control-SFGs would be even larger. In any considerations, the SMGs have a higher overdensity fraction than the bulk of main-sequence galaxy population.

More quantitatively, as shown in Figure~\ref{fig:5.0}, the fraction of CO/\CIIline{} spec-$z$ SMGs in Category E is 71.15\% --- which is comparable to the fraction for radio galaxies at $z \sim 1 - 2$~\citep{2014ApJ...792..114C}. All SMGs samples have higher overdensity fractions than SFG controls, in excess by \(\geq 3\sigma\) in categories A--C --- establishing SMGs as statistically significant tracers. The CO/\CIIline{} spec-$z$ SMGs perform even better: \(\geq 3\sigma\) higher across all categories, with a ~33--fold excess in Category A, confirming that they are significant overdensity tracers. 

The causality of SMG triggering in overdense environments may still be an open question, but they should have physical connections including the higher accretion rate of gas in overdense, large halos, and the higher merging/interacting rate in (proto)cluster environments which can lead to more efficient gas inflowing and starbursting.

Nevertheless, we also notice that the overdensity fraction remains below 100\% even within the CO/\CIIline{} spec-$z$ SMG sample. This suggests that SMG activity is not exclusively triggered by dense environments, but also by environment-independent mechanisms such as major mergers. We discuss this further in Section~\ref{sec:Mergers or secular evolution}.

\begin{figure*}[ht]
\centering
\begin{minipage}{0.95\linewidth}
    \centering
    \includegraphics[width=1\textwidth, trim=0 7mm 0 0]{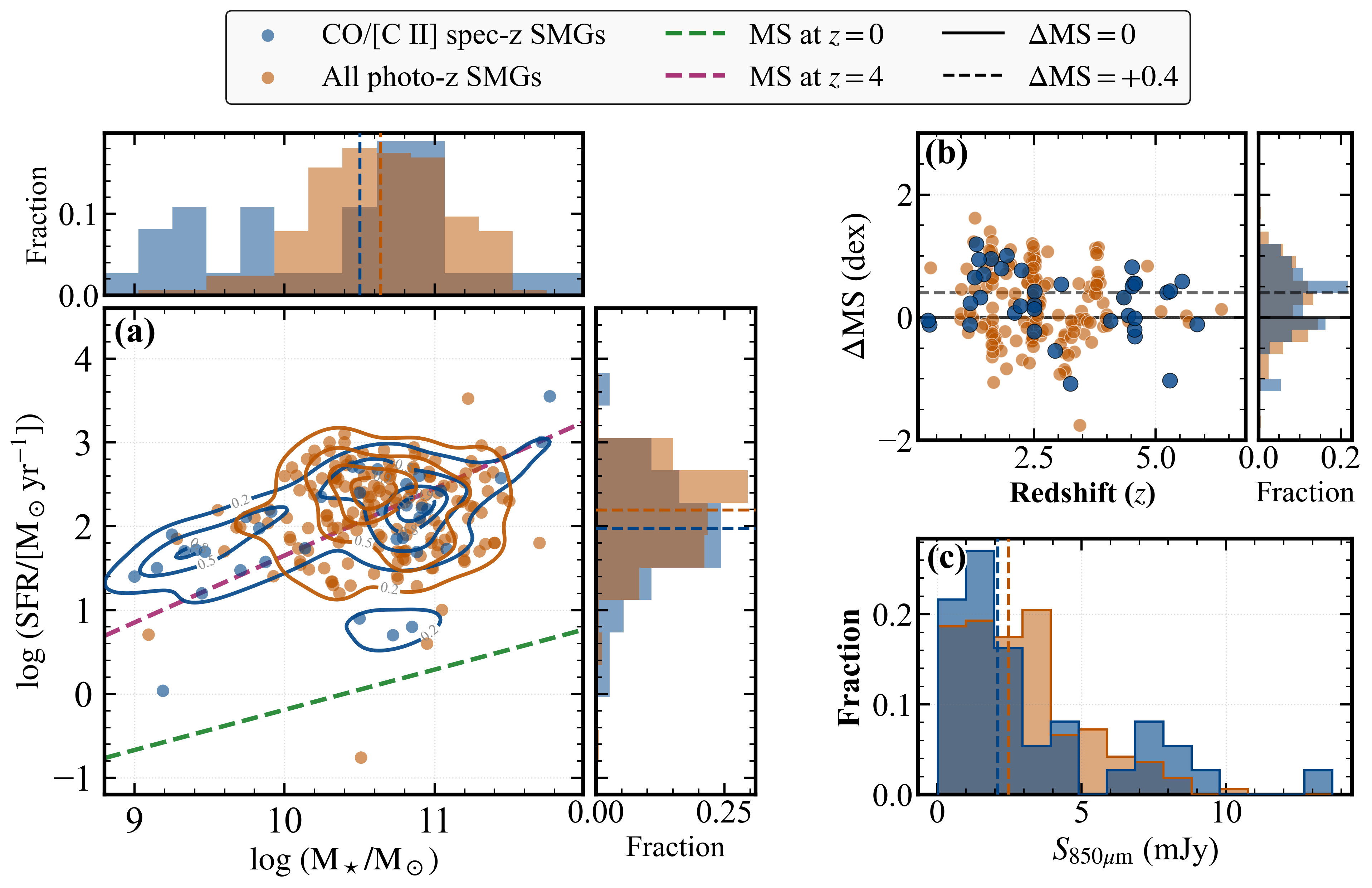}
    \begin{center}
    \end{center}
\end{minipage}
\caption{Physical properties of the SMG samples. The blue and orange dashed lines in the marginal histograms of panel (a) and in panel (c) indicate the median values of the CO/\CIIline{} spec-$z$ and photo-$z$ SMG distributions, respectively. \textbf{Left panel~(a):} SFR-$\rm{M_\star}$ relations. The blue and orange points represent the CO/\CIIline{} spec-$z$ SMGs and the full photo-$z$ SMG sample, respectively. The corresponding solid contours indicate the two-dimensional density distributions of the data points. The top and right subpanels show the normalized distributions of stellar mass and SFR. The green and magenta dashed lines represent the star-forming main-sequence at $z=0$ and $z=4$, respectively. \textbf{Top right panel~(b):} The offset from the main sequence, $\Delta\text{MS}$ as a function of redshift. The right subpanel displays the normalized distribution of $\Delta\text{MS}$. The solid black line marks $\Delta\text{MS} = 0$, while the dashed black line marks $\Delta\text{MS} = +0.4$, which is typically used as the threshold for starburst galaxies. \textbf{Bottom right panel~(c):} Normalized fraction distributions of the 850 $\mu$m flux density ($S_{850\mu\text{m}}$) for the two SMG samples. The vertical dashed lines indicate the median flux densities for each respective sample.
\label{fig:MS_Mass_Flux}}
\end{figure*}

\subsection{Understanding the different overdense fraction in the CO/\CIIline{} spec-z and photo-z SMG samples} \label{sec:Understanding the different overdense fraction in the spec-z SMG and photo-z SMG samples.}

Although the COSMOS2025 catalog is one of the best high-$z$ samples covered by almost 40 bands, the photometric redshifts can still be uncertain for overdensity studies. To verify whether the photo-$z$ selection is biased or not, we compare the 166/281 photo-$z$ SMGs associated with all detected overdensities against the CO/\CIIline{} spec-$z$ SMGs~(37/52); their $\rm{M}_\star$--SFR relations are presented in Figure~\ref{fig:MS_Mass_Flux}~panel~(a). In the main panel, the blue and orange points represent the CO/\CIIline{} spec-$z$ SMGs and all photo-$z$ SMGs, respectively (see Appendix~\ref{sec:Redshift Sources for the All SMGs Sample}, Table~\ref{tab:ztype} for the complete list of photo-$z$ matching sources). The correspondingly colored normalized histograms in the subpanels show the SFR and stellar mass distributions for the two samples.

As indicated in Figure~\ref{fig:MS_Mass_Flux}~panel~(a), both samples exhibit a primary peak distribution around $\log (\rm{M}_\star/\rm{M}_\odot) \sim 10.5$--$11$ and $\log (\rm{SFR}/[\rm{M}_\odot~\rm{yr}^{-1}]) \sim 2$--$3$. However, the CO/\CIIline{} spec-$z$ SMG sample displays a distinct secondary peak to the left, located at approximately $\log (\rm{M}_\star/\rm{M}_\odot) \sim 9$--$10$ and $\log \text{SFR} \sim 1.5$--$2$. Considering the entire samples, the median stellar mass of the CO/\CIIline{} spec-$z$ SMGs is $\log (\rm{M}_\star/\rm{M}_\odot) = 10.50^{+0.42}_{-1.06}$ (where the sub/superscripts represent the 16th/84th percentiles, respectively), compared to $\log (\rm{M}_\star/\rm{M}_\odot) = 10.64^{+0.50}_{-0.44}$ for the photo-$z$ SMGs. A two-sample Kolmogorov-Smirnov (K-S) test performed on the full distributions yields a $p$-value of $0.0071$. This statistically confirms that the overall distributions are inherently different, a divergence driven primarily by the excess population of low-mass galaxies that comprise the secondary peak in the CO/\CIIline{} spec-$z$ sample. The survival of these low-mass, actively star-forming systems in overdense regions is theoretically consistent with \cite{Ghaffari2026}, which demonstrate that environmental quenching is highly inefficient at lower masses ($\log(\rm{M}_\star/\rm{M}_\odot) < 10.5$), allowing them to maintain robust star formation even within dense group environments.

Figure~\ref{fig:MS_Mass_Flux}~panel~(b) presents the deviation from the star-forming main sequence ($\Delta$MS) as a function of redshift ($z$) to further quantify their star-forming modes. A two-sample K--S test finds no statistically significant difference between the two $\Delta_{\rm MS}$ distributions ($p=0.2945$). Both distributions exhibit substantial intrinsic scatter, with their 16th--84th percentile ranges spanning $\gtrsim 0.9$ dex. The CO/[C\,\textsc{ii}]-detected spec-$z$ SMGs do have a somewhat higher median $\Delta_{\rm MS}$ than the photo-$z$ SMGs, at $0.32^{+0.45}_{-0.46}$ dex compared with $0.22^{+0.60}_{-0.56}$ dex, respectively. Here, the values quoted after the medians represent the asymmetric scatter between the 16th and 84th percentiles. Thus, although the median $\Delta_{\rm MS}$ values differ modestly, the large scatter and substantial overlap between the two distributions are consistent with the non-significant K--S test result. Consequently, although a seemingly larger fraction of the CO/\CIIline{} spec-$z$ galaxies (45.9\%, 17/37) exceeds the starburst threshold of $\Delta\text{MS} > +0.4$ dex compared to the photo-$z$ SMGs (38.6\%, 64/166), this apparent excess is not a physical distinction, but rather an artifact of small-number statistics~($N=37$).

To evaluate the effect of sub-millimeter flux densities, we use the best-fit spectral energy distributions (SEDs) from A$^3$COSMOS. We extract the observed-frame best-fit 850~$\mu$m flux densities $S_{850\mu \rm{m}}$ for all 449 SMGs~(with median $S_{850\mu \rm{m}} = 2.37_{-1.49}^{+2.98}$~mJy, and mean $3.04 \pm 2.32$~mJy). As shown in Figure~\ref{fig:MS_Mass_Flux}~panel~(c), the median flux of the CO/\CIIline{} spec-$z$ SMGs is $2.10^{+5.43}_{-1.44}$ mJy, slightly lower than the $2.47^{+2.58}_{-1.64}$ mJy median of the photo-$z$ SMGs. Similarly to the $\Delta$MS analysis, a K-S test ($p = 0.3425$) confirms that this modest difference in dust continuum emission lacks statistical significance.

In summary, while visual inspections might suggest that the CO/\CIIline{} spec-$z$ and the photo-$z$ SMGs differ across multiple properties, rigorous statistical testing isolates the stellar mass as the only fundamentally distinct physical parameter between the two samples. Ultimately, disentangling the genuine physical mass-environment relation from these complex observational biases remains challenging. Future studies with significantly larger, homogeneously selected spec-$z$ SMG samples are required.

\begin{figure}[!htbp]
    \centering
    \includegraphics[width=1\linewidth]{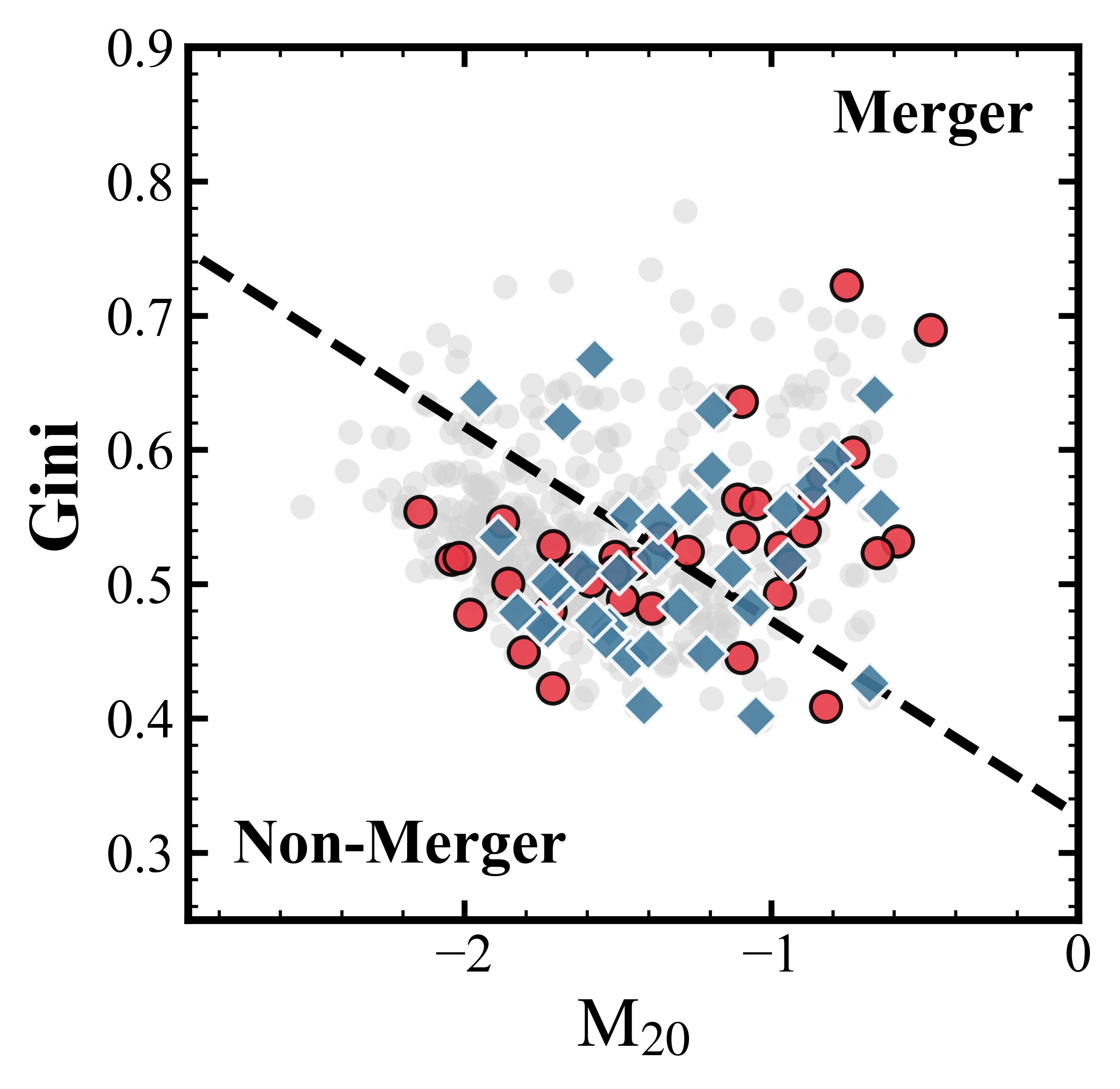}
    \caption{Gini-M$_{20}$ relation for the 447 SMGs based on their JWST NIRcam F277W band images. Red circles and blue diamonds represent the brightest SMGs~($S_{850\mu m} \geq$ 5.36 mJy, N=72) residing within overdense regions and non-overdense environments, respectively. The underlying light gray points indicate the distribution of the remaining, fainter SMG population. The black dashed line denotes the merger criterion we adopted from~\mbox{\cite{2008ApJ...672..177L}}. Galaxies located above this boundary are classified as mergers, whereas those below are classified as non-mergers.\label{fig:11}}
\end{figure}

\begin{figure*}[!ht]
    \centering
    \includegraphics[width=1\linewidth]{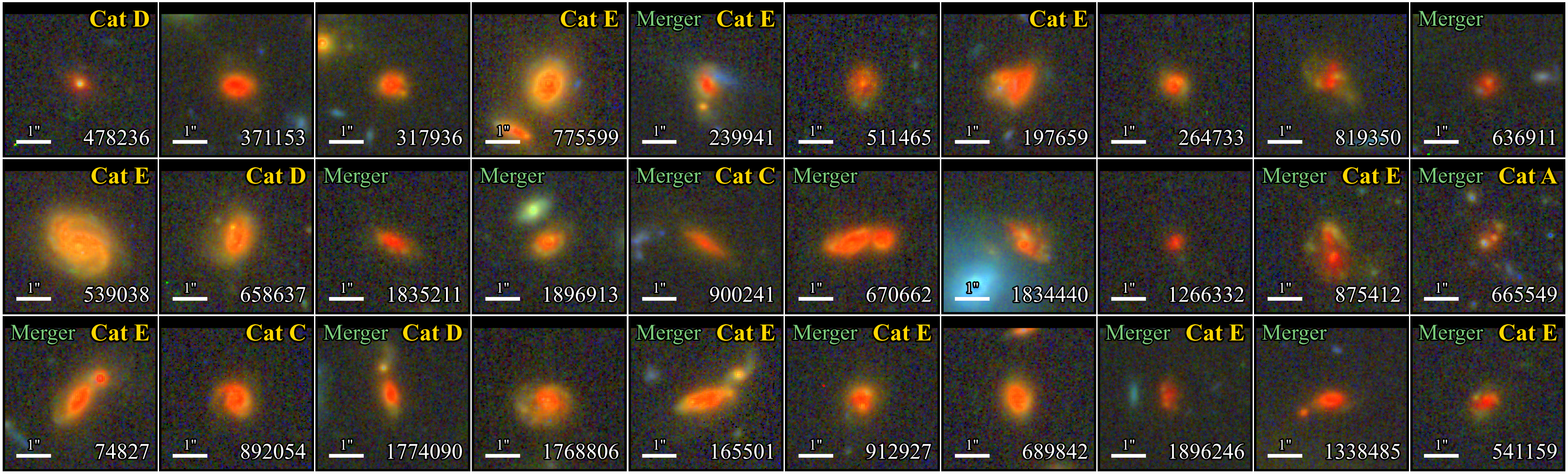}
    \caption{JWST RGB images of the 30 brightest SMGs with a 850 $\mu$m flux $S_{850\mu m} \geq 5.36$ mJy. Each panel presents a $5'' \times 5''$ field of view, with a solid white scale bar in the bottom-left corner representing $1''$. The morphological classification derived from the Gini-M$_{20}$ analysis is indicated in the top-left corner of each cutout, distinguishing between merger (green text) and non-merger systems. The overdensity identification category~(A to E, see Table~\ref{tab:table1} for details) of SMGs shown in the top-right corner. The corresponding source ID from the COSMOS2025 catalog is provided in the bottom-right corner of each image. The complete RGB images~(contains all 72 brightest SMGs) are available at: \url{https://github.com/chance-deng/PPM-analysis-for-COSMOS-Web-field.}\label{fig:placeholder}}  
\end{figure*}

\subsection{SMG formation mechanisms: merger, environment, or secular evolution?} \label{sec:Mergers or secular evolution}

We find that even among the CO/\CIIline{} spec-$z$ SMGs sample with the loosest criterion category (Category E), only 71.15\% are associated with overdensity structures. We also find that, when examining the brightest SMGs ($S_{850\,\mu\rm{m}}$ within the top 16th-percentile, $\gtrsim 5.3$~mJy), their overdensity fraction is even lower than the bulk of SMGs in Category E. 

This naturally raises the question of how these exceptionally bright SMGs trigger their extreme starburst activity in non-overdense environments. Since high gas accretion rates at the overdense nodes of LSS cannot fully explain our statistics, we suggest that major mergers may serve as an alternative mechanism for triggering these starburst SMGs in less-dense regions.

We analyzed morphologies of all 449 SMGs using JWST images from the COSMOS-Web survey. Because SMGs are intrinsically very red, they are typically faint in the shorter-wavelength F115W and F150W bands. Consequently, only the F277W and F444W bands were viable for our study. Between the two, we selected the F277W band because it offers superior angular resolution. We therefore performed the morphological analysis using \verb|Statmorph (Version: 0.6.0)|~\citep{2019MNRAS.483.4140R} on the JWST NIRCam F277W images for the all SMGs. This choice of redder, high-resolution NIRCam filters (e.g., F277W) is a well-established strategy in recent COSMOS-Web morphological studies to effectively trace rest-frame optical structures while minimizing biases from severe dust attenuation \citep{Ghaffari2026}. After running the program, the Gini-M$_{20}$~\citep{Conselice2003,Lotz2004,Lotz2008} parameter results for 447 sources were finally obtained. 

To evaluate the morphological properties of our sample, we adopt the empirical Gini--$\rm M_{20}$ classification boundary from \cite{2008ApJ...672..177L}. The Gini--$\rm M_{20}$ diagram is a widely used method for distinguishing galaxies with potential merger activity from non-merging systems based on two metrics. The Gini coefficient ($G$) quantifies the inequality of a galaxy's relative flux distribution; $G=1$ indicates that all light is concentrated in a single pixel, whereas $G=0$ represents a perfectly uniform distribution. The $\rm M_{20}$ parameter measures the normalized second-order spatial moment of the brightest 20\% of a galaxy's pixels. Galaxies featuring bright, clumpy, and spatially dispersed fragments exhibit higher $\rm M_{20}$ values, while those with less dispersed light distributions yield lower values. By combining these two metrics, this empirical boundary effectively distinguishes highly disturbed, merger-like candidates from relatively undisturbed systems (Figure~\ref{fig:11}). Based on this diagnostic, the brightest SMGs exhibit a substantially enhanced merger fraction compared to the general population. Specifically, we find that 47.2\% (17/36, red circles) of the brightest SMGs residing in overdense regions and 50.0\% (18/36, blue diamonds) of their non-overdense counterparts are classified as mergers. In contrast, the underlying, fainter SMG population ($N=375$, light gray points) displays a lower merger fraction of 28.8\%. This baseline fraction is broadly consistent with typical sub-millimeter samples. For instance, \cite{Chan2025} analyze 125 SMGs in the COSMOS field using a similar Gini--$\rm M_{20}$ approach and report an elevated merger rate in their bright group (27.5\%) relative to their faint group (21.3\%). Although they adopt a simplified power-law relation ($S_{850\mu\text{m}}/S_{\text{obs}} = (\nu_{850\mu\text{m}}/\nu_{\text{obs}})^{3.8}$) to extrapolate fluxes across different observing bands, their observed trend of higher merger fractions in brighter sources is well aligned with our results. This morphological dichotomy suggests that the triggering mechanisms for extreme starbursts scale strongly with sub-millimeter luminosity. For the general SMG population, the $\sim$29\% merger fraction indicates that their star formation is likely sustained by a diverse combination of physical processes, potentially including secular evolution, minor interactions, and steady cold gas accretion within large-scale environments. Conversely, the high merger rates ($\sim$49\%) observed among the brightest SMGs, regardless of their large-scale environmental density, suggest that the most extreme dust-obscured starbursts are predominantly driven by local major mergers or violent disk instabilities~(\citealt{Dekel2020}).

In Figure \ref{fig:placeholder}, we present JWST RGB images of the 30 brightest SMGs ($S_{850\mu \rm m} \ge 5.36$ mJy) to visually investigate their morphological properties. Each cutout displays a $5^{\prime\prime} \times 5^{\prime\prime}$ field of view centered on the source, annotated with its Gini-M${20}$ morphological classification, overdensity category, and unique COSMOS2025 source ID. Overall, these high-resolution images reveal substantial morphological diversity within the brightest SMG population. Crucially, visual inspection demonstrates that even among the SMGs formally classified as non-mergers by the Gini-M${20}$ analysis, almost all sources are clumpy and/or exhibit disturbed morphologies with multiple components that are strongly indicative of ongoing or recent merger activity. A notable exception to this trend is COSMOS2025 ID 539038, which displays a relatively clear, undisturbed disk-like structure. This widespread prevalence of disturbed/clumpy morphologies, independent of large-scale environmental effect, suggests that major mergers or significant local gravitational instability act as the primary triggering mechanism for the most extreme dust-obscured starbursts.

\begin{figure}[htbp]
    \centering
    \includegraphics[width=1\linewidth]{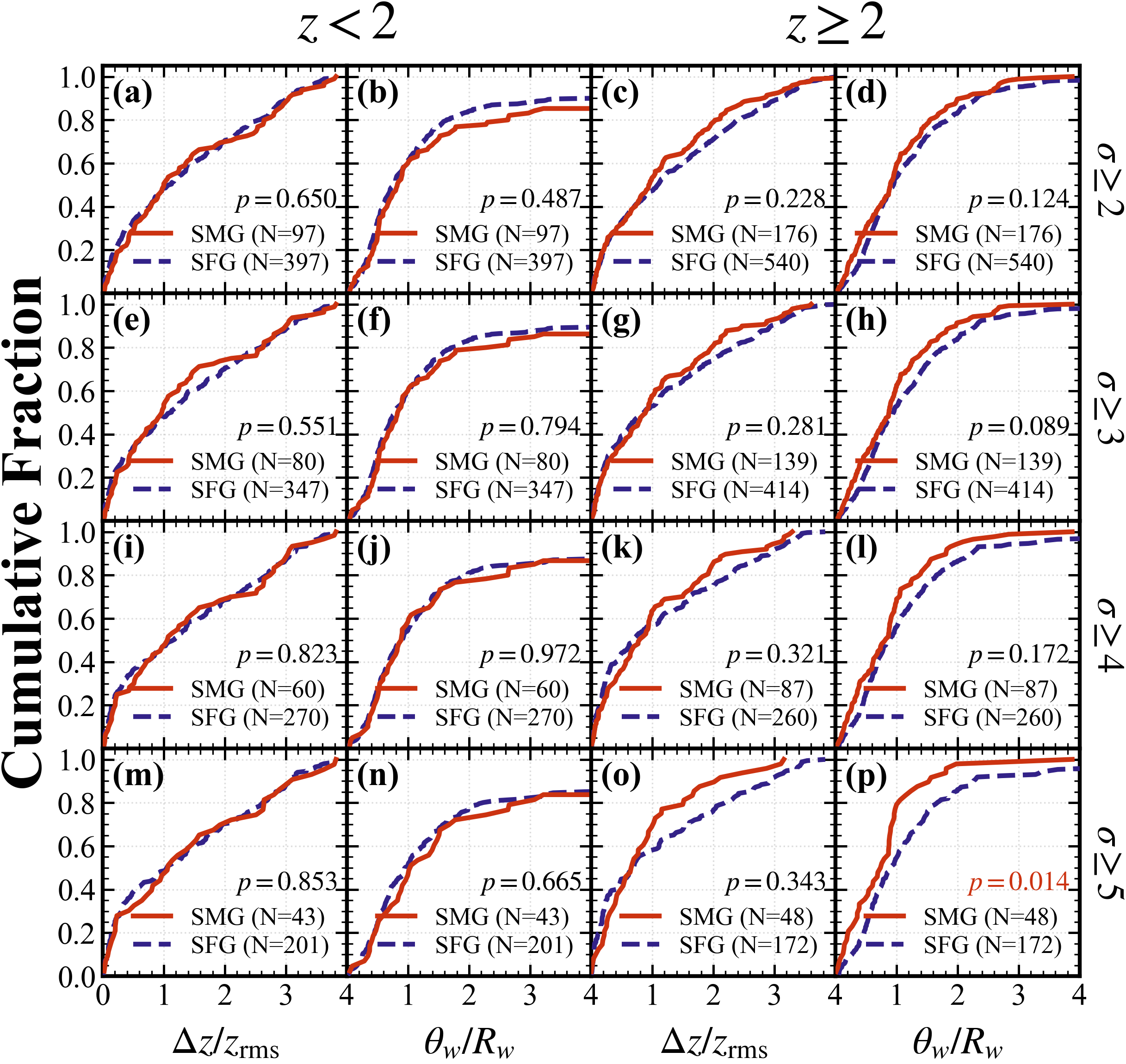}
    \caption{CDFs of Redshift Compactness ($\Delta z/z_{\rm rms}$; columns 1 and 3) and Projection spatial compactness ($\theta_w/R_w$; columns 2 and 4) for SMGs (red solid lines) and mass-redshift matched SFGs (purple dashed lines) within overdense regions. The panels are organized by redshift—$z < 2$ (left two columns) and $z \geq 2$ (right two columns)—and by overdensity significance thresholds increasing from $\sigma \geq 2$ to $\sigma \geq 5$ (top to bottom rows). The $p$-values from two-sample K-S tests are indicated in each panel. At high redshift and extreme density (panel p: $z \geq 2, \sigma \geq 5$), SMGs exhibit significantly higher projection spatial compactness than SFGs ($p = 0.014$), with $\sim 80\%$ of SMGs residing within the overdensity core ($\theta_w/R_w \leq 1$).\label{fig:region}}
\end{figure}

\begin{figure*}[htbp]
    \centering
    \includegraphics[width=0.99\linewidth]{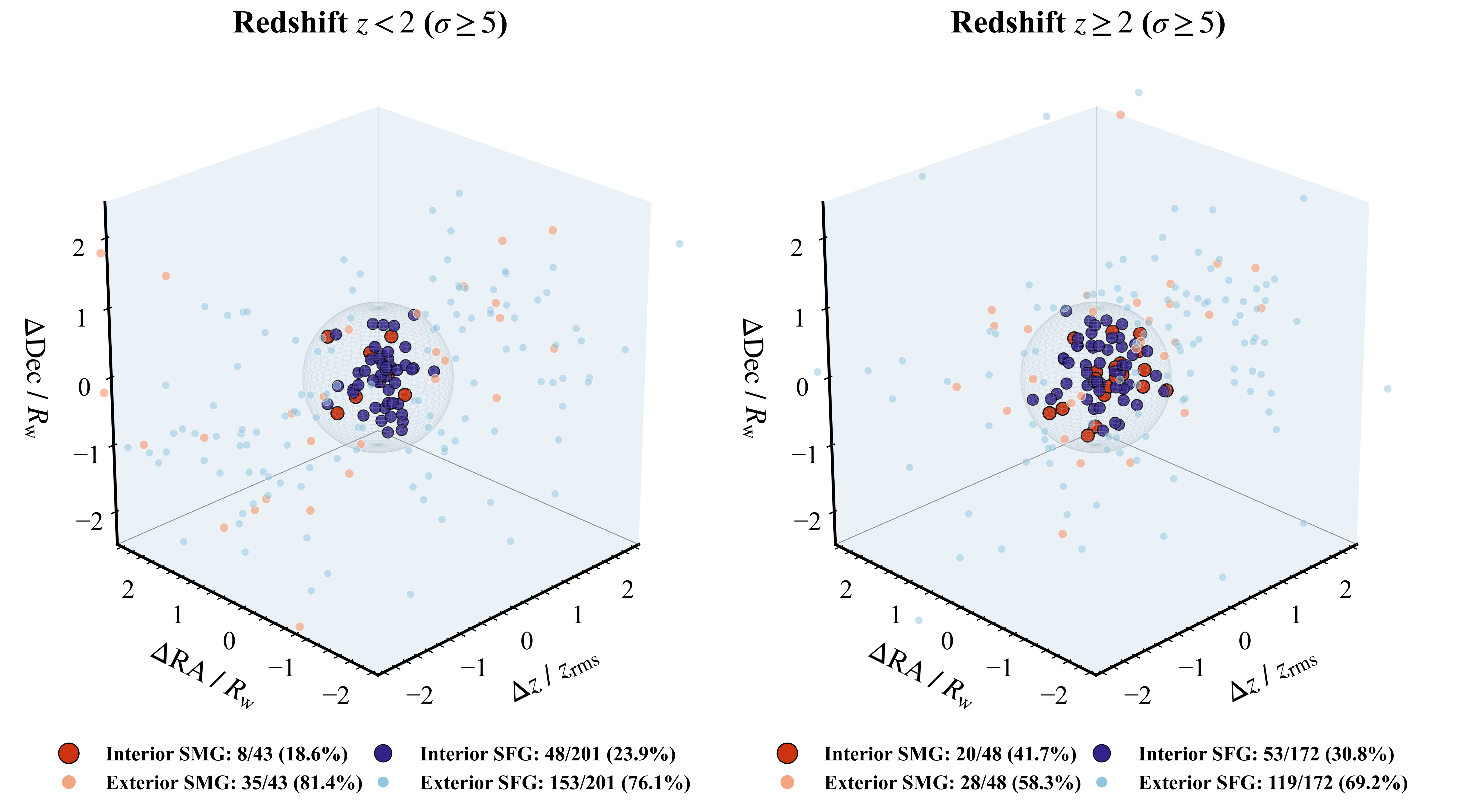}
    \caption{Three-dimensional normalized phase-space distributions of SMGs and SFGs around high significance ($\sigma \geq 5$) overdensity candidates. The left and right panels show the distributions for candidates at low redshift ($z < 2$) and high redshift ($z \geq 2$), respectively. The spatial offsets ($\Delta\mathrm{RA}$, $\Delta\mathrm{Dec}$) and line-of-sight redshift differences ($\Delta z$) are normalized by the overdensity peak estimated radius ($R_{\rm w}$) and redshift dispersion ($z_{\rm rms}$) of each overdensity. The central semi-transparent sphere denotes the core region ($r = \sqrt{(\Delta\mathrm{RA}/R_{\rm w})^2 + (\Delta\mathrm{Dec}/R_{\rm w})^2 + (\Delta z/z_{\rm rms})^2} \leq 1$). Large dark red and dark blue circles represent interior SMGs and SFGs residing within this core, while small light red and light blue circles indicate exterior SMGs and SFGs located in the outskirts or field environments ($r > 1$). The legends at the bottom present the raw counts and fractions of galaxies in each category.\label{fig:3D_distribution_Interior}}
\end{figure*}

\subsection{Distribution of SMGs within overdensity candidates: interior or exterior?} \label{sec:interior or exterior}

We are naturally curious about whether SMGs can directly trace the core regions of overdensity candidates. So we discussed the spatial distribution of all SMGs and mass-redshift matched SFGs. We evaluate the compactness of SMGs and mass-redshift matched SFGs within overdense regions across two dimensions: Redshift Compactness ($\Delta z / z_{\rm rms}$), which reflects the degree of overdense distribution along the line-of-sight; and Projection spatial compactness ($\theta_w / R_w$), which characterizes the spatial overdensity in the projected sky plane. We employ the two-sample K-S test to quantify the statistical significance of the differences between the SMG and SFG distributions. This non-parametric test is sensitive to differences in both the shape and the median of the cumulative distribution functions (CDFs). As illustrated in Figure~\ref{fig:region}, the two leftmost columns represent the low-redshift regime ($z < 2$), while the two rightmost columns correspond to the high-redshift regime ($z \geq 2$). The four rows from top to bottom correspond to regions with overdensity significance thresholds of $\sigma \geq 2$, $\sigma \geq 3$, $\sigma \geq 4$, and $\sigma \geq 5$, respectively.

\begin{figure}[htbp]
    \centering
    \includegraphics[width=0.98\linewidth]{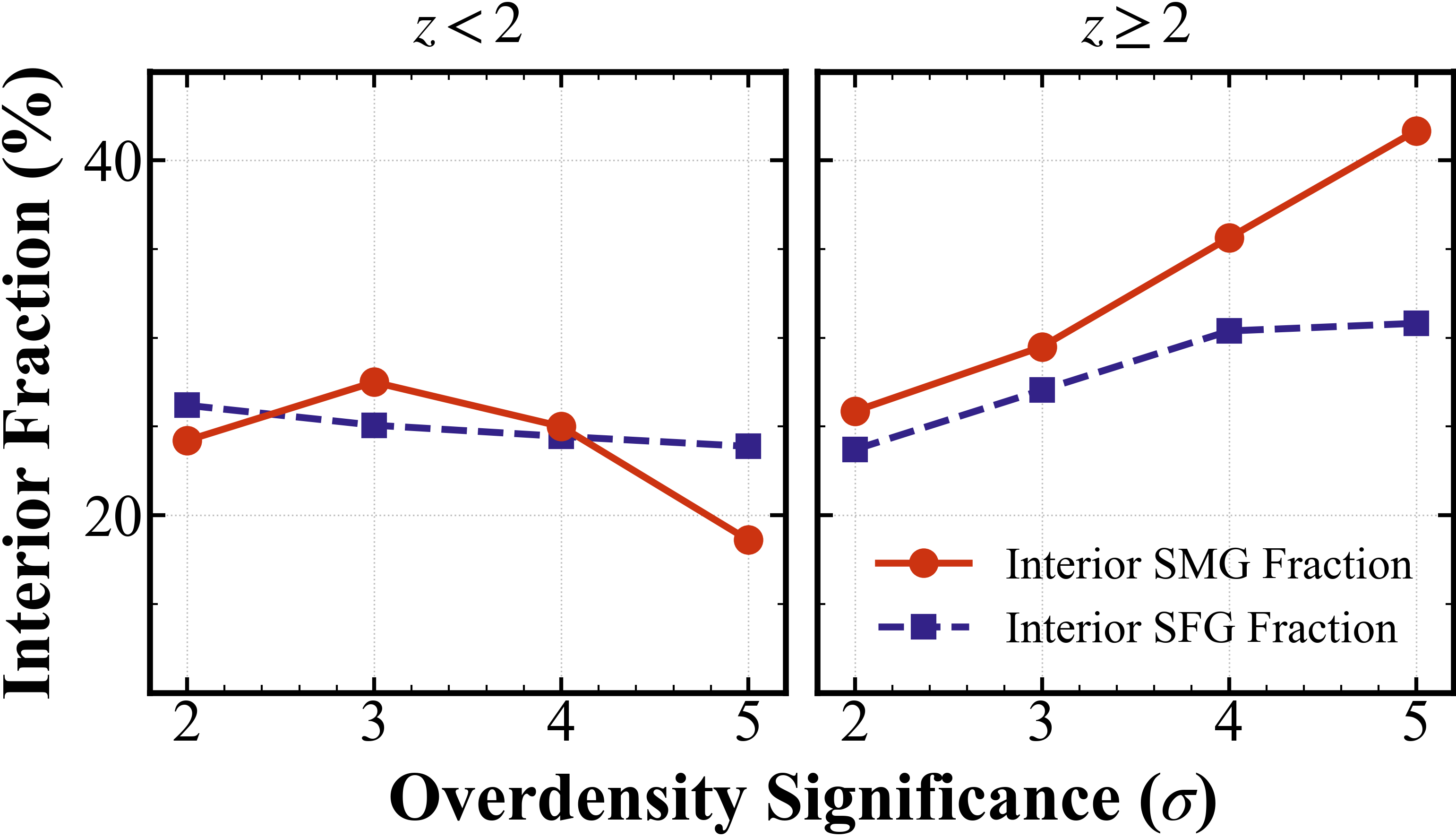}
    \caption{Evolution of the interior fraction of SMGs and SFGs as a function of overdensity significance. The left and right panels display the trends for structures at low redshift ($z < 2$) and high redshift ($z \geq 2$), respectively. The horizontal axis represents the cumulative overdensity significance ($\sigma$) thresholds (i.e., $\sigma \geq 2$, $\sigma \geq 3$, up to $\sigma \geq 5$), while the vertical axis denotes the fraction of galaxies residing within the normalized core region ($r \leq 1$). The solid red lines with circles and the dashed blue lines with squares indicate the interior fractions of SMGs and SFGs, respectively. At $z \geq 2$, the interior fraction of SMGs rises steeply with increasing cumulative overdensity significance, reaching over 40\% at $\sigma \geq 5$, which is significantly higher than that of the SFG population.\label{fig:Interior_fraction}}
\end{figure}

\begin{figure*}[htbp]
    \centering
    \includegraphics[width=0.98\linewidth]{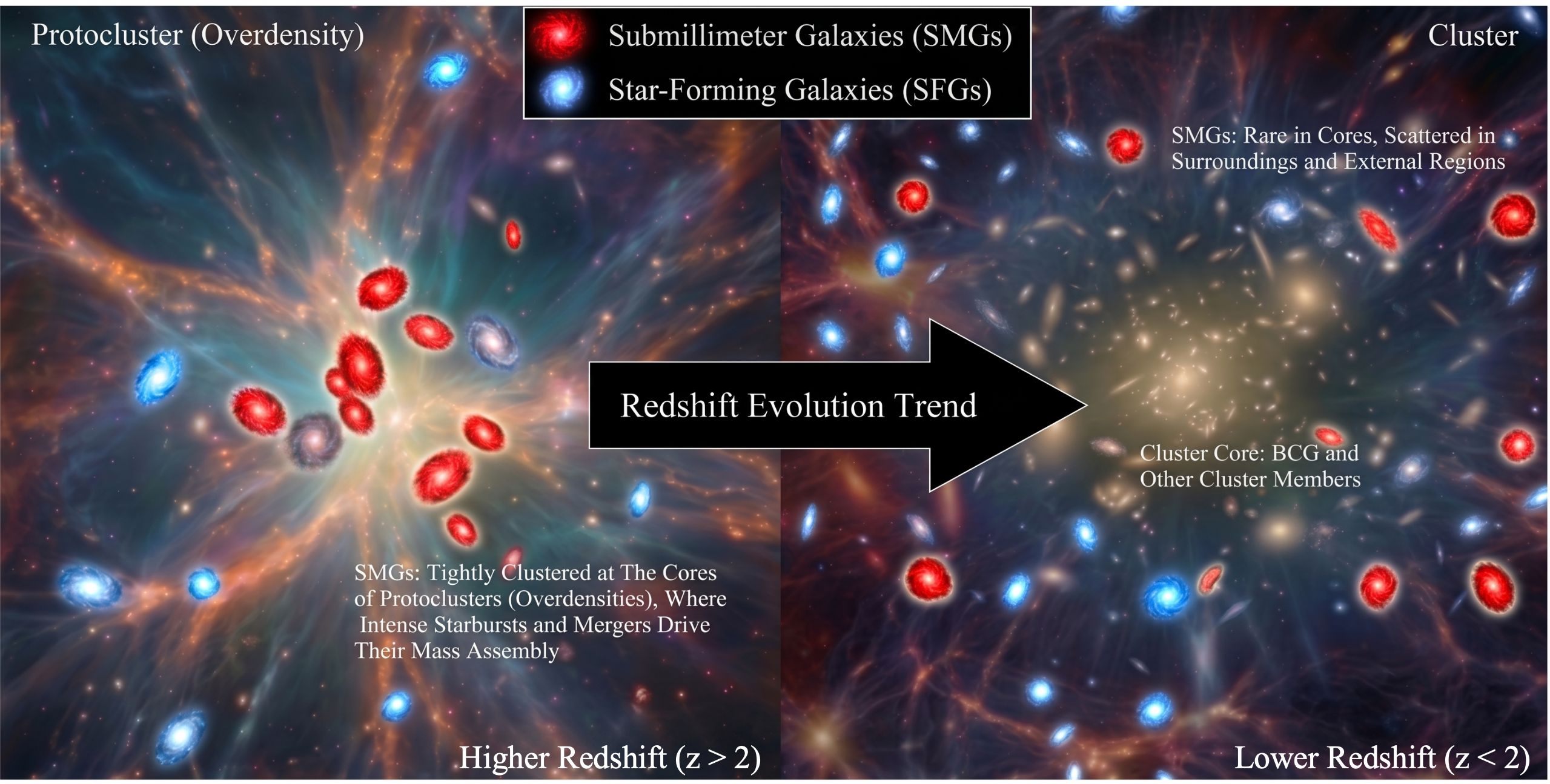}
    \caption{Schematic illustration of the redshift evolution of the spatial distribution of SMGs and SFGs in cosmic large-scale overdense structures. Red symbols mark SMGs, while blue/gray symbols represent SFGs. The left panel depicts the high-redshift ($z>2$) protocluster (overdensity) environment, where SMGs are tightly clustered at the cores of overdensities, as intense starbursts and mergers drive their mass assembly. The right panel shows the evolved galaxy cluster at lower redshift ($z<2$), where SMGs are rare in the cluster core (dominated by the brightest cluster galaxy and other cluster members) and are instead scattered in the structure’s surrounding and external regions.\label{fig:trends}}
\end{figure*}

Focusing on the high-redshift~($z \geq 2$) subsamples, we find that in the projection spatial compactness distribution for $z \geq 2$ and $\sigma \geq 5$ (panel p), the $p$-value reaches 0.014. This indicates that SMGs are more centrally concentrated than typical SFGs in extremely high-density environments at high redshift, supporting the view that SMGs serve as excellent tracers of protocluster cores. For the $z < 2$ groups, the spatial and redshift distributions of SMGs and SFGs remain highly consistent regardless of the significance level (with $p$-values ranging from 0.487 to 0.665). This suggests that the environmental impact on the distribution of these two galaxy populations converges at later cosmic epochs.

In the $z < 2$ subsamples shown on the left side of the Figure~\ref{fig:region} panels, the red solid lines representing SMGs and the purple dashed lines representing SFGs exhibit very similar morphologies, with no significant deviation observed. However, across all significance thresholds, the $p$-values for the high-redshift samples remain consistently lower than those of their low-redshift counterparts. 

In the most extreme environments at high redshift ($z \geq 2, \sigma \geq 5$), Figure~\ref{fig:region} panel (p) reveals a significant divergence in the spatial distributions of the two galaxy populations ($p = 0.014$). At a normalized projection spatial compactness of $\theta_w / R_w = 1$—which defines the characteristic boundary of the overdensity structure—approximately 80\% of the SMG population resides within the interior, whereas only $\sim 60\%$ of the mass-matched SFGs are similarly concentrated.

To better describe the distribution of SMGs within overdense environments across the full 3D space. We construct a dimensionless, 3D normalized phase space combining the relative spatial and redshift distances. This approach allows for a more intuitive comparison of the distribution preferences of SMGs across overdensity structures of varying sizes and redshifts.
\begin{enumerate}
   \item \textbf{Celestial plane:} We convert the differences in right ascension and declination ($\Delta \mathrm{RA}, \Delta \mathrm{Dec}$) between the galaxies and the overdensity centers into angular distances. A cosine declination correction ($\Delta \mathrm{RA} \cdot \cos(\mathrm{Dec})$) is applied to ensure an undistorted projection. Subsequently, we divide the angular offsets ($\theta_w$) by the estimated radius of the structure, $R_w$, to obtain the normalized coordinates, $dx$ and $dy$.
   \item \textbf{Line-of-sight direction:} We directly divide the redshift difference of the galaxies, $\Delta z$, by the characteristic redshift dispersion of the structure, $z_{\rm rms}$, to yield the normalized coordinate $dz$.
   \item \textbf{Core region definition:} Within this dimensionless 3D space, we define a sphere with a radius $r = \sqrt{dx^2 + dy^2 + dz^2} \leq 1.0$ from the center as the interior core of the overdensity structure.
\end{enumerate}
Figure~\ref{fig:3D_distribution_Interior} presents the three-dimensional normalized phase-space distributions of SMGs and SFGs around highly significance ($\sigma \geq 5$) overdensity candidates. The left and right panels show the distributions for candidates at low redshift ($z < 2$) and high redshift ($z \geq 2$), respectively. Large dark red and dark blue circles represent interior SMGs and SFGs residing within this core ($r \leq 1$), while small light red and light blue circles indicate exterior SMGs and SFGs located in the outskirts ($r > 1$). The fraction of interior SMGs increases remarkably from 18.6\% at $z < 2$ to 41.7\% at $z \geq 2$, indicating that SMGs are more strongly concentrated in the densest regions at higher redshifts.

As shown in Figure~\ref{fig:Interior_fraction}, by extracting overdensity structures across varying significance thresholds ($\sigma \geq 2, 3, 4$, and $5$) and calculating the fraction of core members relative to the total sample for each condition, we observe how the overdensity properties of SMGs evolve with increasing overdensity significance. As the significance threshold tightens (from $\sigma \geq 2$ to $\sigma \geq 5$), the interior fraction of SMGs exhibits a steady upward trend, particularly in the $z \geq 2$ panel. This reinforces the premise that high-$\sigma$ candidates indeed represent the most extreme overdense structures in the universe, and that SMGs serve as excellent probes for tracing these genuine environments. \cite{Casey2016} proposes that the progenitors of massive galaxy clusters undergo simultaneous, short-lived bursts of star formation well before virialization at $z>2$. Our statistical finding shows that the SMG core fraction rises sharply at $z \geq 2$ and correlates positively with overdensity significance suggests that extreme starburst activity is intrinsically linked to the rapid assembly phase of dense protocluster cores in the early universe.

In summary, at high redshifts ($z \geq 2$), the interior fraction of SMGs is significantly higher than that of typical SFGs, as evidenced by both the visual density in the 3D scatter plots and the absolute values in the trend lines (e.g., at $\sigma \geq 5$, the SMG fraction reaches $\sim 41.7\%$, compared to $\sim 30.8\%$ for SFGs). This indicates that during the protocluster phase ($z \geq 2$), SMGs exhibit a strong preference for the highest-density environments. However, as the universe evolves to $z < 2$, the interior fraction of SMGs experiences a precipitous drop (plummeting to below $20\%$), falling even below that of ordinary SFGs. Consequently, we illustrate a conceptual diagram depicting the redshift evolution of the spatial distributions of SMGs and SFGs within cosmic large-scale overdense structures, as shown in Figure~\ref{fig:trends}. SMGs act as the ``drivers'' within the cores of dense protoclusters in the early universe; however, in the late universe, the cores of mature galaxy clusters transform into their ``graveyards''.

\section{summary} \label{sec:summary}

In this paper, we performed cross-matching using the A$^3$COSMOS catalog~\citep{2019ApJS..244...40L,2024A&A...685A...1A}, the COSMOS2025 catalog~\citep{Shuntov2026}, and the largest compilation of spectroscopic redshift in COSMOS~\citep{Khostovan2026} and literature~(including \citealt{2013ARAA..51..105C, 2018ApJ...867...92S, 2019ApJ...880...15K, 2020AA...641A.155V, 2025AA...699A..80H}; a full reference list is provided in D. Liu et al., in prep.). We finally obtained a sample of 449 unbiased and unique SMGs in the COSMOS field~(including 168 SMGs with spectroscopic redshift, 52 of which have CO or \CIIline{} 158$\mu$m line detections). Additionally, we randomly selected 449$\times$5 sets of mass matched, 449$\times$5 sets of redshift matched and 449$\times$5 sets of mass-redshift matched unbiased SFG samples from the star-forming main sequence. We applied the robust Poisson Probability Method \cite[PPM;][]{2014ApJ...792..113C, 2019A&A...623A..48C, Calvi2023} to all these samples to search for megaparsec large-scale overdensity structures, and categorized our PPM-detected overdensity candidates into five tiers (Categories A --E) based on their detection significance, overdensity strength, and spatial-redshift alignment with the tracer SMGs/SFGs. This tiered framework allows us to systematically evaluate selection thresholds and isolate the most robust structures (Categories A and B, $\delta \geq 5, \sigma \geq 5$).

The main scientific results are summarized as follows:

\begin{enumerate}
    \item We find excellent overall agreement between the PPM-estimated overdensity redshifts ($z_{\rm ov}$) and the tracer SMG redshifts ($z_{\rm SMG}$). The baseline sample of candidates (Category E) exhibits a redshift matching accuracy of $\sigma(\Delta z / (1+z)) = 0.047$ with negligible bias. In contrast, all of our most strictly defined Category A and B overdensities achieve an exceptional accuracy of $\sigma(\Delta z / (1+z)) = 0.011$. We also conducted classification and statistical analysis on all samples. The results show that the fraction of CO/\CIIline{} spec-$z$ SMGs in all categories is significantly higher, and the overall detection capability of each sample for overdensity structures, ranked from strongest to weakest~(Category A-E), is: CO/\CIIline{} spec-$z$ SMGs (strongest, 11.5-71.2\%) $>$ all spec-$z$ SMGs~(3.6-64.3\%) $>$ all SMGs~(1.3-61.0\%) $>$ control SFGs~(M$_{\star}$-$z$ matched, 0.3-59.4\%). This result supports that CO/\CIIline{} spec-$z$ SMGs can effectively trace overdensity structures and further confirms the superiority of CO/\CIIline{} spec-$z$ SMGs as excellent signposts for protoclusters.

    \item We validate the reliability of our PPM-identified Category A and B overdensity candidates by cross-matching them with a comprehensive compilation of known protoclusters in the COSMOS field. The PPM method successfully reproduces well-documented, dense structures at high redshifts, including the CL J1001+0220 ($z \sim 2.5$) and AzTEC-3 ($z \sim 5.3$) protoclusters. This analysis reveals that 63.6\% of SMG-associated and 44.2\% of SFG-associated candidates are previously known structures, with the recovery rate peaking at 78.9\% for the highest-confidence Category A sources. Furthermore, 100\% (51/51) of our candidates at $z \leq 3.7$ successfully associate with galaxy groups in the recent COSMOS-Web AMICO groups catalog~\citep{Toni2025} with tight spatial and redshift offsets ($\overline{\rm sep} = 0.175~{\rm pMpc}$, $\overline{|\Delta z|} = 0.058$). Finally, our pipeline perfectly recovers established all substructures within the Hyperion proto-supercluster and two known spectroscopically confirmed groups. Collectively, these validations confirm that our PPM pipeline accurately identifies genuine overdense structures, establishing the unmatched candidates as robust new discoveries.
    
    \item After checking duplicated sources in the literature, we find 26 new overdensity candidates at $z \sim 0.8-5.3$ out of 54 most reliable Category A and B sources identified by our PPM analysis. \(\sim 78\%\) (42/54) of our overdensity candidates host at least two spectroscopic redshift (spec-$z$) member galaxies, \(\sim 54\%\) (14/26) of the new overdensity candidates contain at least two spec-$z$ member galaxies. The 3D spatial distribution analysis maps corresponding to these candidates are all included in Appendix~\ref{sec:Protocluster candidates catalog}. We present a catalog containing parameters such as their coordinates, redshifts, core sizes estimated via PPM analysis, and richness, which provides abundant research directions for subsequent studies and observations.
    
    \item We calculated the CO/\CIIline{} line luminosities and molecular gas masses of SMG subsamples. Results of these 47 SMGs (including line luminosities, $\mathrm{M_{H_2}}$, spectroscopic redshifts, and PPM-derived overdensity significance) are in Appendix \ref{sec:spectroscopy match SMG} (Tables~\ref{tab:CO} and \ref{tab:cII}). We find that the $\rm M_{\rm H_2}$ exhibits no overall correlation with the environmental density (Pearson correlation coefficients $r = -0.014$); instead, $\rm M_{\rm H_2}$ correlates significantly with redshift ($r = 0.337$), matching \cite{2021ApJ...909...56L}. We also find positive correlations between molecular gas mass and significance for z $>$ 2.55 and z $<$ 1.99~(consistent with the findings of~\citealt{Calvi2023}), but a slight negative one for 1.99 $<$ z $<$ 2.55.

    \item We compare the physical properties and environmental dependencies of the CO/\CIIline{} spec-$z$ and photo-$z$ SMG samples. Rigorous statistical testing confirms that while their star-forming modes ($\Delta$MS) and sub-millimeter fluxes ($S_{850\mu\rm{m}}$) exhibit no significant differences, the spec-$z$ sample contains a statistically significant overabundance of low-mass galaxies. A mass-dependent environmental effect may be a physical driver for the elevated overdensity fractions observed in the CO/\CIIline{} spec-$z$ sample, but the small number statistics still prohibits in-depth understanding of the difference.

    \item We analyze the morphological properties of the SMG sample using high-resolution JWST/NIRCam F277W imaging. The Gini-M20 diagnostic reveals that while the general SMG population maintains a baseline merger fraction of approximately 29\%, the brightest SMGs~($S_{850\mu m} \geq 5.36$ mJy) exhibit a significantly enhanced merger fraction of roughly 49\%. Crucially, this elevated merger rate remains consistent regardless of whether the sources reside in overdense or field environments. Visual inspections further corroborate this finding, demonstrating that almost all of the exceptionally bright SMGs—even those formally classified as non-mergers—display clumpy and/or disturbed structures indicative of recent interactions or gravitational instability. Consequently, we conclude that while typical SMG star formation is likely sustained by a combination of secular processes and steady environmental gas accretion, the most extreme dust-obscured starbursts are predominantly driven by local major mergers or violent disk instabilities.
    
    \item By comparing the redshift compactness ($\Delta z / z_{\rm rms}$) and projection spatial compactness ($\theta_w / R_w$) of SMGs and mass-redshift matched SFGs, we find that they are highly consistent in the low-redshift ($z < 2$) regime. However, at high redshift ($z \geq 2$), particularly in extremely high-density environments ($\sigma \geq 5$), SMGs exhibit a significant central concentration ($p = 0.014$). Approximately 80\% of SMGs reside within the interior of overdense structures, supporting the view that SMGs serve as excellent tracers of protocluster cores. By constructing a dimensionless, 3D normalized phase space that combines relative spatial and redshift distances, we conclude that in the extremely high-density environments of the early universe ($z \geq 2$), SMGs are more significantly concentrated toward protocluster cores compared to control SFGs. This demonstrates that they act as the central ``driver'' in the rapid assembly of early dense structures.
    
\end{enumerate}

Our study shows that SMGs can statistically trace overdensity structures, CO/\CIIline{} SMGs exhibit stronger overdensities than control SFG sample, and both SMG subsamples show higher overdensities than the control SFG sample, with SMGs preferentially residing in the interior regions of high-redshift, high-overdensity structures, and both high-overdensity environments and galaxy mergers are major channels for forming brightest SMGs. We further examined a series of relations, including the $\mathrm{M_{H_2}}$--significance relation and the $\mathrm{M_{H_2}}$--Redshift relation of SMGs, as well as potentially associated structures on larger scales---extending over 500 Mpc along the line of sight. These conclusions still require further investigation by combining additional samples and numerical simulations, which will be further investigated in our later work. 

With the PPM method and the state-of-the-art COSMOS-Web JWST photo-$z$ catalog, we have identified 31 previously underexplored new high-overdensity~($\delta\geq5$), high-significance~($\sigma\geq5$) candidates, with $\sim54\%$ of these new candidates hosting at least two spectroscopic-redshift member galaxies. All results from our PPM analysis of these samples are made publicly available at: {\url{https://github.com/chance-deng/PPM-analysis-for-COSMOS-Web-field}}. The dataset includes sources up to a redshift of $z\sim8$, with a total of more than 3000 overdensity candidates that meet at least the Category E criteria (54 meeting the most reliable Categories A\&B). 
This dataset serves as a statistic sample with the largest-sample of SMGs for studying the formation and evolution of large-scale cosmic overdensity structures. Although it still lacks full spectroscopic information, the on-going JWST grism observations (e.g., COSMOS-3D; \citealt{Meyer2025}) will soon provide the critical spectroscopic validation in the near future.

\begin{acknowledgments}

This work is supported by the Strategic Priority Research Program of the Chinese Academy of Sciences, grant No. XDB0800400.
This project has received funding from the European Union’s Horizon 2020 research and innovation programme under the Marie Sk\l{}odowska-Curie grant agreement No 101148925.
\end{acknowledgments}

\software{astropy~\citep{2013A&A...558A..33A,2018AJ....156..123A,2022ApJ...935..167A}, Scipy~\citep{Virtanen2020}, Matplotlib~\citep{4160265}, Numpy~\citep{Van_Der_Walt2011}, Pandas~\citep{The_pandas_development_Team2022}, Topcat~\citep{taylor2017topcatworkingdataworking}, Photutils~\citep{larry_bradley_2024_13989456}, Source Extractor~\citep{1996A&AS..117..393B}, Statmorph~\citep{2019MNRAS.483.4140R}, SAOImage DS9~\citep{Joye2003}}

\clearpage
\appendix{}

\section{Redshift-Available Sources for the All SMGs Sample}\label{sec:Redshift Sources for the All SMGs Sample}

We use the latest JWST-based photometric redshifts from COSMOS2025~\citep{Shuntov2026} and the largest compilation of spectroscopic redshift in COSMOS~\citep{Khostovan2026} to update the prior redshifts for A$^3$COSMOS SMGs. Furthermore, we also cross-match with the latest compilation of high-$z$ CO/\CIIline{} line-emitting galaxies~(including \citealt{2013ARAA..51..105C, 2018ApJ...867...92S, 2019ApJ...880...15K, 2020AA...641A.155V, 2025AA...699A..80H}; a full reference list is provided in D. Liu et al., in prep.) for secure redshifts, the summary of all SMGs sample shown as Table~\ref{tab:ztype}.

\input{table3}

\onecolumngrid

\section{Cross matching spectroscopy redshift member galaxies and galaxy groups}\label{sec:Spec-z member of overdensity candidates}

We check the latest COSMOS Spectroscopic Redshift Compilation Catalog by \cite{Khostovan2026}, searching for spec-$z$ counterparts for each of our new overdensity candidates. The spec-$z$ cross-matching results are presented in Table \ref{table4}. This table also contains the cross-matching results of COSMOS-Web galaxy group~($z \leq 3.7$; \citealt{Toni2025}).

\input{table4.tex}

\section{Categories A\&B overdensity candidate~(Beacon:SFG)}\label{sec:Category A and B overdensity candidate (Beacon:SFG)}

Here we present a total of 43 overdensity candidates (PC 12 --54) detected using the PPM with SFGs as beacons in Table~\ref{table:sfg}. These candidates belong to Category A and B, and are associated with high overdensities ($\delta \geq 5$) and high significance ($\sigma \geq 5$). As shown in Table 2, we list the IDs of their associated SFGs, any cross-matched known overdensities (provided as table notes), the redshifts of the beacons and the overdensities ($z$ and $z_{\rm{ov}}$, respectively), the overdensity richness ($N_{\textrm{select}}$), the significance ($\sigma$), the maximum radius at 1\% $\delta$ ($R_{w}$), the peak overdensity values ($\delta$), and the peak positions.

\input{table5.tex}
\clearpage

\section{Validating the PPM analysis with prior positions of known overdensities in Hyperion and massive galaxy groups in COSMOS-Web} \label{sec:PPM Reproduction Results}

\begin{figure*}[ht]
    \centering   
\begin{minipage}{1\linewidth}
    \centering
    \includegraphics[width=0.95\textwidth]{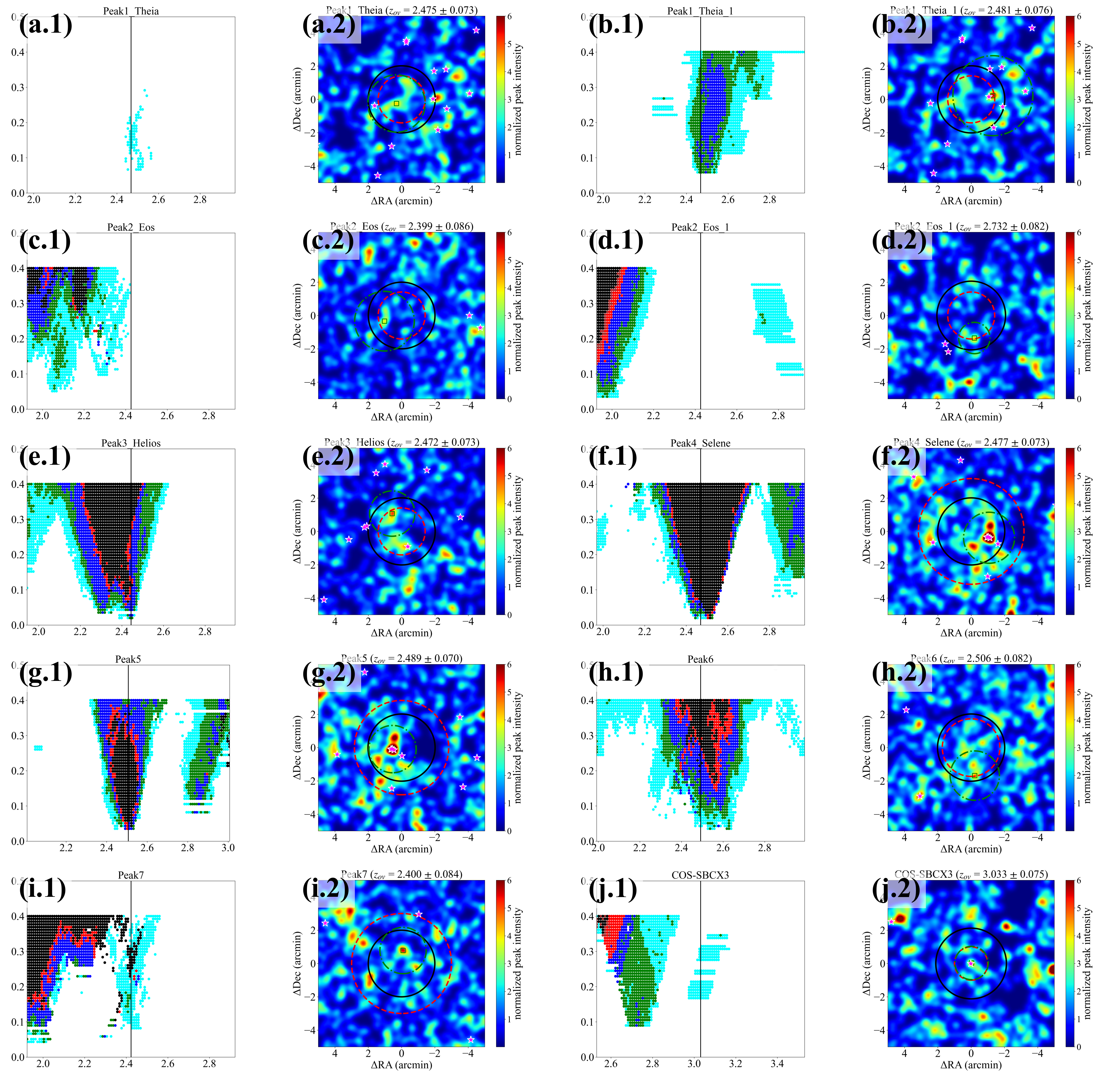}
    \begin{center}
    \end{center}
\end{minipage}
\caption{The sequence of labels (a) --(j) corresponds to the two substructures of \textit{Theia} (a and b), the two substructures of \textit{Eos} (c and d), \textit{Helios} (e), \textit{Selene} (f), \textit{Peak 5} (g), \textit{Peak 6} (h), \textit{Peak 7} (i), and the group COS-SBCX3 (j). For each designated structure, the left sub-panels denoted with ".1" present the PPM plots, while the right sub-panels denoted with ".2" show the corresponding 2D density maps, and the magenta pentagrams represent the SMGs within the redshift range. Other peaks in the PPM plots not aligned with the primary target's redshift correspond to independent foreground and background structures along the same line of sight. For detailed information regarding the symbols and color coding, see Figure~\ref{fig:4.0}. Images of COS-SBC4 are available at: \url{https://github.com/chance-deng/PPM-analysis-for-COSMOS-Web-field}.}\label{fig8}
\end{figure*}

We further validate our PPM analysis by adopting the positions of previously known overdensity peaks in large-scale structures~(LSSs) as priors. This approach additionally tests whether the PPM effectively recovers overdense substructures within complex LSSs. The Hyperion proto-supercluster is a spectacular and extremely massive structure in the COSMOS field at $z \sim 2.45$. \citet{2018AA...619A..49C} use Voronoi tessellation in overlapping redshift slices to identify seven galaxy density peaks within Hyperion. They estimate a total mass of $\sim 4.8 \times 10^{15}\ \rm M_\odot$ spanning a comoving volume of approximately $60 \times 60 \times 150\ \rm Mpc^3$, which makes it one of the largest and most massive structures discovered in the early universe. Multiple studies over the past decade confirm this well-known protocluster, and individual structures within this region are consistently identified as overdensities \citep{Diener2013,2014ApJ...792..114C,Casey2015,2015ApJ...808...37C,2016ApJ...828...56W,2016ApJ...831..181L,2018AA...619A..49C,2022ApJ...941..134H, Toni2025}. We therefore input the positions and redshifts of the seven density peaks within the redshift range $2.40 < z < 2.50$ identified by \citet{2018AA...619A..49C}---designated \textit{Theia}, \textit{Eos}, \textit{Helios}, \textit{Selene}, \textit{Peak 5}, \textit{Peak 6}, and \textit{Peak 7}---as priors into our PPM pipeline, and we present the results below:

\begin{enumerate}
    \item \textit{Theia}---We note that this is the most massive and extended peak in the Hyperion proto-supercluster, consisting of the proto-cluster PCL1002, which is centered on an X-ray source detected by \cite{Casey2015}, and the three proto-groups (D13a, D13b, D13d) discovered by \cite{Diener2013}. Consequently, we do not detect a significant overdense structure at the centroid coordinates provided by \citet{2018AA...619A..49C}. Therefore, we adopt the central coordinates of PCL1002 (which encloses the three proto-groups D13a, D13b, and D13d) alongside the central coordinates of D13d, corresponds to the secondary substructure of \textit{Theia} previously identified by \citet{2018AA...619A..49C}--as prior positions. Running the PPM pipeline a second time with these inputs reveals robust overdense structures in both cases. The derived parameters are $\sigma = 4.03$, $\delta = 4.72$, and $z_{ov} = 2.481 \pm 0.076$ for the former, and $\sigma = 2.56$, $\delta = 3.10$, and $z_{ov} = 2.475 \pm 0.073$ for the latter.
    \item \textit{Eos}---This peak consists of two substructures. \citet{2018AA...619A..49C} demonstrate that increasing the overdensity threshold distinctly separates the two components. Therefore, similar to \textit{Theia}, the centroid actually lies between the two substructures. The projected physical separation between these two peaks is approximately $2\sqrt{2}$ pMpc, corresponding to an angular separation of $\sim 5.7 \arcmin$  at $z = 2.43$. Because our PPM pipeline specifically requires an overdensity signal to initiate within a maximum search radius of $r \le 2.2 \arcmin$ from the prior position (as detailed in Section~\ref{sec:method}), using this centroid naturally yields no significant overdensity detection. Rather than a lack of robustness, this non-detection demonstrates the localized sensitivity of our method, which requires physical proximity and correctly avoids artificially bridging massive empty gaps. Indeed, when we separately input the projected overdensity centers of the two physical substructures back into the PPM pipeline, we detect highly robust overdensities in both cases: $\sigma = 2.52$ and $\delta = 2.77$ at $z_{ov} = 2.399 \pm 0.085$, and $\sigma = 2.61$ and $\delta = 3.16$ at $z_{ov} = 2.732 \pm 0.082$, respectively.
    \item \textit{Helios}---The shape of this peak is compact, making it difficult to clearly resolve into independent substructures even at higher density thresholds. Numerous studies using various methodologies confirm the presence of a robust overdensity at this location (\citealt{Diener2013, 2014ApJ...792..114C, 2016ApJ...831..181L, 2014ApJ...782L...3C, 2015ApJ...808...37C}). Our PPM pipeline successfully detects a significant overdensity here: $\sigma = 4.79$ and $\delta = 4.10$ at $z_{ov} = 2.472 \pm 0.074$.
    \item \textit{Selene}---It consists of a main body containing the majority of the mass and an extended ``tail'' on the plane of the sky. The centroid of the main body perfectly coincides with the protocluster D13c discovered by \citet{Diener2013}. Our PPM pipeline perfectly reproduces this structure, yielding $\sigma = 8.00$ and $\delta = 12.85$ at $z_{ov} = 2.477 \pm 0.074$. Furthermore, the 2D density map clearly reveals the ``tail'' extending northwest from the center (see Figure~\ref{fig8}, panel f.2).
    \item \textit{Peak 5}---Located at the highest redshift among all the peaks ($z=2.507$), this directly corresponds to the structure CL~1001+0220 discovered by \citet{2016ApJ...828...56W}. The PPM pipeline perfectly recovers this structure: $\sigma = 8.00$ and $\delta = 11.80$ at $z_{ov} = 2.489 \pm 0.070$.
    \item \textit{Peak 6}---Its morphology is approximately circular. It is closely related to \textit{Eos} in physical structure, and it merges completely with \textit{Eos} when the system density threshold decreases ~\citep{2018AA...619A..49C}. Our PPM pipeline successfully detects a significant overdensity here: $\sigma = 8.00$ and $\delta = 4.61$ at $z_{ov} = 2.400 \pm 0.084$.
    \item \textit{Peak 7}---This is a lower-mass peak with a regular shape on the plane of the sky~\citep{2018AA...619A..49C}. The PPM pipeline perfectly recovers this structure: $\sigma = 8.00$ and $\delta = 3.52$ at $z_{ov} = 2.506 \pm 0.082$. 
\end{enumerate}
The PPM method successfully reproduced all overdensity peaks in the Hyperion large-scale structure. In addition, we perform the same analysis at the locations of two other spectroscopically confirmed massive galaxy groups~(COS-SBCX3 and COS-SBC4, \citealt{Sillassen2024}) in the COSMOS-Web field, and we successfully detect overdense structures in both cases. In Figures~\ref{fig8}, we present the PPM plots and the 2D density maps for 7 overdensity peaks (namely \textit{Theia}, \textit{Eos}, \textit{Helios}, \textit{Selene}, \textit{Peak5}, \textit{Peak 6}, and \textit{Peak 7}), provided in the study by \cite{2018AA...619A..49C} and the group COS-SBCX3 which is provided by~\cite{Sillassen2024}. From these reproduced results, we obtain mean output parameters of $\bar{\sigma} = 5.02$ and $\bar{\delta} = 5.06$, which verified our Category B thresholds of \(\sigma \geq 5\) and \(\delta \geq 5\) are suitable to reliably recover real overdensity candidates.

\section{3D spatial distribution maps of Category A\&B sources} \label{sec:Protocluster candidates catalog}

\begin{figure*}[ht]
    \centering  
\begin{minipage}{0.92\linewidth}
    \centering
    \hspace{-0cm}
    \includegraphics[width=\textwidth]{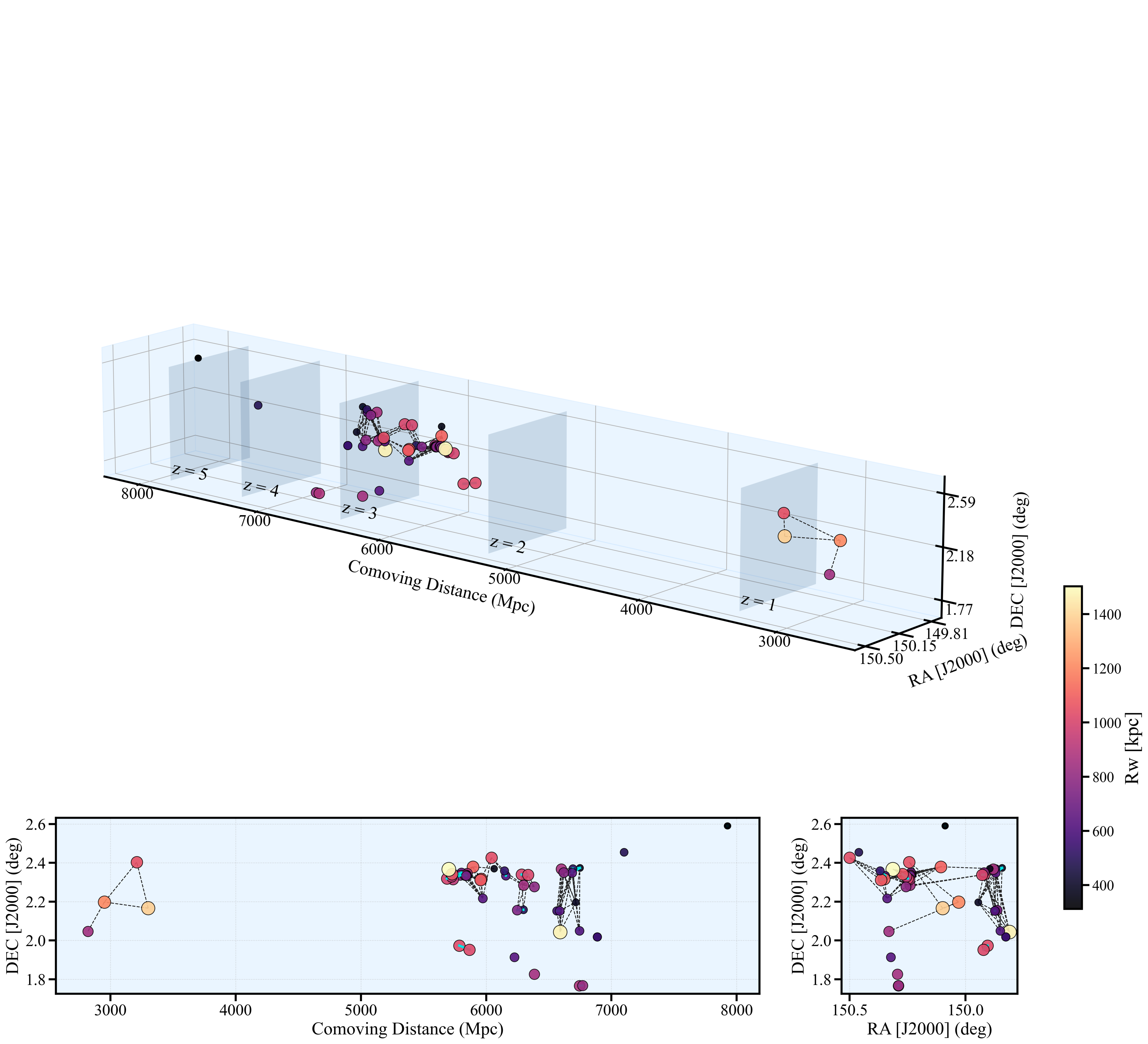}
\end{minipage}
\caption{The 3D distribution of all the protocluster candidates meet Category A \& B. The unit of the z-axis is comoving distance (cMpc), and we have marked the positions of redshifts 1-5 on it. The color and size of each candidate varies according to its $R_{W}$ value. For the two sources connected by the black dashed line: their distance on the projection plane (RA, DEC plane) is $\leq$ 10 pMpc (i.e., the classic size of a protocluster), and their redshift difference is less than twice the maximum $z_{\rm{rms}}$ value between them ($\Delta z \leq 2z_{\rm{rms}}$). For the two sources connected by the thick cyan solid line: their distance on the projection plane is $\leq$ 1 pMpc, and their redshift difference is $\Delta z \leq z_{\rm{rms}}$. \textbf{The complete figure set (4 images) showing breakdowns by specific redshift ranges is available in the online journal.}\label{fig:3d_distribution}}
\end{figure*}

The 3D spatial distribution maps corresponding to Category A and B sources identified by our PPM analysis are presented as a complete figure set in Figure~\ref{fig:3d_distribution}. As shown in the bottom panel of Figure~\ref{fig:3d_distribution}, the large-scale structures at $z \sim 2.5$ have a volume of hundreds of megaparsecs on each side—resulting in a total volume of $\sim 10^5\ \mathrm{Mpc^3}$—and extend over $500\ \mathrm{Mpc}$ along the line of sight. In contrast, the large-scale structures at $z \sim 3.3$ are significantly more compact.

\figsetstart
\figsetnum{19} 
\figsettitle{3D distribution of Category A and B overdensity candidates at different redshifts}

\figsetend

\clearpage

\input{table6}
\clearpage

\input{table7}
\clearpage


\bibliography{bibliography}{}
\bibliographystyle{aasjournalv7}



\end{document}

%% file: table1.tex
\begin{deluxetable}{lccccc}
\tablecaption{Overdensity Identification Categories \label{tab:table1}}
\tablewidth{\textwidth}
\tablehead{
    \colhead{Parameter / Category} & \colhead{A} & \colhead{B} & \colhead{C} & \colhead{D} & \colhead{E\tablenotemark{a}}
}
\startdata
\cutinhead{Identification Criteria}
Significance ($\sigma$)                                 & $\geq 5\sigma$   & $\geq 5\sigma$   & $\geq 4\sigma$     & $\geq 3\sigma$     & $\geq 2\sigma$ \\
Spatial peak offset ($\theta_{\rm ov}$)                 & $\leq R_{w}$     & $\leq R_{w}$     & $\leq 2 R_{w}$     & $\leq 2 R_{w}$     & \nodata \\
Overdensity ($\delta$)                                  & $\geq 7$         & $\geq 5$         & $\geq 4$           & $\geq 3$           & \nodata \\
Redshift peak offset ($|z_{\rm ov}-z_{\rm beacon}|$)    & $\leq z_{\rm rms}$ & $\leq z_{\rm rms}$ & $\leq 2 z_{\rm rms}$ & $\leq 3 z_{\rm rms}$ & $\leq 0.3$ \\
\cutinhead{Cumulative Source Counts\tablenotemark{b}}
CO/\CIIline{} spec-$z$ SMGs                             & 6  & 6  & 7   & 17   & 37 \\
All spec-$z$ SMGs                                       & 6  & 8  & 14  & 37   & 71 \\
All SMGs                                                & 6  & 11 & 40  & 105  & 273 \\
All unique SFGs                                         & 13 & 43 & 268 & 987 & 3005 \\
\enddata
\tablecomments{
Five criteria categories for the robustness of our overdensity identification, ranked from the most confident (Category A) to the least confident (Category E). The five parameters from the PPM calculation used as criteria are: the significance of the overdensity ($\sigma$), the redshift value of the overdensity peak ($z_{\rm ov}$), the overdensity redshift uncertainty ($z_{\rm rms}$ dispersion), the overdensity value ($\delta$), and the projected radius ($R_{w}$, defined as the radius where the peak overdensity drops to 1\%). More details for each parameter can be found in Section \ref{sec:method}.
}
\tablenotetext{a}{Adopts the same criteria as in \cite{Calvi2023}.}
\tablenotetext{b}{Categories A through E are in an inclusive relationship; each subsequent category includes all sources from the preceding ones (e.g., the 40 SMGs in Category C include the 6 from A and the 11 from B).}
\end{deluxetable}

%% file: table2.tex
\startlongtable
\begin{deluxetable*}{l c c c c c c c c}
\tablewidth{\textwidth} 
\tablecaption{%
    Overdensity candidates in categories A \& B from this work (Beacon:SMG)
    \label{table:3}
}
\tablehead{%
    \colhead{Name (SMG:ID)} & \colhead{Peak coordinates} & \colhead{$z$} & \colhead{$N_{\mathrm{select}}$} & \colhead{$z_{\rm{ov}}$} & \colhead{${R_{w} [\mathrm{kpc}]}$} & \colhead{$\sigma$} & \colhead{$\delta$} & \colhead{Category} \\
    \colhead{(1)} & \colhead{(2)} & \colhead{(3)} & \colhead{(4)} & \colhead{(5)} & \colhead{(6)} & \colhead{(7)} & \colhead{(8)} & \colhead{(9)}
}
\startdata
\textit{(SMGs)} & & & & & & & & \\
PC 1 (SMG:674217){$^{a,b,c,d}$} & 10:00:57.12 +02:20:12.48 & 2.503\tablenotemark{$^\ast$} & 287 & \(2.476 \pm 0.069\) & 766 & 8.0$\sigma$ & 12.26 & A \\
PC 2 (SMG:676246){$^{a,b,c,d}$} & 10:00:57.02 +02:20:12.48 & 2.504\tablenotemark{$^\ast$} & 289 & \(2.476 \pm 0.069\) & 760 & 8.0$\sigma$ & 12.15 & A \\
PC 3 (SMG:547511){$^{a,b,c,d}$} & 10:00:57.19 +02:20:09.96 & 2.507\tablenotemark{$^\ast$} & 237 & \(2.467 \pm 0.071\) & 871 & 8.0$\sigma$ & 12.83 & A \\
PC 4 (SMG:673258){$^{a,b,c,d}$} & 10:00:57.07 +02:20:11.04 & 2.513\tablenotemark{$^\ast$} & 291 & \(2.475 \pm 0.070\) & 760 & 8.0$\sigma$ & 13.01 & A \\
PC 5 (SMG:845982) & 10:00:57.48 +02:17:02.76 & 2.941\tablenotemark{$^\ast$} & 579 & \(2.997 \pm 0.083\) & 843 & 8.0$\sigma$ & 5.07 & B \\
PC 6 (SMG:195685) & 09:59:29.76 +02:09:09.36 & 3.251 & 150 & \(3.211 \pm 0.086\) & 623 & 7.9$\sigma$ & 5.15 & B \\
PC 7 (SMG:858393)\tablenotemark{$^e$} & 09:59:31.42 +02:22:05.88 & 3.267\tablenotemark{$^\ast$} & 293 & \(3.343 \pm 0.085\) & 865 & 6.9$\sigma$ & 5.96 & B \\
PC 8 (SMG:921875) & 09:59:22.24 +02:22:25.20 & 3.441 & 47 & \(3.441 \pm 0.079\) & 382 & 5.4$\sigma$ & 6.78 & B \\
PC 9 (SMG:488700) & 09:59:18.38 +02:01:07.14 & 3.615 & 175 & \(3.584 \pm 0.088\) & 549 & 5.3$\sigma$ & 5.22 & B \\
PC 10 (SMG:667211)\tablenotemark{$^f$} & 10:00:21.17 +02:35:26.16 & 5.293\tablenotemark{$^\ast$} & 22 & \(5.349 \pm 0.073\) & 312 & 5.2$\sigma$ & 24.79 & A \\
PC 11 (SMG:665549)\tablenotemark{$^f$} & 10:00:21.22 +02:35:27.96 & 5.298\tablenotemark{$^\ast$} & 23 & \(5.362 \pm 0.083\) & 315 & 5.1$\sigma$ & 20.05 & A \\
\enddata
\tablecomments{(1) Name (SMG:ID): Name of the (proto)cluster candidates (SMGs are used as beacons, with their IDs derived from the A$^3$COSMOS~(Data version 20220606) catalogs); (2) peak coordinates: Projected coordinates of the overdensity candidate peaks; (3) $z$: Redshift of SMGs, values marked with the superscript ({$^\ast$}) are spectroscopic redshifts; all others are photometric redshifts from COSMOS2025; (4) $N_{\rm{select}}$: Richness estimated from PPM analysis; (5) $z_{\rm{ov}}$: Redshift of the overdensity peak from PPM analysis; (6) $R_{w}$: The position where the peak density drops to one percent, i.e., the estimated size of the core; (7) $\sigma$: Overdensity significance from PPM analysis; (8) $\delta$: Peak overdensity value; (9) Category: Criteria classified according to different standards (see Table \ref{tab:table1} for details).}
\tablerefs{%
Known overdensity sources:
\tablenotetext{$a$}{CL~J1001+0220 \citep{2016ApJ...828...56W};}
\tablenotetext{$b$}{zCOSMOS-ID 26~\citep{Diener2013};}
\tablenotetext{$c$}{Hyperion-P 4~\citep{2018AA...619A..49C};}
\tablenotetext{$d$}{Hyperion-P 5~\citep{2018AA...619A..49C};}
\tablenotetext{$e$}{Elent\'{a}ri-P 8~\citep{Forrest2023};}
\tablenotetext{$f$}{AzTEC-3 protocluster \citep{2011Natur.470..233C}.} 
}
\end{deluxetable*}

%% file: table3.tex
\begin{deluxetable}{l c c}[ht!]
\tablecaption{Summary of Redshift-available Sources for the All SMGs Sample \label{tab:ztype}}
\tablehead{
\colhead{Redshift-available Source} & \colhead{Number of Sources} & \colhead{Fraction (\%)}
}
\startdata
Spectroscopic (Spec-$z$) & 168 & 37.4 \\
COSMOS2025 Photo-$z$ & 271 & 60.4 \\
\cite{Jin2018} Photo-$z$ & 4 & 0.9 \\
COSMOS2020 Classic Photo-$z$ & 5 & 1.1 \\
\cite{Davidzon2017} Photo-$z$ & 1 & 0.2 \\
\hline
Total & 449 & 100.0 \\
\enddata
\tablerefs{COSMOS2025 Photo-$z$~\citep{Shuntov2026}; COSMOS2020 Classic Photo-$z$~\citep{Weaver2022}; Spec-$z$~(\citealt{Khostovan2026,2013ARAA..51..105C, 2018ApJ...867...92S, 2019ApJ...880...15K, 2020AA...641A.155V, 2025AA...699A..80H}, a full reference list is provided in D. Liu et al., in prep.)}
\end{deluxetable}

%% file: table4.tex
\startlongtable
\begin{deluxetable*}{l@{\extracolsep{\fill}}ccccccccc}
\tablecaption{Overdensity candidate in category A \& B matching spec-$z$ member galaxies and COSMOS-Web galaxy groups~($z \leq 3.7$).\label{table4}}
\tablewidth{\textwidth}
\tablehead{
    \colhead{Name (Beacon:ID)} & \colhead{$R_{\rm PPM}$} & \colhead{$z_{\rm ov}$} & \colhead{$z_{\rm rms}$} & \colhead{$N_{\rm projected}$} & \colhead{$N_{\rm spec-z~member}$} & \colhead{Group ID ($\lambda$)} & \colhead{S/N$_{\rm group}$} & \colhead{Sep} & \colhead{$|\Delta z|$} \\
    \colhead{} & \colhead{(arcmin)} & \colhead{} & \colhead{} & \colhead{} & \colhead{} & \colhead{} & \colhead{} & \colhead{(Mpc/$h$)} & \colhead{} \\
    \colhead{(1)} & \colhead{(2)} & \colhead{(3)} & \colhead{(4)} & \colhead{(5)} & \colhead{(6)} & \colhead{(7)} & \colhead{(8)} & \colhead{(9)} & \colhead{(10)}
}
\startdata
\textit{(SMGs)} & & & & & & & & & \\
PC 1 (SMG:674217)\tablenotemark{$^{a,b,c,d}$}  & 2.24 & 2.476 & 0.069 & 488 & 47 & 202 (3)  & 3.92 & 0.08 & 0.086  \\
PC 2 (SMG:676246)\tablenotemark{$^{a,b,c,d}$}  & 2.24 & 2.476 & 0.069 & 495 & 46 & 202 (3)  & 3.92 & 0.08 & 0.086  \\
PC 3 (SMG:547511)\tablenotemark{$^{a,b,c,d}$}  & 2.00 & 2.467 & 0.071 & 406 & 44 & 202 (3)  & 3.92 & 0.09 & 0.077  \\
PC 4 (SMG:673258)\tablenotemark{$^{a,b,c,d}$}  & 2.24 & 2.475 & 0.070 & 494 & 46 & 202 (3)  & 3.92 & 0.08 & 0.085  \\
PC 5 (SMG:845982)\tablenotemark{$^e$}        & 3.74 & 2.997 & 0.083 & 937 & 16 & 437 (4)  & 3.94 & 0.19 & 0.017  \\
PC 6 (SMG:195685)                            & 2.00 & 3.211 & 0.086 & 183 & 0  & 1312 (3) & 2.83 & 0.01 & 0.041  \\
PC 7 (SMG:858393)\tablenotemark{$^e$}        & 2.83 & 3.343 & 0.085 & 391 & 8  & 117 (2)  & 4.51 & 0.61 & 0.057  \\
PC 8 (SMG:921875)                            & 1.00 & 3.441 & 0.079 & 43  & 3  & 117 (2)  & 4.51 & 0.11 & 0.041  \\
PC 9 (SMG:488700)                            & 3.00 & 3.584 & 0.088 & 295 & 0  & 142 (2)  & 3.69 & 0.05 & 0.026  \\
PC 10 (SMG:667211)\tablenotemark{$^f$}       & 2.24 & 5.349 & 0.073 & 83  & 2  & -      & -    & -    & -      \\
PC 11 (SMG:665549)\tablenotemark{$^f$}       & 2.24 & 5.362 & 0.083 & 83  & 2  & -      & -    & -    & -      \\
\textit{(SFGs)} & & & & & & & & & \\
PC 12 (SFG:170718)                           & 2.45 & 0.841 & 0.087 & 259 & 48 & 216 (4)  & 3.35 & 0.31 & 0.011  \\
PC 13 (SFG:61114)                            & 2.65 & 0.931 & 0.081 & 427 & 85 & 36 (4)   & 5.46 & 0.06 & 0.009  \\
PC 14 (SFG:466489)                           & 3.32 & 0.915 & 0.090 & 558 & 72 & 70 (2)   & 4.87 & 0.06 & 0.015  \\
PC 15 (SFG:99562)                            & 2.24 & 0.926 & 0.082 & 309 & 45 & 369 (3)  & 3.92 & 0.03 & 0.006  \\
PC 16 (SFG:729398)\tablenotemark{$^{a,b,c,d}$} & 2.00 & 2.451 & 0.080 & 390 & 40 & 202 (3)  & 3.92 & 0.09 & 0.061  \\
PC 17 (SFG:503164)                           & 3.16 & 2.396 & 0.086 & 754 & 24 & 249 (4)  & 3.72 & 0.18 & 0.146  \\
PC 18 (SFG:731169)\tablenotemark{$^{a,b,c,d}$} & 3.32 & 2.429 & 0.086 & 904 & 58 & 202 (4)  & 3.92 & 0.08 & 0.039  \\
PC 19 (SFG:734147)\tablenotemark{$^{a,b,c,d}$} & 3.00 & 2.404 & 0.088 & 796 & 31 & 202 (4)  & 3.92 & 0.08 & 0.014  \\
PC 20 (SFG:684645)\tablenotemark{$^{a,b,c,d}$} & 2.12 & 2.435 & 0.087 & 389 & 40 & 202 (4)  & 3.92 & 0.08 & 0.045  \\
PC 21 (SFG:243115)                           & 1.41 & 2.500 & 0.078 & 55  & 2  & 91 (2)   & 3.82 & 0.02 & 0.040  \\
PC 22 (SFG:689114)\tablenotemark{$^{a,b,c,d}$} & 2.24 & 2.471 & 0.070 & 493 & 45 & 202 (3)  & 3.92 & 0.08 & 0.081  \\
PC 23 (SFG:689718)\tablenotemark{$^{a,b,c,d}$} & 2.24 & 2.487 & 0.068 & 466 & 44 & 202 (3)  & 3.92 & 0.09 & 0.097  \\
PC 24 (SFG:688366)\tablenotemark{$^{a,b,c,d}$} & 2.83 & 2.484 & 0.069 & 716 & 48 & 202 (3)  & 3.92 & 0.08 & 0.094  \\
PC 25 (SFG:689117)\tablenotemark{$^{a,b,c,d}$} & 2.24 & 2.473 & 0.070 & 494 & 47 & 202 (3)  & 3.92 & 0.08 & 0.083  \\
PC 26 (SFG:687390)\tablenotemark{$^{a,b,c,d}$} & 2.24 & 2.480 & 0.065 & 500 & 44 & 202 (3)  & 3.92 & 0.09 & 0.090  \\
PC 27 (SFG:689023)\tablenotemark{$^{a,b,c,d}$} & 2.65 & 2.468 & 0.073 & 635 & 49 & 202 (3)  & 3.92 & 0.08 & 0.078  \\
PC 28 (SFG:241038)                           & 2.24 & 2.503 & 0.082 & 153 & 4  & 91 (2)   & 3.82 & 0.03 & 0.037  \\
PC 29 (SFG:645076)\tablenotemark{$^{g,h,i}$} & 2.45 & 2.495 & 0.075 & 640 & 36 & 118 (3)  & 4.18 & 0.03 & 0.005  \\
PC 30 (SFG:734673)                           & 2.00 & 2.640 & 0.086 & 166 & 4  & 521 (2)  & 3.29 & 0.02 & 0.050  \\
PC 31 (SFG:734686)                           & 2.45 & 2.638 & 0.087 & 297 & 4  & 521 (2)  & 3.29 & 0.00 & 0.052  \\
PC 32 (SFG:719482)\tablenotemark{$^j$}       & 2.24 & 2.582 & 0.094 & 282 & 10 & 1537 (4) & 2.57 & 0.57 & 0.092  \\
PC 33 (SFG:560232)                           & 2.24 & 2.684 & 0.086 & 186 & 0  & 300 (6)  & 3.92 & 0.58 & 0.166  \\
PC 34 (SFG:589013)\tablenotemark{$^k$}       & 2.45 & 2.725 & 0.083 & 640 & 36 & 575 (2)  & 2.91 & 0.05 & 0.165  \\
PC 35 (SFG:505154)\tablenotemark{$^l$}       & 3.46 & 2.816 & 0.084 & 629 & 10 & 209 (2)  & 4.56 & 0.18 & 0.064  \\
PC 36 (SFG:736972)\tablenotemark{$^l$}       & 2.65 & 2.721 & 0.087 & 417 & 7  & 521 (2)  & 3.29 & 0.03 & 0.031  \\
PC 37 (SFG:371514)\tablenotemark{$^m$}       & 2.00 & 2.936 & 0.082 & 162 & 3  & 166 (4)  & 4.58 & 0.33 & 0.006  \\
PC 38 (SFG:7939)                             & 1.41 & 2.893 & 0.075 & 91  & 2  & 127 (1)  & 4.91 & 0.06 & 0.013  \\
PC 39 (SFG:585672)                           & 2.45 & 2.922 & 0.083 & 315 & 5  & 67 (4)   & 4.75 & 0.45 & 0.112  \\
PC 40 (SFG:8088)                             & 1.41 & 2.941 & 0.080 & 89  & 3  & 127 (1)  & 4.91 & 0.09 & 0.061  \\
PC 41 (SFG:585657)                           & 2.98 & 2.980 & 0.082 & 222 & 6  & 67 (3)   & 4.75 & 0.44 & 0.170  \\
PC 42 (SFG:723370)                           & 3.00 & 2.985 & 0.080 & 507 & 15 & 437 (3)  & 3.94 & 0.17 & 0.005  \\
PC 43 (SFG:359770)\tablenotemark{$^m$}       & 2.83 & 3.047 & 0.083 & 275 & 0  & 1010 (5) & 3.18 & 0.34 & 0.043  \\
PC 44 (SFG:7858)                             & 1.41 & 3.233 & 0.086 & 93  & 0  & 1312 (3) & 2.83 & 0.06 & 0.063  \\
PC 45 (SFG:206664)                           & 3.00 & 3.336 & 0.088 & 271 & 0  & 312 (3)  & 4.02 & 0.21 & 0.056  \\
PC 46 (SFG:587023)                           & 2.83 & 3.292 & 0.087 & 399 & 5  & 1561 (2) & 2.66 & 0.30 & 0.102  \\
PC 47 (SFG:583968)\tablenotemark{$^e$}       & 2.65 & 3.358 & 0.084 & 329 & 4  & 117 (2)  & 4.51 & 0.60 & 0.042  \\
PC 48 (SFG:587399)\tablenotemark{$^e$}       & 1.73 & 3.365 & 0.081 & 137 & 3  & 117 (2)  & 4.51 & 0.58 & 0.035  \\
PC 49 (SFG:56143)                            & 2.65 & 3.343 & 0.082 & 388 & 0  & 1204 (2) & 3.10 & 0.51 & 0.067  \\
PC 50 (SFG:352015)                           & 1.73 & 3.438 & 0.079 & 91  & 0  & 1445 (3) & 2.82 & 0.23 & 0.018  \\
PC 51 (SFG:250840)                           & 1.21 & 3.449 & 0.082 & 51  & 0  & 591 (3)  & 3.45 & 0.03 & 0.021  \\
PC 52 (SFG:352238)                           & 2.24 & 3.476 & 0.081 & 140 & 1  & 1445 (3) & 2.82 & 0.22 & 0.056  \\
PC 53 (SFG:203990)                           & 3.00 & 3.588 & 0.087 & 295 & 0  & 142 (2)  & 3.69 & 0.04 & 0.022  \\
PC 54 (SFG:561931)                           & 2.45 & 3.866 & 0.087 & 242 & 0  & -      & -    & -    & -      \\
\enddata
\tablecomments{(1) Name (Beacon:ID): Name of the (proto)cluster candidates (SMGs or SFGs are used as beacons, with their IDs derived from the A$^3$COSMOS~(Date version 20220606) and COSMOS2025 catalogs respectively); (2) $R_{\rm PPM}$: the estimate of the overdensity candidate's projected radius; (3) $z_{ov}$: Redshift of the overdensity candidates from PPM analysis; (4) $z_{\rm rms}$: The redshift uncertainty; (5) $N_{\rm projected}$: The number of projected associated galaxies for each overdensity candidate; (6) $N_{\rm spec-z~member}$: The number of spec-$z$ member galaxies\citep{Khostovan2026} for each overdensity candidate; (7) Group ID ($\lambda$): The ID of the matched COSMOS-Web galaxy group catalog~($z \leq 3.7$; \citealt{Toni2025}), with the total number of matched groups indicated in parentheses; (8) S/N$_{\rm group}$: The signal-to-noise ratio of the matched COSMOS-Web galaxy group, including background and group contributions~(\citealt{Toni2025}); (9) Sep [Mpc/$h$]: the spatial separation between the category A \& B overdensity candidate and matched COSMOS-Web galaxy group; (10) $|\Delta z|$: the absolute redshift difference between the category A \& B overdensity candidate and matched COSMOS-Web galaxy group.}
\tablerefs{
Known overdensities sources:
\tablenotetext{$a$}{CL~J1001+0220 \citep{2016ApJ...828...56W};}
\tablenotetext{$b$}{zCOSMOS-ID 26~\citep{Diener2013};}
\tablenotetext{$c$}{Hyperion-P 4~\citep{2018AA...619A..49C};}
\tablenotetext{$d$}{Hyperion-P 5~\citep{2018AA...619A..49C};}
\tablenotetext{$e$}{Elent\'{a}ri-P 8~\citep{Forrest2023};}
\tablenotetext{$f$}{AzTEC-3 protocluster \citep{2011Natur.470..233C};}
\tablenotetext{$g$}{PCL1002~\citep{Casey2015};}
\tablenotetext{$h$}{zCOSMOS-ID 30~\citep{Diener2013};}
\tablenotetext{$i$}{Hyperion-P 1~\citep{2018AA...619A..49C};}
\tablenotetext{$j$}{Hyperion-P 6~\citep{2018AA...619A..49C};}
\tablenotetext{$k$}{zCOSMOS-ID 42~\citep{Diener2013};}
\tablenotetext{$l$}{RO-1001 group~\citep{Daddi2021};}
\tablenotetext{$m$}{zCOSMOS-ID 41~\citep{Diener2013};}
}
\end{deluxetable*}

%% file: table5.tex
\startlongtable
\begin{deluxetable*}{l c c c c c c c c}
\tablewidth{\textwidth}
\tablecaption{%
    Overdensity candidates in categories A \& B from this work (Beacon: SFG)
    \label{table:sfg}
}
\tablehead{%
    \colhead{Name (SFG:ID)} & \colhead{Peak coordinates} & \colhead{$z_{bc}$} & \colhead{$N_{\mathrm{select}}$} & \colhead{$z_{\rm{ov}}$} & \colhead{${R_{w} [kpc]}$} & \colhead{$\sigma$} & \colhead{$\delta$} & \colhead{Category} \\
    \colhead{(1)} & \colhead{(2)} & \colhead{(3)} & \colhead{(4)} & \colhead{(5)} & \colhead{(6)} & \colhead{(7)} & \colhead{(8)} & \colhead{(9)}
}
\startdata
\textit{(SFGs)} & & & & & & & & \\
PC 12 (SFG:170718) & 10:01:21.24 +02:03:14.21 & 0.812 & 1238 & \( 0.841 \pm 0.087 \) & 844 & 6.5$\sigma$ & 5.37 & B \\
PC 13 (SFG:61114) & 10:00:06.40 +02:12:12.16 & 0.861 & 1256 & \( 0.931 \pm 0.081 \) & 1209 & 8.0$\sigma$ & 5.19 & B \\
PC 14 (SFG:466489) & 10:00:53.24 +02:23:49.97 & 0.962 & 1402 & \( 0.915 \pm 0.090 \) & 1054 & 5.1$\sigma$ & 5.14 & B \\
PC 15 (SFG:99562) & 10:00:22.43 +02:11:54.54 & 0.999 & 912 & \( 0.926 \pm 0.082 \) & 1384 & 6.9$\sigma$ & 5.25 & B \\
PC 16 (SFG:729398){$^{a,b,c,d}$} & 10:00:57.17 +02:20:09.54 & 2.382 & 282 & \( 2.451 \pm 0.080 \) & 995 & 6.4$\sigma$ & 11.97 & A \\
PC 17 (SFG:503164) & 10:01:22.31 +02:21:26.85 & 2.394 & 713 & \( 2.396 \pm 0.086 \) & 1502 & 7.4$\sigma$ & 5.18 & B \\
PC 18 (SFG:731169){$^{a,b,c,d}$} & 10:00:57.01 +02:20:09.79 & 2.412 & 729 & \( 2.429 \pm 0.086 \) & 927 & 6.0$\sigma$ & 9.54 & A \\
PC 19 (SFG:734147){$^{a,b,c,d}$} & 10:00:57.03 +02:20:09.86 & 2.416 & 659 & \( 2.404 \pm 0.088 \) & 1063 & 7.1$\sigma$ & 5.83 & B \\
PC 20 (SFG:684645){$^{a,b,c,d}$} & 10:00:57.00 +02:20:10.34 & 2.422 & 320 & \( 2.435 \pm 0.087 \) & 927 & 5.7$\sigma$ & 10.36 & A \\
PC 21 (SFG:243115) & 09:59:36.79 +01:57:40.04 & 2.464 & 111 & \( 2.500 \pm 0.078 \) & 1032 & 5.0$\sigma$ & 5.32 & B \\
PC 22 (SFG:689114){$^{a,b,c,d}$} & 10:00:57.12 +02:20:12.45 & 2.470 & 289 & \( 2.471 \pm 0.070 \) & 767 & 8.0$\sigma$ & 12.89 & A \\
PC 23 (SFG:689718){$^{a,b,c,d}$} & 10:00:57.14 +02:20:11.38 & 2.479 & 263 & \( 2.487 \pm 0.068 \) & 703 & 5.9$\sigma$ & 11.97 & A \\
PC 24 (SFG:688366){$^{a,b,c,d}$} & 10:00:57.07 +02:20:10.06 & 2.486 & 392 & \( 2.484 \pm 0.069 \) & 696 & 7.8$\sigma$ & 11.86 & A \\
PC 25 (SFG:689117){$^{a,b,c,d}$} & 10:00:57.11 +02:20:11.23 & 2.505 & 288 & \( 2.473 \pm 0.070 \) & 743 & 8.0$\sigma$ & 13.18 & A \\
PC 26 (SFG:687390){$^{a,b,c,d}$} & 10:00:57.22 +02:20:11.67 & 2.511 & 282 & \( 2.480 \pm 0.065 \) & 703 & 8.0$\sigma$ & 12.31 & A \\
PC 27 (SFG:689023){$^{a,b,c,d}$} & 10:00:57.09 +02:20:11.33 & 2.517 & 383 & \( 2.468 \pm 0.073 \) & 861 & 8.0$\sigma$ & 12.58 & A \\
PC 28 (SFG:241038) & 09:59:36.61 +01:57:40.38 & 2.534 & 231 & \( 2.503 \pm 0.082 \) & 1022 & 5.1$\sigma$ & 5.12 & B \\
PC 29 (SFG:645076){$^{e,f,g}$} & 10:00:26.00 +02:22:33.60 & 2.559 & 255 & \( 2.495 \pm 0.075 \) & 1103 & 5.8$\sigma$ & 5.98 & B \\
PC 30 (SFG:734673) & 10:01:23.58 +02:19:18.25 & 2.613 & 192 & \( 2.640 \pm 0.086 \) & 1075 & 6.1$\sigma$ & 7.59 & A \\
PC 31 (SFG:734686) & 10:01:23.73 +02:19:20.35 & 2.617 & 270 & \( 2.638 \pm 0.087 \) & 1103 & 7.2$\sigma$ & 8.34 & A \\
PC 32 (SFG:719482){$^{h}$} & 10:01:18.30 +02:12:54.36 & 2.628 & 175 & \( 2.582 \pm 0.094 \) & 631 & 5.4$\sigma$ & 6.34 & B \\
PC 33 (SFG:560232) & 10:01:53.35 +02:25:23.81 & 2.692 & 254 & \( 2.684 \pm 0.086 \) & 1041 & 5.7$\sigma$ & 5.45 & B \\
PC 34 (SFG:589013){$^{i}$} & 09:59:34.18 +02:22:10.11 & 2.711 & 235 & \( 2.725 \pm 0.083 \) & 357 & 5.2$\sigma$ & 5.51 & B \\
PC 35 (SFG:505154){$^{j}$} & 10:01:29.40 +02:20:26.93 & 2.790 & 535 & \( 2.816 \pm 0.084 \) & 532 & 7.8$\sigma$ & 5.74 & B \\
PC 36 (SFG:736972){$^{j}$} & 10:01:23.63 +02:19:24.93 & 2.801 & 325 & \( 2.721 \pm 0.087 \) & 603 & 6.9$\sigma$ & 5.34 & B \\
PC 37 (SFG:371514){$^{k}$} & 10:01:17.51 +01:53:32.07 & 2.867 & 204 & \( 2.936 \pm 0.082 \) & 629 & 6.7$\sigma$ & 5.03 & B \\
PC 38 (SFG:7939) & 09:59:28.02 +02:09:35.66 & 2.890 & 110 & \( 2.893 \pm 0.075 \) & 748 & 5.4$\sigma$ & 5.65 & B \\
PC 39 (SFG:585672) & 09:59:40.30 +02:20:04.90 & 2.926 & 296 & \( 2.922 \pm 0.083 \) & 997 & 8.0$\sigma$ & 5.04 & B \\
PC 40 (SFG:8088) & 09:59:27.52 +02:09:34.76 & 2.939 & 106 & \( 2.941 \pm 0.080 \) & 489 & 5.2$\sigma$ & 5.22 & B \\
PC 41 (SFG:585657) & 09:59:40.26 +02:20:05.91 & 2.978 & 193 & \( 2.980 \pm 0.082 \) & 1013 & 6.3$\sigma$ & 5.40 & B \\
PC 42 (SFG:723370) & 10:00:57.73 +02:17:06.23 & 3.030 & 377 & \( 2.985 \pm 0.080 \) & 784 & 7.8$\sigma$ & 5.09 & B \\
PC 43 (SFG:359770){$^{k}$} & 10:01:12.84 +01:51:04.21 & 3.031 & 302 & \( 3.047 \pm 0.083 \) & 853 & 5.2$\sigma$ & 5.91 & B \\
PC 44 (SFG:7858) & 09:59:29.37 +02:09:17.90 & 3.226 & 190 & \( 3.233 \pm 0.086 \) & 557 & 5.4$\sigma$ & 5.43 & B \\
PC 45 (SFG:206664) & 09:59:15.44 +02:02:34.79 & 3.256 & 299 & \( 3.336 \pm 0.088 \) & 1470 & 6.3$\sigma$ & 5.66 & B \\
PC 46 (SFG:587023) & 09:59:40.39 +02:22:01.97 & 3.288 & 245 & \( 3.292 \pm 0.087 \) & 761 & 6.3$\sigma$ & 5.23 & B \\
PC 47 (SFG:583968){$^{l}$} & 09:59:31.38 +02:22:09.04 & 3.366 & 235 & \( 3.358 \pm 0.084 \) & 645 & 7.9$\sigma$ & 6.80 & B \\
PC 48 (SFG:587399){$^{l}$} & 09:59:31.03 +02:22:07.92 & 3.373 & 122 & \( 3.365 \pm 0.081 \) & 503 & 6.1$\sigma$ & 5.88 & B \\
PC 49 (SFG:56143) & 09:59:47.39 +02:11:37.27 & 3.401 & 135 & \( 3.343 \pm 0.082 \) & 387 & 5.3$\sigma$ & 6.73 & B \\
PC 50 (SFG:352015) & 10:01:09.77 +01:46:08.99 & 3.437 & 106 & \( 3.438 \pm 0.079 \) & 888 & 5.2$\sigma$ & 7.76 & A \\
PC 51 (SFG:250840) & 09:59:22.64 +02:03:19.71 & 3.437 & 59 & \( 3.449 \pm 0.082 \) & 603 & 5.4$\sigma$ & 5.90 & B \\
PC 52 (SFG:352238) & 10:01:10.41 +01:46:09.46 & 3.467 & 172 & \( 3.476 \pm 0.081 \) & 823 & 8.0$\sigma$ & 7.47 & A \\
PC 53 (SFG:203990) & 09:59:19.09 +02:01:09.27 & 3.615 & 222 & \( 3.588 \pm 0.087 \) & 549 & 5.3$\sigma$ & 5.24 & B \\
PC 54 (SFG:561931) & 10:01:50.60 +02:27:08.18 & 3.901 & 138 & \( 3.866 \pm 0.087 \) & 487 & 5.2$\sigma$ & 5.60 & B \\
\enddata
\tablecomments{(1) Name (SFG:ID): Name of the (proto)cluster candidates (SFGs are used as beacons, with their IDs derived from the COSMOS2025 catalog); (2) peak coordinates: Projected coordinates of the peaks; (3) $z_{bc}$: Photometric redshift of SFGs from the COSMOS2025 catalog ; (4) $N_{\rm{select}}$: Richness estimated from PPM analysis; (5) $z_{\rm{ov}}$: Redshift of the overdensity peak from PPM analysis; (6) $R_{w}$: The position where the peak density drops to one percent, i.e., the estimated size of the core; (7) $\sigma$: Clustering significance from PPM analysis; (8) $\delta$: Peak overdensity value; (9) Category: Criteria classified according to different standards (see Table \ref{tab:table1} for details).
}
\tablerefs{%
Known overdensities sources:
\tablenotetext{$a$}{CL~J1001+0220 \citep{2016ApJ...828...56W};}
\tablenotetext{$b$}{zCOSMOS-ID 26~\citep{Diener2013};}
\tablenotetext{$c$}{Hyperion-P 4~\citep{2018AA...619A..49C};}
\tablenotetext{$d$}{Hyperion-P 5~\citep{2018AA...619A..49C};}
\tablenotetext{$e$}{PCL1002~\citep{Casey2015};}
\tablenotetext{$f$}{zCOSMOS-ID 30~\citep{Diener2013};}
\tablenotetext{$g$}{Hyperion-P 1~\citep{2018AA...619A..49C};}
\tablenotetext{$h$}{Hyperion-P 6~\citep{2018AA...619A..49C};}
\tablenotetext{$i$}{zCOSMOS-ID 42~\citep{Diener2013};}
\tablenotetext{$j$}{RO-1001 group~\citep{Daddi2021};}
\tablenotetext{$k$}{zCOSMOS-ID 41~\citep{Diener2013};}
\tablenotetext{$l$}{Elent\'{a}ri-P 8~\citep{Forrest2023};}
}
\end{deluxetable*}

%% file: table6.tex
\begin{sidewaystable}[!htbp]
\section{SMG CO/\CIIline{} subsamples}\label{sec:spectroscopy match SMG}

We calculate the CO(1--0) and \CIIline{} line luminosity-based molecular gas mass (M$_{\mathrm{H_2}}$) for all 47 SMGs, long with their spectroscopic redshifts, emission line information, IDs from the literature, referenced studies, and the overdensity significance ($\sigma$) calculated by us based on PPM analysis, are presented in Tables~\ref{tab:CO} and~\ref{tab:cII}.

\caption{Properties of the SMGs subsamples with the CO$(J\rightarrow J-1)$ emission line.}
\centering
\label{tab:CO}
\footnotesize
\renewcommand{\arraystretch}{0.8}
\begin{tabular*}{\linewidth}{@{\extracolsep{\fill}} lcccccccr @{}}
\hline\hline
ID(s) & $z_{{spec}}$ & Line & Excitation ratio & $S_{\rm{CO}(J\rightarrow{}J-1)}\Delta v$ & $L^{\prime}_{\rm{CO}(J\rightarrow{}J-1)}$ & $\rm{M_{H_{2}}}$ & significance & Reference(s) \\
 &  &  &  & (Jy km s$^{-1}$) & $(10^{10}\rm{K km s^{-1} pc^2})$ & ($10^{10}\rm{M}_{\odot}$) & ($\sigma$) &  \\
(1) & (2) & (3) & (4) & (5) & (6) & (7) & (8) & (9) \\
\hline
AzTEC-C159 & 4.565 & CO(2--1) & 0.90 & $0.105\pm0.019$ &  $2.17\pm0.39$ & $1.93\pm0.35$ & 3.02 & {\cite{2018AA...615A..25J}} \\
COSBO11/ & 1.8275 & CO(2--1) & 0.90 & $1.33\pm0.083$ &  $5.86\pm0.37$ & $5.21\pm0.33$ & 4.58 & \cite{2008AA...491..173A}, \\
SDSS\_J100038.01+0208 &  &  &  &  &   &  &  & \cite{2013ARAA..51..105C} \\
MAMBO-9/ & 5.85 & CO(6--5) & 0.27 & $0.48\pm0.03$ & $1.61\pm0.10$ & $4.76\pm0.30$ & 3.13 & \cite{2019ApJ...887...55C} \\
ID85001929 &  &  &  &  &   &  &  & \\
AS2COS0023.1/ & 4.341 & CO(4--3) & 0.34 & $1.9\pm0.1$ &  $9.08\pm0.48$ & $21.4\pm1.12$ & 3.58 & \cite{2024ApJ...961..226L} \\
AzTEC-1 &  &  &  &  &   &  &  & \\
AS2COS0028.1/ & 3.0966 & CO(3--2) & 0.63 & $0.9\pm0.2$ &  $4.43\pm0.98$ & $5.63\pm1.25$ & -- & \cite{2024ApJ...961..226L} \\
AzTEC-8 &  &  &  &  &   &  &  & \\
AS2COS0066.1 & 3.2492 & CO(4--3) & 0.34 & $1.6\pm0.3$ &  $4.80\pm0.90$ & $11.3\pm2.12$ & 3.21 & \cite{2024ApJ...961..226L} \\
Herschel-COSMOS-4 & 2.104 & CO(1--0) & 1.00 & $0.185\pm0.055$ & $4.21\pm1.25$ & $3.37\pm1.00$ & 5.60 & \cite{2019ApJ...880...15K} \\
Herschel-COSMOS-7 & 2.24 & CO(1--0) & 1.00 & $0.236\pm0.059$ & $6.00\pm1.50$ & $4.80\pm1.20$ & 5.04 & \cite{2019ApJ...880...15K} \\
Herschel-COSMOS-10/ & 2.934 & CO(1--0) & 1.00 & $0.141\pm0.031$ & $5.71\pm1.26$& $4.57\pm1.00$ & 2.77 & \cite{2019ApJ...880...15K} \\
Cassata-Gal3 &  &  &  &  &   &  &  & \\
Herschel-COSMOS-11 & 1.941 & CO(1--0) & 1.00 & $0.368\pm0.074$ & $7.24\pm1.46$ & $5.79\pm1.16$ & 5.05 & \cite{2019ApJ...880...15K} \\
Cassata-Gal2 & 3.0557 & CO(4--3) & 0.34 & $0.5\pm0.11$ &  $1.35\pm0.30$ & $3.19\pm0.70$ & 3.48 & \cite{2020ApJ...891...83C} \\
Cassata-Gal4 & 3.4315 & CO(4--3) & 0.34 & $1.1\pm0.12$ &  $3.61\pm0.39$ & $8.49\pm0.93$ & 8.00 & \cite{2020ApJ...891...83C} \\
PACS-819 & 1.4449 & CO(2--1) & 0.90 & $1.1\pm0.07$ &  $3.13\pm0.20$ & $2.78\pm0.18$ & -- & \cite{2015ApJ...812L..23S,2018ApJ...867...92S}\\
PACS-830 & 1.461 & CO(2--1) & 0.90 & $1.18\pm0.1$ &  $3.42\pm0.29$ & $3.04\pm0.26$ & 5.61 & \cite{2015ApJ...812L..23S,2018ApJ...867...92S}\\
PACS-197 & 1.6016 & CO(2--1) & 0.90 & $1.086\pm0.088$ &  $3.75\pm0.30$ & $3.33\pm0.27$ & -- & \cite{2018ApJ...867...92S} \\
ID-31880 & 1.3999 & CO(2--1) & 0.90 & $0.9241\pm0.1247$ &  $2.47\pm0.33$ & $2.20\pm0.30$ & 4.55 & \cite{2020AA...641A.155V} \\
ID-51670 & 1.6168 & CO(2--1) & 0.90 & $0.7075\pm0.114$ &  $2.49\pm0.40$ & $2.21\pm0.36$ & 6.57 & \cite{2020AA...641A.155V} \\
ID-41210 & 1.3145 & CO(2--1) & 0.90 & $0.8086\pm0.0691$ &  $1.92\pm0.16$ & $1.70\pm0.15$ & 2.63 & \cite{2020AA...641A.155V} \\
ID-2993 & 1.1917 & CO(2--1) & 0.90 & $0.9073\pm0.1312$ &  $1.78\pm0.26$ & $1.58\pm0.23$ & 5.43 & \cite{2020AA...641A.155V} \\
ID-48136 & 1.1831 & CO(2--1) & 0.90 & $2.0664\pm0.1529$ &  $3.99\pm0.30$ & $3.55\pm0.26$ & 5.47 & \cite{2020AA...641A.155V} \\
ALMA05 & 0.323 & CO(3--2) & 0.63 & $8.54\pm0.89$ & $0.509\pm0.053$ & $0.646\pm0.067$ & 3.99 & \cite{2017MNRAS.471.2124L} \\
ALMA07 & 0.354 & CO(3--2) & 0.63 & $14.86\pm1.55$ & $1.07\pm0.11$ & $1.36\pm0.14$ & 4.60 & \cite{2017MNRAS.471.2124L} \\
ALMA14 & 0.488 & CO(3--2) & 0.63 & $1.19\pm0.15$ & $0.168\pm0.021$ & $0.213\pm0.027$ & 2.82 & \cite{2017MNRAS.471.2124L} \\
cid\_346 & 2.2198 & CO(3--2) & 0.63 & $0.632\pm0.069$ &  $1.76\pm0.19$ & $2.23\pm0.24$ & 5.95 & \cite{2021AA...646A..96C} \\
ID-20010161 & 5.051 & CO(5--4) & 0.36 & $0.33\pm0.06$ &  $1.27\pm0.23$ & $2.83\pm0.52$ & 4.18 & \cite{2019ApJ...887..144J}\\
ID-5340 & 3.815 & CO(4--3) & 0.34 & $0.9\pm0.06$ & $3.50\pm0.23$ & $8.25\pm0.55$ & 2.54 & \cite{2022AA...665A...3J}\\
ID-9316 & 4.072 & CO(4--3) & 0.34 & $0.89\pm0.03$ & $3.84\pm0.13$ & $9.05\pm0.30$ & 3.00 & \cite{2022AA...665A...3J}\\
CL1001-130891 & 2.513 & CO(1--0) & 1.00 & $0.109\pm0.01$ &  $3.39\pm0.31$ & $2.71\pm0.25$ & 8.00 & \cite{2016ApJ...828...56W,2018ApJ...867L..29W} \\
CL1001-130933 & 2.504 & CO(1--0) & 1.00 & $0.021\pm0.003$ &  $0.648\pm0.093$ & $0.518\pm0.074$ & 8.00 & \cite{2016ApJ...828...56W,2018ApJ...867L..29W} \\
CL1001-ID4/ & 2.504 & CO(1--0) & 1.00 & $0.0926\pm0.0126$ &  $2.86\pm0.39$ & $2.29\pm0.31$ & 8.00 & \cite{2021ApJ...913..110C}\\
$[Wang2016:130949]$ &  &  &  &  &   &  &  & \\
CL1001-ID5/ & 2.507 & CO(1--0) & 1.00 & $0.0374\pm0.0051$ &  $1.16\pm0.16$ & $0.926\pm0.126$ & 8.00 & \cite{2021ApJ...913..110C}\\
$[Casey2013:850.82]$/ &  &  &  &  &   &  &  & \\
$[Wang2016:130359]$ &  &  &  &  &   &  &  & \\
\hline
\end{tabular*}
\par
\vspace{6pt}
\begin{minipage}{\linewidth}
\footnotesize{\textbf{Notes.}(1)~Source ID(s). (2)~CO-based spectroscopic redshift. (3)~CO$(J\rightarrow J-1)$ transition. (4)~Excitation ratio, $r_{J1}$ = $L^{\prime}_{\rm{CO}(J\rightarrow{}J-1)}/L^{\prime}_{\rm{CO}(1-0)}$ = 1.0, 0.9, 0.63, 0.34, 0.36 and 0.27 for J = 1, 2, 3, 4, 5 and 6 from \cite{2021MNRAS.501.3926B}. (5)~Velocity integrated CO$(J\rightarrow J-1)$ flux. (6)~CO$(J\rightarrow J-1)$ line luminosity. (7)~The molecular H$_{2}$ gas mass estimated assuming a CO-to-H$_{2}$ conversion factor $\alpha_{\mathrm{CO}}$ = 0.8$\rm{M}_{\odot}$$(\rm{K km s^{-1} pc^2})^{-1}$. (8)~Overdensity significance calculated by PPM. (9)~Reference(s) for the z$_{spec}$, Line and Velocity integrated CO$(J\rightarrow J-1)$ flux.}
\end{minipage}
\end{sidewaystable}

%% file: table7.tex
\begin{sidewaystable}[!htbp]
\caption{Properties of the SMGs subsamples with the \CIIline~$158\mu m$ emission line.}
\label{tab:cII}
\renewcommand{\arraystretch}{0.8}
\begin{tabular*}{\linewidth}{@{\extracolsep{\fill}} lccccccr @{}}
\hline\hline
ID(s) & $z_{{spec}}$ & Line & $S_{\mathrm{\CIIline}} \Delta v$ & $L_{\mathrm{\CIIline}}$ & $\rm{M_{H_{2}}}$ & $\sigma$ & Reference(s) \\
 &  &  & (Jy km s$^{-1}$) & ($10^{8}\rm{L}_{\odot}$) & ($10^{10}\rm{M}_{\odot}$) &  &  \\
(1) & (2) & (3) & (4) & (5) & (6) & (7) & (8)\\
\hline
DEIMOS\_COSMOS\_434239 & 4.4883 & CII158$\mu$m  & $2.314\pm0.273$ & $15.1\pm1.8$ & $4.52\pm0.53$ & -- & \cite{2020AA...643A...2B} \\
DEIMOS\_COSMOS\_493583 & 4.5134 & CII158$\mu$m  & $0.703\pm0.102$ & $4.61\pm0.67$ & $1.38\pm0.20$ & 3.09 & \cite{2020AA...643A...2B} \\
DEIMOS\_COSMOS\_494763 & 5.2337 & CII158$\mu$m  & $0.633\pm0.073$ & $5.21\pm0.60$ & $1.56\pm0.18$ & 3.24 & \cite{2020AA...643A...2B} \\
DEIMOS\_COSMOS\_539609 & 5.1818 & CII158$\mu$m  & $0.652\pm0.085$ & $5.29\pm0.69$ & $1.59\pm0.21$ & 2.50 & \cite{2020AA...643A...2B} \\
DEIMOS\_COSMOS\_627939 & 4.5341 & CII158$\mu$m  & $1.173\pm0.102$ & $7.75\pm0.67$ & $2.33\pm0.20$ & 3.18 & \cite{2020AA...643A...2B} \\
vuds\_cosmos\_5100969402 & 4.5785 & CII158$\mu$m & $0.832\pm0.087$ & $5.58\pm0.58$ & $1.67\pm0.18$ & 3.02 & \cite{2020AA...643A...2B} \\
vuds\_cosmos\_5101209780 & 4.5701 & CII158$\mu$m & $1.17\pm0.25$ & $7.83\pm1.67$ & $2.35\pm0.50$ & 5.86 & \cite{2020AA...643A...2B} \\
CRISTAL-02/ & 5.294 & CII158$\mu$m & $2.58\pm0.21$ & $21.6\pm1.8$ & $6.48\pm0.53$ & 5.17 & \cite{2025AA...699A..80H} \\
DEIMOS\_COSMOS\_848185/ &  &  &  &  &   &  &  \\
HZ6/ &  &  &  &  &   &  & \\
LBG-1 &  &  &  &  &   &  & \\
CRISTAL-04a/ & 4.52 & CII158$\mu$m & $0.86\pm0.08$ & $5.66\pm0.53$ & $1.70\pm0.16$ & 3.38 & \cite{2025AA...699A..80H} \\
vuds\_cosmos\_5100822662 &  &  &  &  &   &  & \\
CRISTAL-11/ & 4.439 & CII158$\mu$m & $0.97\pm0.08$ & $6.20\pm0.51$ & $1.86\pm0.15$ & 4.15 & \cite{2025AA...699A..80H} \\
DEIMOS\_COSMOS\_630594 &  &  &  &  &   &  & \\
CRISTAL-24/ & 4.546 & CII158$\mu$m & $5.7\pm0.42$ & $37.8\pm2.8$ & $11.3\pm0.84$ & 5.50 & \cite{2025AA...699A..80H} \\
DEIMOS\_COSMOS\_873756 &  &  &  &  &   &  & \\
CRISTAL-25/ & 4.573 & CII158$\mu$m & $3.35\pm0.43$ & $22.4\pm2.88$ & $6.73\pm0.86$ & 4.96 & \cite{2025AA...699A..80H} \\
vuds\_cosmos\_5101218326 &  &  &  &  &   &  & \\
S\_D\_C\_848185/ & 5.2965 & CII158$\mu$m & $11.57\pm0.65$ & $97.0\pm5.5$ & $29.1\pm1.6$ & 5.14 & \cite{2021AA...646A..76L} \\
S848185/ &  &  &  &  &   &  & \\
AzTEC-3 &  &  &  &  &   &  & \\
COSBO11-West/ & 1.82821 & CII158$\mu$m & $2.3\pm0.6$ & $3.28\pm0.86$ & $0.984\pm0.257$ & 3.53 & \cite{2025ApJ...987...61R} \\
J1000-satellite &  &  &  &  &   &  & \\
COS-3018555981 & 6.854 & CII158$\mu$m & $0.39\pm0.05$ & $4.79\pm0.61$ & $1.44\pm0.18$ & 7.84 & \cite{2018Natur.553..178S}\\
HZ9 & 5.541 & CII158$\mu$m & $1.95\pm0.07$ & $17.5\pm0.6$ & $5.25\pm0.19$ & 3.28 & \cite{2017ApJ...850..180J}(kinematics) \\
\hline
\end{tabular*}
\par
\vspace{6pt}
\begin{minipage}{\linewidth}
{\textbf{Notes.} (1)~Source ID(s). (2)~\CIIline-based spectroscopic redshift. (3)~Emission line. (4)~Velocity integrated CII158$\mu$m flux. (5)~CII158$\mu$m line luminosity. (6)~The molecular H$_{2}$ gas mass estimated assuming a \CIIline-to-H$_{2}$ conversion factor $\alpha_{\mathrm{\CIIline{}}}$ = 30$\rm{M}_{\odot}/\rm{L}_{\odot}$. (7)~Overdensity significance calculated by PPM. (8)~Reference(s) for the z$_{spec}$, Line and Velocity integrated CII158$\mu$m flux.}
\end{minipage}
\end{sidewaystable}